%% file: main-lipics.tex
\documentclass[a4paper,USenglish,cleveref,autoref,thm-restate, pdfa]{lipics-v2021}

\pdfoutput=1

\usepackage{pifont}
\usepackage{makecell}
\usepackage{mathtools}
\usepackage{stmaryrd}
\usepackage{tikz}
\usepackage{booktabs}

\nolinenumbers

\newtheorem{defn}{Definition}
\newtheorem{prop2}[theorem]{Theorem}
\newtheorem{sdefn}[defn]{Definition}
\newtheorem{slemma}[lemma]{Lemma}
\newtheorem{spropn}[theorem]{Theorem}
\newenvironment{sproof}{\begin{proof}}{\end{proof}}

\newcommand{\recrel}{intensional relation}
\newcommand{\restrictcollect}{RC}
\newcommand{\alias}{\alpha}
\newcommand{\rquery}{RQuery}
\newcommand{\rexpr}{RExpr}

\newcommand{\figurefontXLsize}{\fontsize{8pt}{9.6pt}\selectfont
}

\newcommand{\system}{TyQL}

\newcommand{\calc}{\texorpdfstring{$\lambda_{RQL}$}{lambda-RQL}}
\newcommand{\lsd}{\texorpdfstring{$\text{LSD-Datalog}^\neg$}{LSD-Datalog}}

\newcommand{\ir}{\texorpdfstring{$\lambda_{IR}$}{lambda-IR}}
\newcommand{\irenv}{\Pi}
\newcommand{\irtypingenv}{}
\newcommand{\tonrlsd}{\textbf{to-NRLSD}}
\newcommand{\datalogns}{$\text{Datalog}^{\neg s}$}
\newcommand{\rr}{P1}
\newcommand{\mono}{P2}
\newcommand{\mr}{P3}
\newcommand{\lin}{P4}
\newcommand{\set}{P5}
\newcommand{\cf}{P6}
\newcommand{\total}{6}
\newcommand{\totalF}{six}
\newcommand{\judgement}[1]{\vdash_{\kern-3pt\raisebox{0pt}{\scalebox{0.8}{$\scriptscriptstyle {#1}$}}}}

\newcommand{\rrJudge}{\judgement{\rr{}}}
\newcommand{\monoJudge}{\judgement{\mono{}}}
\newcommand{\mrJudge}{\judgement{\mr{}}}
\newcommand{\linJudge}{\judgement{\lin{}}}
\newcommand{\setJudge}{\judgement{\set{}}}
\newcommand{\cfJudge}{\judgement{\cf{}}}

\definecolor{dkgreen}{rgb}{0.0,0.6,0}
\definecolor{dkblue}{rgb}{0.0,0.0,0.8}
\definecolor{gray}{rgb}{0.5,0.5,0.5}
\definecolor{mauve}{rgb}{0.58,0,0.82}
\definecolor{darkred}{RGB}{120,0,45}
\definecolor{darkgreen}{RGB}{34,85,34}
\newcommand{\highlight}[1]{\textcolor{darkred}{#1}}
\definecolor{recursiveref}{rgb}{0.99, 0.58, 0.20}
\definecolor{sourceref}{rgb}{0.1, 0.7, 0.6}

\newif\ifborders
\bordersfalse

\ifborders
\newcommand{\bordercolor}{\color{red}}
  \newcommand{\borderrule}{0.4pt}
  \newcommand{\figbox}[1]{\fcolorbox{red}{white}{#1}}
\else
\newcommand{\bordercolor}{\color{black}}  \newcommand{\borderrule}{0pt}
  \newcommand{\figbox}[1]{#1}  \fi

\newcolumntype{L}[1]{>{\raggedright\arraybackslash}m{#1}}
\newcolumntype{C}[1]{>{\centering\arraybackslash}m{#1}}

\lstdefinestyle{tablescalastyle}{
  language=scala,
  xleftmargin=0mm,
  xrightmargin=0mm,
  aboveskip=0mm,
  belowskip=-3mm,
  fontadjust=false,
  columns=[c]fixed,
  keepspaces=true,
  tabsize=1,
  basicstyle=\scriptsize\ttfamily,
  commentstyle=\itshape\color{gray},
  keywordstyle=\bfseries\color{blue},
  mathescape=true,
  captionpos=b,
  framerule=0pt,
  numbersep=0mm,
  numbers=none,
  numberstyle=\tiny,
  showstringspaces=false,
  frame=single,
  backgroundcolor={},
  morekeywords={extension, given},
  otherkeywords={<:, <:<, &, =:=, =>},
  emphstyle={\color{recursiveref}},
  emphstyle=[2]{\color{sourceref}},
  emph=[2]{edges, base},
  emph=[1]{restrictedFix},
}

\lstdefinestyle{errorstyle}{
  xleftmargin=0mm,
  xrightmargin=0mm,
  aboveskip=0mm,
  belowskip=0mm,
  fontadjust=false,
  columns=[c]fixed,
  keepspaces=true,
  tabsize=1,
  basicstyle=\scriptsize\ttfamily\color{red},
  commentstyle=\itshape\color{gray},
  keywordstyle=\ttfamily\color{red}\underbar,
  mathescape=true,
  captionpos=b,
  framerule=0pt,
  numbersep=0mm,
  numbers=none,
  numberstyle=\tiny,
  showstringspaces=false,
  frame=single,
  backgroundcolor={},
  morekeywords={TyQL, ScalaSQL}
}

\lstdefinestyle{tablesqlstyle}{
  language=SQL,
  basicstyle={\scriptsize\ttfamily},
  xleftmargin=0mm,
  xrightmargin=0mm,
  aboveskip=0mm,
  belowskip=-3mm,
  breakatwhitespace=true,
  breaklines=true,
  classoffset=0,
  columns=flexible,
  commentstyle=\color{gray},
  framexleftmargin=0em,
  frameshape={}{}{}{},
  keywordstyle=\color{dkblue},
  numbers=none,
  numberstyle=\tiny\color{gray},
  showstringspaces=false,
  stringstyle=\color{mauve},
  tabsize=1,
  backgroundcolor={},
  morekeywords={TEXT, WITH},
  emphstyle={\color{recursiveref}},
  emphstyle=[2]{\color{sourceref}},
  emph=[2]{BasicParts,SubParts, Edges, Parents, Base},
  literate={'}{{\textquotesingle}}1,
  }

\lstdefinelanguage{RQL}{
  morekeywords={
    map,flatMap,filter,aggregate,groupBy,fix,
    union,unionAll,sum,avg,max,min, distinct, stringContains, stringConcat, count, table, False, True
  },
  sensitive=true,
  morestring=[b]",
  morecomment=[l]{\#},
}

\lstdefinestyle{lrqlnum}{
  language=RQL,
  numbers=left,
  numberblanklines=false,
  basicstyle=\scriptsize\itshape\fontfamily{lmr}\selectfont,
  keywordstyle=\bfseries\upshape,
  tabsize=1,
  literate={\ \ \ \ }{\ \ }1
               {->}{{$\rightarrow$}}2
               {==}{{$==$}}3
  ,
  columns=flexible,
  keepspaces=true,
  showstringspaces=false,
  xleftmargin=0mm,
  xrightmargin=0mm,
  aboveskip=0mm,
  belowskip=-3mm,
  backgroundcolor={},
  commentstyle=\color{gray},
  emphstyle={\color{recursiveref}},
  emphstyle=[2]{\color{sourceref}}, emph=[2]{
    BasicParts,SubParts, Edges, Parents, numbers, shares, control, assign, dereference, empty, newT, assign, store, loadT, friends, organizers, counts, term, lits, vars, abs, app, baseData, baseCtrl, base, edge, addressOf, assbl, basic, readOp, writeOp, jumpOp, baseHPT
  },
  stringstyle=\color{mauve},
}

\lstdefinestyle{lrql}{
  language=RQL,
  basicstyle=\scriptsize\itshape\fontfamily{lmr}\selectfont,
  keywordstyle=\bfseries\upshape,
  tabsize=1,
  literate={\ \ \ \ }{\ \ }1
               {->}{{$\rightarrow$}}2
               {==}{{$==$}}3
  ,
  columns=flexible,
  keepspaces=true,
  showstringspaces=false,
  xleftmargin=0mm,
  xrightmargin=0mm,
  aboveskip=0mm,
  belowskip=-3mm,
  backgroundcolor={},
  commentstyle=\color{gray},
  emphstyle={\color{recursiveref}},
  emphstyle=[2]{\color{sourceref}}, emph=[2]{
    BasicParts,SubParts, Edges, Parents, numbers, shares, control, assign, dereference, empty, newT, assign, store, loadT, friends, organizers, counts, term, lits, vars, abs, app, baseData, baseCtrl, base, edge, addressOf, assbl, basic, readOp, writeOp, jumpOp, baseHPT
  },
  stringstyle=\color{mauve},
}

\lstdefinestyle{lrql-ex}{
  language=RQL,
basicstyle=\scriptsize\itshape\fontfamily{lmr}\selectfont,
  keywordstyle=\bfseries\upshape,
  tabsize=1,
  literate=
{->}{{$\rightarrow$}}1
    {==}{{$==$}}2
  ,
  columns=flexible,
  keepspaces=true,
  showstringspaces=false,
  xleftmargin=0mm,
  xrightmargin=0mm,
  aboveskip=0mm,
  belowskip=-3mm,
  backgroundcolor={},
  commentstyle=\color{gray},
  emphstyle={\color{recursiveref}},
  emphstyle=[2]{\color{sourceref}}, emph=[2]{
    BasicParts,SubParts, Edges, Parents, numbers, shares, control, assign, dereference, empty, newT, assign, store, loadT, friends, organizers, counts, term, lits, vars, abs, app, baseData, baseCtrl, base, edge, addressOf, assbl, basic, readOp, writeOp, jumpOp, baseHPT
  }, stringstyle=\color{mauve},
}
\lstdefinelanguage{IR}{
  morekeywords={
    query,letrec,in,table,agg,filter,map,flatMap, False, True
  },
  sensitive=true,
  morestring=[b]",
}

\lstdefinestyle{ir}{
  language=IR,
  basicstyle=\scriptsize\itshape\ttfamily,
  keywordstyle=\bfseries\upshape,
  tabsize=1,
  literate={->}{{$\rightarrow$}}2
               {==}{{$==$}}3
               {t-ref}{{$\textbf{t\text{-}ref}$}}1
               {rt-ref}{{$\textbf{rt\text{-}ref}$}}1
               {r-table}{{$\textbf{r\text{-}table}$}}1
               {rec-query}{{$\textbf{rec\text{-}query}$}}1
               {a_1}{{$\alpha_1$}}2
               {r-table(a_1)}{{$\color{recursiveref}{\textbf{r-table}(\alpha_1)}$}}1
               {a_2}{{$\alpha_2$}}1
               {a_3}{{$\alpha_3$}}1
               {a_4}{{$\alpha_4$}}1
               {a_5}{{$\alpha_5$}}1
               {a_6}{{$\alpha_6$}}1
               {a_7}{{$\alpha_7$}}1
  ,
  columns=flexible,
  keepspaces=true,
  showstringspaces=false,
  xleftmargin=0mm,
  xrightmargin=0mm,
  aboveskip=0mm,
  belowskip=-3mm,
  backgroundcolor={},
  commentstyle=\color{gray},
  emphstyle={\color{recursiveref}},
  emphstyle=[2]{\color{sourceref}}, emph=[2]{Parents, Edges},
  stringstyle=\color{mauve},
}

\lstdefinestyle{datalog}{
  basicstyle=\scriptsize\ttfamily,
  keywordstyle=\bfseries,
  tabsize=1,
  morecomment=[l]{\#},
  literate={:-}{{$\text{:-}$}}2
               {==}{{$==$}}3
               {p_1}{{$p_1$}}1
               {p_2}{{$p_2$}}1
               {p_3}{{$p_3$}}1
               {p_4}{{$p_4$}}1
               {a_1}{{$\color{recursiveref}{\alpha_1}$}}1
               {"A"}{{\color{mauve}{"A"}}}3
               {"B"}{{\color{mauve}{"B"}}}3
  ,
  columns=flexible,
  keepspaces=true,
  showstringspaces=false,
  xleftmargin=0mm,
  xrightmargin=0mm,
  aboveskip=0mm,
  belowskip=-3mm,
  backgroundcolor={},
  commentstyle=\color{gray},
  emphstyle={\color{recursiveref}},
  emphstyle=[2]{\color{sourceref}}, emph=[2]{Parents, Edges},
  stringstyle=\color{mauve},
}

\title{Language-Integrated Recursive Queries}
\titlerunning{Language-Integrated Recursive Queries}

\author{Anna Herlihy}{EPFL, Lausanne, Switzerland}{anna.herlihy@epfl.ch}{https://orcid.org/0009-0005-8658-9569}{}

    \author{Amir Shaikhha}{University of Edinburgh, Edinburgh, Scotland}{amir.shaikhha@ed.ac.uk}{https://orcid.org/0000-0002-9062-759X}{}

    \author{Anastasia Ailamaki}{EPFL, Lausanne, Switzerland}{anastasia.ailamaki@epfl.ch}{https://orcid.org/0000-0002-9949-3639}{}

    \author{Martin Odersky}{EPFL, Lausanne, Switzerland}{martin.odersky@epfl.ch}{https://orcid.org/0009-0005-3923-8993}{}

\authorrunning{A. Herlihy et al.}

\Copyright{Anna Herlihy, Amir Shaikhha, Anastasia Ailamaki, and Martin Odersky}



  \relatedversion{The full version of the paper that includes the appendix is available on arXiv.}
  \relatedversiondetails[linktext={arXiv:2504.02443}]{Extended Version}{https://arxiv.org/abs/2504.02443}

\funding{
  This research was funded in part by the Swiss National Science Foundation (SNSF), grant number TMAG-2\_209506/1, ``Capabilities for Typing Resources and Effects'' and by the Swiss State Secretariat for Education, Research and Innovation (SERI) in the framework of the Prodasys project, approved by the European Research Council (ERC) as an Advanced Grant.
  This work was partly funded by a gift from RelationalAI and the Google Research Scholar Program.
}

\acknowledgements{We thank the anonymous reviewers, Oliver Bra\v{c}evac, and Guillaume Martres for the insightful feedback. }

\EventEditors{Robbert Krebbers and Alexandra Silva}
\EventNoEds{2}
\EventLongTitle{40th European Conference on Object-Oriented Programming (ECOOP 2026)}
\EventShortTitle{ECOOP 2026}
\EventAcronym{ECOOP}
\EventYear{2026}
\EventDate{June 29--July 3, 2026}
\EventLocation{Brussels, Belgium}
\EventLogo{}
\SeriesVolume{372}
\ArticleNo{5}

\begin{document}

\maketitle

\begin{abstract}
Performance-critical applications, including large-scale program analyses, graph analyses, and distributed system analyses, rely on fixed-point computations. The introduction of recursion using the WITH RECURSIVE keyword in SQL:1999 extended the ability of relational database systems to handle fixed-point computations, unlocking significant performance advantages by allowing computation to move closer to the data. Yet, with recursion, SQL becomes a Turing-complete programming language with new correctness and safety risks.

Full SQL lacks a fixed semantics, as the SQL specification is written in natural language with ambiguities that database vendors resolve in divergent ways. As a result, reasoning about the correctness of recursive SQL programs must rely on isolated, composable properties of queries rather than wrestling a unified formal model out of a language with notoriously inconsistent implementations across systems. To address these challenges, we propose a calculus, \calc{}, that derives properties from embedded recursive queries using the host-language type system and, depending on the database backend, rejects queries that may lead to the three classes of recursive query errors: runtime database exceptions, incorrect results, and nontermination. Queries that respect all properties are guaranteed to find the minimal fixed point in a finite number of steps.
We introduce \system{}, a practical implementation in Scala for safe, recursive language-integrated query. \system{} uses modern type system features of Scala 3, namely Named-Tuples and type-level pattern matching, to ensure query portability and safety.
\system{} shows no performance penalty compared to SQL queries expressed as embedded strings while enabling a three-order-of-magnitude speedup over non-recursive SQL.
\end{abstract}

\section{Introduction}
Fixed-point workloads are computed by repeatedly applying a query over the result of the previous iteration, until the result no longer changes.
SQL:1999~\cite{sql99} introduced the \texttt{WITH RECURSIVE} keyword, enabling fixed-point workloads to be executed within relational database management systems (RDBMS).
Recursive SQL unlocks significant performance gains, in part by pushing computation closer to the data~\cite{withrecursive-goto}. Ideally, applications with fixed-point computations would be able to leverage the billions of dollars invested in optimizing RDBMS, utilizing cutting-edge research and engineering advances without needing to be retrofitted into specialized, domain-specific systems, such as Datalog engines.

Despite excellent performance, the use of recursion within SQL is widely regarded as ``powerful and versatile but notoriously hard to grasp and master''~\cite{fixation}.
The difficulty is exacerbated by the lack of a well-defined formal semantics, as the SQL specification is written in natural language and is inconsistently followed across RDBMS implementations. Each database vendor adopts its own interpretation, leading to differences in evaluation order, type coercion, null handling, and even fundamental query behavior~\cite{saneql}. As a result, there is no single type-safe or semantically well-defined compilation target for SQL, only a patchwork of best-effort approximations that either make simplifying assumptions or target a limited subset of SQL~\cite{formalsql1991, formalsql2017, qex2010, hottsql, verifiedsql2010, sqlnulls2022}, none widely applied industrially.

Language-integrated query addresses inconsistencies across RDBMS from within the programming language, providing safety and portability
by allowing users to write their query once and have the language compiler type-check and specialize it for multiple backends.
The most successful language-integrated query library is the LINQ framework for .NET~\cite{linqxml}, and the concepts have been widely adopted across modern programming languages and databases~\cite{slick, quill, jslinq, diesel}. Yet neither LINQ nor other language-integrated SQL libraries support recursion.
With the addition of \texttt{WITH RECURSIVE}, SQL becomes a Turing-complete programming language, and with that comes a host of new concerns orthogonal to classic language-integrated query problems like nested queries or data-type safety.
Moreover, the support and implementation of recursive queries vary more widely across commercial RDBMS compared to other SQL features~\cite{rsql}.
As a result, the most effective and practical way to reason about recursive SQL requires adapting to different RDBMS semantics.

\begin{figure}[t]
\centering
\includegraphics[trim={0.08cm .2cm .08cm 0.11cm}, clip, width=\linewidth]{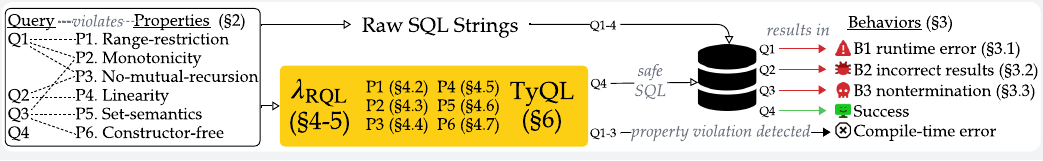}
  \caption{
{
        Overview of recursive query safety.
        Queries Q1--Q3 that violate at least one of properties P1--P\protect\total{} may lead to behaviors B1--B3.
        \system{} rejects unsafe queries; only safe queries generate SQL.
        }
    }\label{fig:overview}
\end{figure}

The key insight of this work is to provide a DBMS-agnostic reasoning framework over recursive queries based on independently applicable and composable mathematical properties of fixed-point computations that determine how queries will behave: \textit{range-restriction}, \textit{monotonicity}, \textit{mutual-recursion}, \textit{linearity}, \textit{set-semantics}, and \textit{constructor-freedom}.
We propose a solution based on language-integrated query that constrains queries at application compile-time, enabling significant performance benefits while generating only the safe subset of SQL that will not trigger runtime exceptions, nontermination, or incorrect results.
{Figure~\ref{fig:overview} provides an overview of our approach.
    Six properties (Section~\ref{sec:background}) determine safe recursion;
    violations lead to unwanted behaviors (Section~\ref{sec:motivation}).
    Queries Q1--Q3 violate properties and exhibit unwanted behaviors on the RDBMS (top path), while Q4 satisfies all properties and succeeds.
    Our approach (bottom path, Sections~\ref{sec:formalization}-\ref{sec:implementation})
    automatically derives the relevant properties and, depending on the backend, rejects unsafe queries at compile-time.
    Only correct SQL is generated and run.}
Figure~\ref{fig:example} shows a classic recursive database query for the transitive closure of a directed graph in \system{}, our language-integrated query library for Scala, and in SQL.
This paper makes the following contributions:
\begin{itemize} \item Previous work in language-integrated query did not support recursive SQL.
    With the addition of a fixed-point operator, SQL becomes a Turing-complete programming language, leading to new and complex challenges for query correctness and safety.
    In Section~\ref{sec:background} we provide background on recursive database queries and in Section~\ref{sec:motivation} we identify the three classes of ways that recursive queries show unsafe behavior: runtime exceptions, incorrect results, and nontermination, and show the mathematical properties of queries that are responsible for each behavior.
    \item  We present a calculus, \calc{}, that automatically checks the \totalF{} properties responsible for these classes of errors at the host-language compile-time and will always generate a single SQL query. Sections~\ref{sec:formalization} and~\ref{sec:semantics} show how \calc{} uses type classes, linear types, and union types to independently encode constraints for each targeted query property so that the generated query is specialized to the semantics of the database backend.
    When fully restricted, \calc{} is guaranteed to find the unique and minimal fixed point in a finite number of steps (Theorem~\ref{theprop}).
    \item We propose \system{}: practical language-integrated queries in Section~\ref{sec:implementation}. \system{} users express their data model using Named-Tuples, enabling efficient implementation within the programming language and straightforward error messages.
\item In Section~\ref{sec:eval} we propose a benchmark of recursive queries adapted from industrial benchmarks and academic works in the domains of recursive SQL, Datalog, and Graph Database Systems. We conduct a survey of modern RDBMS with respect to query behavior.
    We then evaluate \system{} with regard to query coverage and the performance of \system{} with alternative approaches, showing no performance penalty compared to raw SQL strings and a three-order-of-magnitude speedup over state-of-the-art language-integrated query libraries using non-recursive SQL queries with an in-memory embedded database. \end{itemize}

\begin{figure}[t]
  \setlength{\fboxsep}{0pt}
  \begin{minipage}[t]{0.5\linewidth}
    \begin{lstlisting}[style=tablescalastyle, emph={path},
        basicstyle=\footnotesize\ttfamily,
        frame=single,
        framesep=3pt,
        framerule=\borderrule,
        rulecolor=\bordercolor,
        aboveskip=0pt,
        belowskip=0pt,
      ]
case class Edge(x: Int, y: Int)
val edges = Table[Edge]("edges")
edges.fix(path =>
  path.flatMap(p =>
    edges
      .filter(e => p.y == e.x)
      .map(e => (x = p.x, y = e.y))))\end{lstlisting}
\end{minipage}
  \hfill
  \begin{minipage}[t]{0.43\linewidth}
    \begin{lstlisting}[style=tablesqlstyle, emph={path},
      basicstyle=\footnotesize\ttfamily,
      frame=single,
      framesep=3pt,
      framerule=\borderrule,
      rulecolor=\bordercolor,
      aboveskip=0pt,
      belowskip=0pt,
    ]
CREATE TABLE edges (x INT, y INT);
WITH RECURSIVE path AS (
  SELECT * FROM edges
    UNION ALL
  SELECT p.x, e.y FROM path p, edges e
  WHERE p.y = e.x);
SELECT * FROM path\end{lstlisting}
\end{minipage}
  \caption{Recursive query for the transitive closure in \protect\system{} and SQL}
  \label{fig:example}
\end{figure}
  \section{Background on Recursion in Databases}\label{sec:background}
Modern applications demand increasingly sophisticated query capabilities beyond the limits of traditional select-project-join-aggregate queries. Performance-critical industrial applications, including large-scale program analyses, network analyses, artificial intelligence, and distributed system analyses, rely on fixed-point computations~\cite{datalogrust, awsdl, bigdata, logicai, vadalog, drivingdatalog}.

\texttt{WITH RECURSIVE} was added to the SQL standard in 1999~\cite{sql99}.
Prior, hierarchical or recursive relationships (e.g., organizational charts, family trees, etc.) required use of Datalog{~\cite{dl}}, application-side iteration on data extracted from the database, or non-standard SQL extensions, e.g., Oracle's \texttt{CONNECT BY} keyword or procedural extensions like PL/SQL. Yet repeated round-trips between the application and database or context switches between procedural and plain SQL have significant overhead compared to a single recursive query~\cite{withrecursive-goto}.

Recursive queries define \textit{\recrel{}s} (i.e., recursively defined relations) and are composed of a base-case query and a recursive-case query that references the \recrel{}. Figure~\ref{fig:example} {shows an example reachability query that defines an \recrel{} named ``path'': the base case is the ``edges'' relation, and the recursive case joins the ``edges'' and ``path'' relations, producing the transitive closure of all edges}.

Since 1999, databases have added more and more powerful and expressive support for recursion.
Table~\ref{tab:feature-matrix} shows feature support across several modern RDBMS.
\textcolor{red}{\ding{55}} indicates that the feature is unsupported and will throw a relevant error message. \textcolor{orange}{\huge\textbullet} indicates \emph{syntactic} support for a feature, namely, the query will not throw an error but the system may not necessarily have the requisite internal implementation to execute the query successfully.

\begin{table}[t]
\caption{
Support for Recursion on Modern RDBMS.
\textcolor{red}{\ding{55}} Prevented
\textcolor{orange}{\textbullet} Syntactically OK
\textcolor{dkgreen}{\ding{51}} Supported
}\label{tab:feature-matrix}
\centering
\setlength{\tabcolsep}{2pt}
\begin{tabular}{|l|c|c|c|c|c|c|c|
}
\hline
\multicolumn{1}{|c|}{} & \makecell{\texttt{\textbf{WITH}}\\[-2pt]\texttt{\textbf{RECURSIVE}}} & \makecell{\textbf{Range-Rest}\\[-2pt]\textbf{(\rr{})}} & \makecell{\textbf{Agg}\\[-2pt]\textbf{(\mono{})}} & \makecell{\textbf{Mutual}\\[-2pt]\textbf{(\mr{})}} & \makecell{\textbf{Non-linear}\\[-2pt]\textbf{(\lin{})}} & \makecell{\textbf{Set}\\[-2pt]\textbf{(\set{})}}
& \makecell{\textbf{Const-Free}\\[-2pt]\textbf{(\cf{})}}\\ \hline

\textbf{MySQL}
& \textcolor{dkgreen}{\ding{51}} (2017) & \textcolor{dkgreen}{\ding{51}} (2017)
& \textcolor{red}{\ding{55}}
& \textcolor{red}{\ding{55}}  & \textcolor{orange}{\textbullet} (2017) & \textcolor{dkgreen}{\ding{51}} (2017) & \textcolor{dkgreen}{\ding{51}} (2017)
\\ \hline

\textbf{OracleDB}
& \textcolor{dkgreen}{\ding{51}} (2009) & \textcolor{dkgreen}{\ding{51}} (2009)
& \textcolor{red}{\ding{55}}
& \textcolor{red}{\ding{55}}
& \textcolor{red}{\ding{55}}
& \textcolor{red}{\ding{55}}
& \textcolor{dkgreen}{\ding{51}} (2009)
\\ \hline

\textbf{PostgreSQL}
& \textcolor{dkgreen}{\ding{51}} (2009) & \textcolor{dkgreen}{\ding{51}} (2009)
& \textcolor{red}{\ding{55}}
& \textcolor{red}{\ding{55}}
& \textcolor{red}{\ding{55}}
& \textcolor{dkgreen}{\ding{51}} (2009) & \textcolor{dkgreen}{\ding{51}} (2009)
\\ \hline

\textbf{SQL Server}
& \textcolor{dkgreen}{\ding{51}} (2005) & \textcolor{dkgreen}{\ding{51}} (2005)
& \textcolor{red}{\ding{55}}
& \textcolor{red}{\ding{55}}
& \textcolor{orange}{\textbullet} (2008)
& \textcolor{red}{\ding{55}}
& \textcolor{dkgreen}{\ding{51}} (2005)
\\ \hline

\textbf{SQLite}
& \textcolor{dkgreen}{\ding{51}} (2014) & \textcolor{dkgreen}{\ding{51}} (2014)
& \textcolor{red}{\ding{55}}
& \textcolor{red}{\ding{55}} & \textcolor{orange}{\textbullet} (2020) & \textcolor{dkgreen}{\ding{51}} (2014) & \textcolor{dkgreen}{\ding{51}} (2014)
\\ \hline

\textbf{MariaDB}
& \textcolor{dkgreen}{\ding{51}} (2017) & \textcolor{dkgreen}{\ding{51}} (2017)
& \textcolor{red}{\ding{55}}
& \textcolor{dkgreen}{\ding{51}} (2017)
& \textcolor{dkgreen}{\ding{51}} (2017)
& \textcolor{dkgreen}{\ding{51}} (2017)
& \textcolor{dkgreen}{\ding{51}} (2017)
\\ \hline

\textbf{DuckDB}
& \textcolor{dkgreen}{\ding{51}} (2020) & \textcolor{dkgreen}{\ding{51}} (2020)
& \textcolor{dkgreen}{\ding{51}} (2020)
& \textcolor{orange}{\textbullet} (2020)
& \textcolor{orange}{\textbullet} (2020)
& \textcolor{dkgreen}{\ding{51}} (2020)
& \textcolor{dkgreen}{\ding{51}} (2020)
\\ \hline

\end{tabular}
\end{table}

 \begin{figure}[t]
    \centering
    \begin{minipage}{.335\linewidth}
        \centering
\begin{lstlisting}[style=tablesqlstyle, emph={waitFor},
          basicstyle=\footnotesize\ttfamily,
          frame=single,
          framesep=3pt,
          framerule=\borderrule,
          rulecolor=\bordercolor,
          aboveskip=0pt,
          belowskip=0pt,
          breaklines=false
        ]
WITH RECURSIVE WaitFor AS
  (SELECT part, days
  FROM BasicParts
    UNION
  SELECT sp.part, MAX(wf.days)
  FROM SubParts sp, WaitFor wf
  WHERE sp.sub = wf.part
  GROUP BY sp.part);
SELECT * FROM WaitFor
\end{lstlisting}
\subcaption{\textbf{Non-monotonic query}:\\ \centering Bill-of-materials (Q1)}\label{fig:ex-sql-monotonic}
    \end{minipage}
    \hfill
    \begin{minipage}{.28\linewidth}
        \centering
\begin{lstlisting}[style=tablesqlstyle, emph={path},
          basicstyle=\footnotesize\ttfamily,
          frame=single,
          framesep=3pt,
          framerule=\borderrule,
          rulecolor=\bordercolor,
          aboveskip=0pt,
          belowskip=0pt,
          breaklines=false
        ]
WITH RECURSIVE Path AS
  (SELECT * FROM Edges
    UNION
(WITH Path as --Postgres
    (SELECT * FROM Path)
  SELECT p1.x, p2.y
  FROM Path p1, Path p2
  WHERE p1.y = p2.x));
SELECT * FROM Path
\end{lstlisting}
\subcaption{\textbf{Non-linear query}: \\ \centering Transitive closure (Q2)}\label{fig:ex-sql-linear}
    \end{minipage}
    \hfill
    \begin{minipage}{.355\linewidth}
        \centering
\begin{lstlisting}[style=tablesqlstyle, emph={gens},
          basicstyle=\footnotesize\ttfamily,
          frame=single,
          framesep=3pt,
          framerule=\borderrule,
          rulecolor=\bordercolor,
          aboveskip=0pt,
          belowskip=0pt,
          breaklines=false
        ]
WITH RECURSIVE Gens AS
  (SELECT p.ch as nm, 1 as g
  FROM Parents p WHERE p.par='A'
    UNION ALL
SELECT p.ch as nm, g.g+1 as g
  FROM Parents as p, Gens as g
  WHERE p.par = g.nm);
SELECT * FROM Gens
WHERE Gens.g = 2
\end{lstlisting}
 \centering
\subcaption{\textbf{Bag-semantic query: }\\ \centering Same-Generation (Q3)}\label{fig:ex-sql-bag}
    \end{minipage}
    \caption{Examples of Dangerous Recursive Queries}\label{fig:example-sql-3x}
\end{figure}

\textit{Range-restriction}.
The ``Range-Rest'' column in Table~\ref{tab:feature-matrix} indicates that queries must be range-restricted.
The {database theory} literature defines the property of \textit{range-restriction}~\cite{dl} {that, when }applied to recursive SQL, requires the project clause to contain only constants or references to columns present in the FROM clause. {For example, the query \texttt{SELECT z FROM edges} on the \texttt{edges} relation defined in Figure~\ref{fig:example} would be rejected because there is no column \texttt{z}.}
{Range-restriction is a basic syntactic requirement that all RDBMS enforce.}

\textit{Monotonicity}.
The ``Agg'' column indicates support for aggregation operations within the bodies of recursive queries.
{
  For example, Figure~\ref{fig:ex-sql-monotonic} shows a query on the widely-used Bill-of-Materials domain~\cite{rasql} that models items sold by a business that are made out of sub-parts. The relation \texttt{SubParts(part, sub)} models each item a business sells and its sub-parts (and sub-sub-parts, etc.); and the relation \texttt{BasicParts(part, days)} models base parts and how many days it takes to arrive from a supplier. The query determines when a part will be ready, given it is the day the last subpart arrives.
  The \texttt{MAX} aggregation is applied to the \recrel{} \texttt{WaitFor}, therefore the query contains aggregation within the body of the recursive query.
  }
{Aggregation between distinct recursive queries is called \textit{stratified} aggregation. For example, if the \texttt{MAX} operation in the query in Figure~\ref{fig:ex-sql-monotonic} was applied to the \texttt{BasicParts} relation or an \recrel{} defined in a separate \texttt{WITH RECURSIVE} call, the aggregation would be stratified.}
All the RDBMS in Table~\ref{tab:feature-matrix} support stratified aggregation, while DuckDB is more expressive and supports unstratified aggregation.

\textit{Mutual-recursion}.
The ``Mutual'' column indicates support for mutually recursive queries, where two or more \recrel{}s refer to each other in a cyclic dependency,
useful for expressing many static analyses or bidirectional graph traversals.
{For example, the query \texttt{WITH RECURSIVE a AS b, b AS a} defines two \recrel{}s \texttt{a} and \texttt{b} that are mutually recursive.}
Recently, MariaDB added support for mutual recursion, and DuckDB inlines relations such that it is possible to express some mutually recursive queries.

\textit{Linearity}.
The ``Non-linear'' column indicates support for non-linear recursive queries, in which an \recrel{} is referenced more than once within a recursive query.
{For example, Figure~\ref{fig:ex-sql-linear} shows the non-linear version of the query shown in Figure~\ref{fig:example}, as it contains the \recrel{} \texttt{Path} twice in the \texttt{FROM} clause.}
Non-linear queries are particularly useful for program analysis queries~\cite{graspan, flix, flan}. PostgreSQL applies a simple and easily avoided syntactic check, DuckDB does not check, and MariaDB supports non-linear queries.

\textit{Set-semantics}. The ``Set'' column indicates support for the UNION operator to combine the base and recursive-cases, which uses set semantics. Without UNION, the default is UNION ALL, which uses bag (i.e., multiset) semantics.
{
  For example, Figure~\ref{fig:ex-sql-bag} shows a query on a parent-child ancestry database that finds all descendants of a person (\texttt{`A'}) that are of the same generation (2nd). The query is bag-semantic because it uses \texttt{UNION ALL} to combine the base and recursive cases.
}
Some RDBMS, e.g., OracleDB, require the use of bag semantics, while others also allow set semantics.

\textit{Constructor-freedom}. The ``Const-Free'' column indicates support for queries that violate the constructor-freedom property. The {database theory} literature refers to queries that do not contain interpreted functions over infinite domains (e.g., integer arithmetic) as having the \textit{constructor-freedom} property~\cite{datafun}.
{For example, the \texttt{g+1} clause of the example query in Figure~\ref{fig:ex-sql-bag} causes the query to violate constructor-freedom.}
All RDBMS support constructors.

As illustrated in Table~\ref{tab:feature-matrix}, the technical landscape is wide and ever-growing, and no two modern databases support the same set of features.
The lack of alignment among commercial RDBMS renders a single, unified formal semantics untenable as a foundation for correctness checking of recursive queries; practical systems must instead support a flexible and composable framework that adapts to the specific capabilities of each backend.

\section{How Do Recursive Queries ``Go Wrong''?}\label{sec:motivation}
In this section, we classify the problems that arise with recursive queries into three areas based on the emergent database behavior
and define the mathematical query properties responsible for each class of error. As illustrated in Figure~\ref{fig:overview}, the three behaviors targeted are runtime exception (B1),
incorrect results (B2),
and nontermination (B3).
We define \totalF{} properties: range-restriction (\rr{}), monotonicity (\mono{}), no-mutual-recursion (\mr{}), linearity (\lin{}), set-semantics (\set{}), and constructor-freedom (\cf{}), and show how each behavior B1--B3 may result from violating one or more of P1--P\total{}, providing an example (shown in Figure~\ref{fig:example-sql-3x}) for each of the placeholder queries Q1--Q3 in
 Figure~\ref{fig:overview}. Lastly, we discuss the cases where a user may want to selectively apply or relax the restriction of each property.

\subsection{Recursive Query Runtime Exception (B1)}\label{sec:recursion:monotone}

Language-integrated query targets many problems associated with query writing: datatype or schema errors, for example, falsely assuming a table to have a particular column name or data type;
syntactic errors like simple typos;
security vulnerabilities such as SQL injection; and structural mistakes like \texttt{HAVING} without \texttt{GROUP BY}. Without language integration, the RDBMS query compiler will identify these errors and throw exceptions, a runtime error for the application. These types of errors are already well-covered by existing language-integrated query techniques that are complementary to our approach~\cite{tlinq}. However RDBMS query compilers throw recursion-specific exceptions that are not caught by existing techniques.

\textit{Example:} the query in Figure~\ref{fig:ex-sql-monotonic} contains a \texttt{MAX} aggregation in the body of the recursive query.
This constraint is usually checked by the query compiler, which can only be invoked at application runtime.
The emergent behavior is a runtime error for the application, as it must wait for the round-trip time for the query to be sent to the database and the error returned.
Aggregations are restricted because to guarantee that the unique and minimal fixed point will be found in a finite number of steps, operations within recursive queries must be \emph{monotonic} under the ordering of set inclusion. A monotonic query is defined as a query $Q$ and databases $D1$ and $D2$, such that if $D1 \subset D2$ then $Q(D1) \subset Q(D2)$~\cite{amateur}.
{In other words, a} query is monotonic if adding more data to its input does not remove data from its output, and negation operations like \texttt{NOT EXISTS} and aggregations like \texttt{MAX} can violate this property even if they are considered monotonic with respect to other orders.

Most widely-used commercial RDBMS officially support only monotonic operations within recursion, so for the query in Figure~\ref{fig:ex-sql-monotonic} to pass a database query compiler check, the \texttt{MAX} aggregation must be moved out of the recursive query.
Therefore, it is the violation of property \textbf{\mono{}: monotonicity} that leads to behavior \textbf{B1: recursive query runtime exception} on the systems that do not support non-monotonic recursion.
Queries that violate \textbf{\rr{}: range-restriction} will also always cause exceptions.
Some RDBMS query compilers check for mutual or non-linear recursion, so queries that violate property \textbf{\mr{}: no-mutual-recursion} or \textbf{\lin{}: linearity} may throw an exception. In Figure~\ref{fig:overview}, queries that violate \rr{}--P4 and are checked by the query compiler are represented by query Q1.

However, recent advances in recursive query engines have shown that some forms of aggregation~\cite{rasql, datalogo} are permissible within recursion without losing termination guarantees. While this has not yet been implemented widely in commercial systems, it shows when a user may want to ``turn off'' the monotonicity constraint for certain backends.

\subsection{Incorrect Results (B2)}\label{sec:recursion:linear}
The SQL standard defines a \textit{linear query} to be a query that references each \recrel{} once. Around the time that SQL'99 was written, there was a belief that ``most `real life' recursive queries are indeed linear''~\cite{amateur}. While this belief is no longer widely held, this assumption is built into the implementation of many RDBMS.

\textit{Example:} Figure~\ref{fig:ex-sql-linear} shows an example of a non-linear query, where the \recrel{} \texttt{path} is referenced twice in the body of the query. This query is represented by Q2 in Figure~\ref{fig:overview}.
The reason why RDBMS may return incorrect results is an internal optimization that works only for linear queries and is best illustrated with {the example SQL query in Figure~\ref{fig:ex-sql-linear}}.
Given a 3-step input graph, i.e. \texttt{\{(0, 1), (1, 2), (2, 3)\}} this query returns a result with 5 rows (the input edges plus \texttt{\{(0, 2), (1, 3)\}}) in PostgreSQL (v15) and DuckDB (v1.2).
Quickly stepping through the graph shows that \texttt{3} is reachable from \texttt{0}, so the result returned from the RDBMS is incorrect.
The reason for this is {an internal database optimization where at each iteration, the results are computed by only reading data returned by the previous iteration, causing the algorithm to terminate before returning all results for non-linear queries.}
Terminating early may cause the RDBMS to return only partial results, or when nested within an outer query can lead to fully incorrect results.

The SQL specification defines behavior only for queries that are linearly recursive, leaving the behavior of non-linear queries undefined. Some RDBMS attempt to reject non-linear queries by limiting references to the \recrel{}s to only once within the recursive subquery. However this is a purely syntactic restriction, so simple aliasing can evade this check, while other databases do not check query linearity at all and allow non-linear queries to execute and silently return incorrect results. Systems that allow mutual recursion and implement it via inlining will show similar behavior and return incorrect results.
Therefore, it is the violation of properties \textbf{\lin{}: linearity} or \textbf{\mr{}: no-mutual-recursion} that leads to behavior \textbf{B2: incorrect results} on systems that allow non-linear or mutual-recursion.

As there are modern databases that do not perform checks for \lin{} and \mr{}, it is of utmost importance that users be prevented from unknowingly sending queries that will silently return invalid results. However, recently MariaDB added support for mutual and non-linear queries~\cite{mariadb}, so a user may wish to turn off this constraint depending on their RDBMS.

\subsection{Nontermination (B3)}\label{sec:recursion:set}
\texttt{WITH RECURSIVE} expands the expressive power of SQL to be Turing-complete, so it is possible to express infinitely recurring queries. Recursive queries can be computationally expensive, and users may be left unsure if their queries will eventually terminate, given enough time, or never terminate. Assuming the SQL specification is followed, that is, all queries are monotonic and linear, nontermination is a consequence of infinitely growing relations.

\textit{Example:}
The bottom-up evaluation assumes \textit{set semantics} (\set{}), which prevents duplicate tuples from causing the intermediate relations to grow infinitely. Yet RDBMS use bag semantics unless otherwise specified, so a reachability query using bag semantics over data containing cycles will repeatedly re-discover the same paths, leading to infinitely growing relations. Only certain operators like \texttt{DISTINCT} or \texttt{UNION} use set semantics.
For example, the query in Figure~\ref{fig:ex-sql-bag} applied to a dataset that contains a cycle from a data-loading error
e.g., \texttt{\{(A, B), (B, A)\}}, will infinitely recur.
Relations can also grow infinitely from non-duplicate tuples if the domain of the query is not finite. SQL queries are not limited to a finite domain, as column-level ``constructor'' operations like addition or string concatenation can arbitrarily introduce new elements. The query in Figure~\ref{fig:ex-sql-bag} contains the \texttt{+} operator, which can produce values not in the input domain and can cause nontermination.

Therefore, it is the violation of properties \textbf{\set{}: set-semantics}, or \textbf{\cf{}: constructor-freedom} that leads to behavior \textbf{B3: nontermination}. On systems that do not reject non-monotonic queries, property \textbf{\mono{}: monotonicity} can also lead to nontermination.
Nonterminating queries that violate these properties are represented by Q3 in Figure~\ref{fig:overview}.

Set-semantic recursive queries are guaranteed to generate only finite relations, given they are {constructor-free and range-restricted}. Yet not all queries require set semantics to terminate, although given only the query, it is undecidable whether a query will terminate when using less restrictive duplicate checks~\cite{DBLP:conf/slp/MaherR89}. If the data does not contain cycles, then the user may prefer to use bag semantics to avoid wasted time on duplicate elimination. Other RDBMS \emph{require} the use of \texttt{UNION ALL} between the base and recursive case, for example SQL Server and OracleDB, and provide alternative language structures to check for infinite recursion (e.g. \texttt{CYCLE} in OracleDB and PostgreSQL v14+ checks if there is a cycle in the query results). Users may also wish to turn off the constructor-freedom constraint, as not all constructors lead to infinitely growing domains. Given the high penalty of infinite recursion, both on the application and the RDBMS, as other users may see cross-query interference, users must be able to prevent nonterminating queries.

 \section{Safe Recursion with \calc{}}\label{sec:formalization}
In this section, we present the \calc{} calculus for recursive queries.
To avoid generating unsafe SQL queries that cause behaviors B1--B3, we must design our system to reject queries that violate properties P1--P\total{}. Yet, due to the fractured support for recursion in relational databases, as well as the cases where a user may want to relax certain constraints even for a single RDBMS, for our system to be practically and immediately useful it must be possible to pick-and-choose \textit{independently} and \textit{composably} which properties should be constrained at any given time. For example, a user of PostgreSQL could choose to constrain mutual-recursion, linearity, and monotonicity but leave the constructor-freedom and set-semantics properties unconstrained. Another user should be able to allow bag semantics for queries where they are certain there are no cycles in the data, and set semantics for when there are cycles. \textbf{Constraint-independence} informs every aspect of the design and evaluation of our approach.
{We first show how the query properties P1--P\total{} are encoded into {a family of} DSL type systems independently in Sections~\ref{sec:formalization:base}-\ref{sec:formalization:cf} and then show how to extend \calc{} to a 2-level DSL and host language type system in Section~\ref{sec:formalization:embedding}}.

\subsection{Syntax and Base Type System \calc{}}\label{sec:formalization:base}

The base of \calc{} extends T-LINQ~\cite{tlinq}.
T-LINQ, as well as the Nested Relational Calculus (NRC)~\cite{NRC}, established the structural equivalence between SQL-style {\small\texttt{SELECT-FROM-WHERE}} and combinator-style \texttt{flatMap}, \texttt{filter}, \texttt{map}, etc. expressions.  In \system{}, for-comprehensions desugar to combinators, {to keep the syntax similar to the implementation, which is designed to be Scala-like}. We do not include for-comprehensions in \calc{} to keep the syntax minimal. The recursive query syntax is built on previous work on the NRC with a bounded fixpoint~\cite{fix-b}.

\begin{figure}[t]
    \centering
    \small
\setlength{\fboxsep}{0pt}
    \figbox{\begin{minipage}[t]{\textwidth}
        \setlength{\jot}{1.5pt}
          \setlength{\arraycolsep}{0pt}
        \underline{\textbf{Syntax}}\\[2pt]
        % $\begin{array}{@{}>{$}p{1cm}<{$}@{\quad}l@{\ }c@{\ }l@{}}
        $\begin{array}{@{}>{$}p{1cm}<{$}>{$}p{1.3cm}<{$}>{$}p{0.5cm}<{$}>{$}p{11cm}<{$}@{}}
            \textit{(vars)}  & x                                  &     & \\
            \textit{(dbs)}   & \textit{db}                        &     & \\
            \textit{(cnst)} & c                                  & ::= & \textit{number} \mid \textit{boolean} \mid \textit{string} \\
            \textit{(col)}   & \textit{O}                         & ::= & \text{Int} \mid \text{Bool} \mid \text{String} \\
            \textit{(row)}   & \textit{A}, \textit{B}, \textit{E} & ::= & (l_i: \textit{O}_i){_{i=1}^n} \\
            \textit{(res)}   & \textit{R}                         & ::= & \text{Query}[\textit{A}] \mid \text{Agg}[\textit{A}] \\
            \textit{(type)}  & \textit{T}, \textit{V}             & ::= & \textit{O} \mid \textit{A} \mid \textit{R} \mid \textit{T} \rightarrow \textit{V} \mid (\textit{O}_{i}){_{i=1}^n} \mid (l_i \colon \textit{T}_{i}){_{i=1}^n} \\
            \textit{(term)}  & m, q, r, f                         & ::= & c \,|\, (x) \rightarrow m \,|\, (l_i\! =\! m_i){_{i=1}^n} \,|\, m.l \,|\, (m_i){_{i=1}^n} \,|\, m.i \,|\, m\textbf{++}r \,|\, p \,|\, \textbf{table}(\textit{db}) \,|\, \textit{op}\ m \\
            \textit{(cmb)}   & p                                  & ::= & \textbf{map}(q,f)\,|\, \textbf{flatMap}(q,f)\,|\, \textbf{filter}(q,f)\,|\, \textbf{agg}(q,f)\,|\, \textbf{groupBy}(q,f,m,r)
            \\ &&&
            \,|\, \textbf{fix}(q,f) \\
        \end{array}$\\[4pt]
        \underline{\textbf{Syntax $\Sigma$: }\text{Example entries for }\textit{op}}\\[2pt]
        $\begin{array}{@{}>{$}p{1.1cm}<{$}>{$}p{1.3cm}<{$}>{$}p{0.5cm}<{$}>{$}p{8cm}<{$}@{}}
            &\textit{exprOp} & ::= & m + r \mid m\ \&\&\ r \mid \textbf{sum}(m) \mid \dots \\
            &\textit{relOp}  & ::= & \textbf{union}(m,\ r) \mid \text{\textbf{unionAll}}(m,\ r) \mid \dots \\
        \end{array}$
        \end{minipage}}
    \caption{Syntax of \calc{} DSL Types and Terms and Example Contents of Signature $\Sigma$}\label{fig:syntax-sig}
\end{figure}
\begin{figure}[t]
    \small
    \begin{minipage}[t]{.325\linewidth}
\begin{lstlisting}[style=lrql, emph={waitFor},
          basicstyle=\footnotesize\itshape\fontfamily{lmr}\selectfont,
          frame=single,
          framesep=3pt,
          framerule=\borderrule,
          rulecolor=\bordercolor,
          aboveskip=0pt,
          belowskip=0pt,
          breaklines=false
        ]
fix(table(BasicParts),
    (waitFor) ->
        distinct(agg(table(SubParts),
          (sp) -> groupBy(
              filter(waitFor, (wf) ->
                  sp.sub == wf.part),
              (wf) -> sp.part,
              (wf) ->
                  (part=sp.part,
                     days=max(wf.days)),
              (wf) -> true))))._1
\end{lstlisting}
\subcaption{\textbf{Non-monotonic query}}\label{fig:ex-calc-monotonic}
    \end{minipage}\hfill
    \begin{minipage}[t]{.28\linewidth}
\begin{lstlisting}[style=lrql, emph={pathR},
          basicstyle=\footnotesize\itshape\fontfamily{lmr}\selectfont,
          frame=single,
          framesep=3pt,
          framerule=\borderrule,
          rulecolor=\bordercolor,
          aboveskip=0pt,
          belowskip=0pt,
          breaklines=false]
fix(table(Edges),
    (pathR) ->
        distinct(flatMap(pathR,
          (p1) ->  map(
              filter(pathR,
                  (p2) ->
                      p1.y == p2.x),
              (p2) -> (x=p1.x,
                                     y=p2.y)
           ))))._1
\end{lstlisting}
\vspace{1em}
\subcaption{\textbf{Non-linear query}}\label{fig:ex-calc-linear}
    \end{minipage}\hfill
    \begin{minipage}[t]{.38\linewidth}
\begin{lstlisting}[style=lrql, emph={gensR},
          basicstyle=\footnotesize\itshape\fontfamily{lmr}\selectfont,
          frame=single,
          framesep=3pt,
          framerule=\borderrule,
          rulecolor=\bordercolor,
          aboveskip=0pt,
          belowskip=0pt,
          breaklines=false
        ]
filter(fix(filter(map(
  table(Parents),
  (e) -> (nm=e.ch, g=1)),
  (p) -> p.par == "A"),
  (gensR) -> flatMap(table(Parents),
        (p) -> map(
            filter(gensR,
              (g) -> p.par == g.nm),
            (g) -> (nm=p.ch,
                                g=g.g+1))))._1,
  (g) -> g.g == 2)
\end{lstlisting}
\subcaption{\textbf{Bag-semantic query}}\label{fig:ex-calc-bag}
    \end{minipage}
    \caption{\calc{} Representation of Queries from Figure~\ref{fig:example-sql-3x}. {$(a)\rightarrow b$ is a function value.}}\label{fig:example-calc-3x}
\end{figure}

The syntax of the \calc{} DSL is presented in Figure~\ref{fig:syntax-sig}, where $x$ ranges over variables. Database columns are base types \textit{(col)}, rows with Named-Tuples \textit{(row)}, $\text{Query}[A]$ represents a relation or query that returns a collection of rows of type $A$, and $\text{Agg}[A]$ represents an aggregation that returns a single scalar result of type $A$ \textit{(res)}. Tuples and functions are used to construct the combinators \textit{(cmb)} that represent database operations. The combinators follow the precedent set by the NRC, and we add \textbf{fix} to model recursion.
``Column-level expressions'' refer to any expression that can go into the filter or project clause of a query: for \textbf{map}, it would be {\small\texttt{SELECT \underline{a + 1} FROM R}} vs. for \textbf{agg} it would be {\small\texttt{SELECT \underline{sum(a) + 1} FROM R}}. Both {\small\texttt{\underline{a + 1}}} and {\small\texttt{\underline{sum(a) + 1}}} are column-level expressions. ``Query-level expressions'' refer to the full query expression and operations that combine subqueries, e.g. \textbf{union}. The DSL syntax does not include function application. Functions can be passed to the constructs in \textit{cmb} but do not get reduced until normalization {(Section~\ref{sec:semantics:normalization})}.
Figure~\ref{fig:example-calc-3x} shows the three example queries from Figure~\ref{fig:example-sql-3x} expressed in the \calc{} DSL.

\begin{figure}[t]
\setlength{\fboxsep}{0pt}
\figbox{\begin{minipage}{1\textwidth}
  \setlength{\mathindent}{0pt}
  \small
  \setlength{\arraycolsep}{0pt}
    \[\begin{gathered}
      {\setlength{\fboxsep}{2pt}
        \raisebox{2.5ex}{\boxed{\Delta\rrJudge{} m \colon \textit{T}}}}
            \qquad
            \stackrel{\textsc{CONST-D}}{\frac{\Sigma(c) = \textit{O}}{\Delta\rrJudge{} c \colon \textit{O}}}
            \quad
            \stackrel{\textsc{VAR-D}}{\frac{ x \colon \textit{T} \in \Delta}{\Delta\rrJudge{} x \colon \textit{T} }}
            \quad
            \stackrel{\textsc{FUN-D}}{\frac{\Delta, x \colon \textit{T} \rrJudge{} m \colon \textit{V} }{\Delta\rrJudge{} ( x ) \rightarrow m \colon \textit{T} \rightarrow \textit{V} }}
            \quad
            \stackrel{\textsc{TUPLE-D}}{\frac{\Delta\rrJudge{} m_i \colon \textit{T}_i \quad \forall i{_{=1}^n}}{\Delta\rrJudge{} (m_i){_{i=1}^n} \colon (\textit{T}_i){_{i=1}^n}}}
            \\
            \stackrel{\textsc{PROJECT-D}}{\frac{\Delta\rrJudge{} m \colon (\textit{T}_i){_{i=1}^n} \ j \in 1..n}{\Delta\rrJudge{} m.j \colon \textit{T}_j}}
            \quad
            \stackrel{\textsc{TABLE-D}}{\frac{
                \begin{array}{c}
                    \Sigma(\textit{db}) = A \end{array}
            }{
                \Delta\rrJudge{} \textbf{table}(\textit{db}) \colon \text{Query}[A]}}
            \quad
            \stackrel{\textsc{NAMED-TUPLE-D}}{\frac{\Delta\rrJudge{} m_i \colon \textit{T}_i \ \forall i{_{=1}^n}}{\Delta\rrJudge{} (l_i = m_i){_{i=1}^n} \colon (l_i \colon \textit{T}_i){_{i=1}^n}}}
            \\
            \stackrel{\textsc{NAMED-PROJECT-D}}{\frac{
              \begin{array}{c}
                \Delta\rrJudge{} m \colon (l_i \colon \textit{T}_i){_{i=1}^n} \\
                j \in 1..n
              \end{array}
              }{\Delta\rrJudge{} m.l_j \colon \textit{T}_j}}
            \quad
            \stackrel{\textsc{NAMED-CONCAT-D}}{\frac{
                \begin{array}{c}
                    \Delta\rrJudge{} m_1 \colon (l_i \colon \textit{T}_i){_{i=1}^n} \
                    \Delta\rrJudge{} m_2 \colon ({l}_j \colon \textit{V}_j){_{j=n+1}^m}
\\
m \geq n \quad l_i \neq l_j
                \end{array}
            }{
                \Delta\rrJudge{} m_1 \ \textbf{++} \ m_2 \colon (l_i \colon \textit{T}_i, l_j \colon \textit{V}_j){_{i=1}^n,\ _{j=n+1}^m}
            }}
            \quad
            \stackrel{\textsc{OP-D}}{\frac{
                \begin{array}{c}
                    \Delta\rrJudge{} m \colon (\textit{T}_i){_{i=1}^n} \\
                    \Sigma(\textit{op}) = (\textit{T}_i){_{i=1}^n} \rightarrow \textit{T}
                \end{array}
            }{
                \Delta\rrJudge{} \textit{op}\ m \colon \textit{T}
            }}
            \\
           \stackrel{\textsc{MAP-D}}{\frac{
                \begin{array}{c}
                    \Delta\rrJudge{} q \colon \text{Query}[\textit{A}] \\
                    \Delta\rrJudge{} f\colon \textit{A} \rightarrow \textit{B}
                \end{array}
            }{
                \Delta\rrJudge{} \textbf{map}(q,\ f) \colon \text{Query}[\textit{B}]
            }}
            \quad
            \stackrel{\textsc{FILTER-D}}{\frac{
                \begin{array}{c}
                    \Delta\rrJudge{} q \colon \text{Query}[\textit{A}] \\
                    \Delta\rrJudge{} f\colon \textit{A} \rightarrow \text{Bool}
                \end{array}
            }{
                \Delta\rrJudge{} \textbf{filter}(q,\ f) \colon \text{Query}[\textit{A}]
            }}
            \quad
            \stackrel{\textsc{FLATMAP-D}}{\frac{
                \begin{array}{c}
                    \Delta\rrJudge{} q \colon \text{Query}[\textit{A}] \\
                    \Delta\rrJudge{} f\colon \textit{A} \rightarrow \text{Query}[\textit{B}]
                \end{array}
            }{
                \Delta\rrJudge{} \textbf{flatMap}(q,\ f) \colon \text{Query}[\textit{B}]
            }}
            \\[2pt]
            \stackrel{\textsc{AGGREGATE-D}}{\frac{
                \begin{array}{c}
                    \Delta\rrJudge{} q \colon \text{Query}[\textit{A}] \\
                    \Delta\rrJudge{} f\colon \textit{A} \rightarrow \textit{B}
                \end{array}
            }{
                \Delta\rrJudge{} \textbf{agg}(q,\ f) \colon \text{Agg}[\textit{B}]
            }}
            \quad
            \stackrel{\textsc{GROUPBY-D}}{\frac{
                \begin{array}{c}
                    \Delta\rrJudge{} q \colon \text{Query}[\textit{A}] \\
                    \Delta\rrJudge{} g\colon \textit{A} \rightarrow \textit{E} \\
                    \Delta\rrJudge{} s\colon \textit{A} \rightarrow \textit{B} \
                    \Delta\rrJudge{} h\colon \textit{A} \rightarrow \text{Bool}
                \end{array}
            }{
              \Delta\rrJudge{} \textbf{groupBy}(q,g,s,h)\colon \text{Query}[\textit{B}]
            }}
            \quad
            \stackrel{\textsc{FIX-D}}{\frac{
                \begin{array}{c}
                    Q =(\text{Query}[\textit{A}_i]){_{i=1}^n} \\
                    \Delta \rrJudge{} \textit{A}_i\colon (l_j\colon \textit{O}_j){_{j=1}^{m_i}} \
                    \forall i{_{=1}^n} \\
                    \Delta\rrJudge{} q \colon Q \
                    \Delta\rrJudge{} f\colon Q \rightarrow Q
                \end{array}
            }{
                \Delta\rrJudge{} \textbf{fix}(q,\ f) \colon Q
            }}
    \end{gathered}\]
              % \end{array}\]
\end{minipage}
  }
\caption{Typing rules for \protect\calc{} with only range-restricted recursion}\label{fig:unrestricted-dsl}
\end{figure}

Following the convention set by T-LINQ, we assume a signature $\Sigma$ that maps each constant \textit{c}, operator \textit{op},  and database \textit{db} to the corresponding  typing rule. $\Sigma$ is useful to abstract over the large set of operations that share the same typing behavior.
The operations stored in  $\Sigma$ (\textit{op}) can be column-level expressions (\textit{exprOp}, e.g. $+$ or \textbf{sum}), or relation-level expressions (\textit{relOps}, e.g. \textbf{union}).
We use the syntax $\textit{op}\ m$ as a stand-in for all operations, where $m$ is typically a tuple of arguments. Some operations like $+$ or $==$ use infix syntax.

Figure~\ref{fig:unrestricted-dsl} shows the typing rules for the {\calc{}} DSL, which closely follows the type system of T-LINQ, with the addition of $\textbf{fix}$.
The $\textbf{fix}$ function defines $n$ \recrel{}s within a single stratum. For $i\in1..n$, $\textbf{fix}$ takes as arguments a tuple of $n$ base-case queries $(q_i: \text{Query}[A_i])$ and a function $f: (r)\rightarrow s$. The function $f$ takes a tuple of $n$ references to the \recrel{}s being defined $(r_i: \text{Query}[A_i])$ and returns a tuple of $n$ recursive-case definitions $(s_i: \text{Query}[A_i])$. Each $s_i$ is composed from the recursive references $r_i$ (and any relations in scope that are defined outside the body of $\textbf{fix}$) using the terms in \textit{cmb}.

In the next section, we show how to independently identify and prevent violations of properties P1--P\total{}. We start with \rr{} and include it in the base type system as there are no cases where a user would want to violate \rr{}.
{
  To distinguish between the type systems targeting each property, we parameterize the typing judgements with the property.
  For example, judgment $\Delta \judgement{P2} m:T$ states that term $m$ has type $T$ in \calc{} DSL environment $\Delta$ under the type system targeting property \mono{} (and \rr{}, as range-restriction is always enforced).
  We use $\vdash$ to indicate the fully restricted \calc{} type system that enforces all properties.
}

\subsection{\rr{}: Range-restricted Recursion}\label{sec:formalization:rr}
\textit{Range-restriction} is the constraint {that the query's project clause contain only constants or references to columns present in relations in the FROM clause}.
Because this constraint does not depend on the semantics of the database system, it is encoded into the basic \textsc{fix-d} typing rule in Figure~\ref{fig:unrestricted-dsl} by enforcing that the recursive references $r_i$ and the recursive definitions $s_i$ all have the same type, $\text{Query}[A_i]$, and that $A_i$ is a Named-Tuple. As Named-Tuples must have the same key and value types, order, and arity to be considered the same type, this restriction enforces that all variables in the head of the rule (e.g. columns in project) are present in the body of the rule (e.g. recursive definitions returned by $f$).
Note that the typing rules for operations stored in $\Sigma$, for example \textbf{union}, are covered by \textsc{op-d}.

\begin{defn}\label{def:rr} A function $f$ $\Delta\!\rrJudge{}\!f\!:(R_i){_{i=1}^n}\!\rightarrow\!(S_i){_{i=1}^n}$ holds \rr{} if $\forall i\ R_i=S_i$.\end{defn}

\subsection{\mono{}: Monotone Recursion}\label{sec:formalization:mono}
We refer to operations that are monotonic under set inclusion as \textit{nonscalar}: for $n$ inputs they will produce $n$ outputs. Non-monotonic operations are \textit{scalar}.
For example, the query {\small\texttt{SELECT max(a) + 1 FROM R}} is non-monotonic {due to \texttt{max}} and produces a single {result}, while {\small\texttt{SELECT a + 1 FROM R}} is monotonic and \textit{nonscalar} and produces one {result} per row in \texttt{R}.
Both scalar and nonscalar expressions can be constructed from each other: \texttt{+} is nonscalar while \texttt{sum} is scalar {and non-monotonic under set inclusion}, but the expressions {\small\texttt{sum(a + 1)}} or {\small\texttt{sum(a) + 1}} are both valid and should produce an expression of type scalar, while {\small\texttt{SELECT x + (SELECT max(y) FROM R1) FROM R2}} should produce an expression of type nonscalar (as the {non-monotonic} subquery does not change that the full expression is monotonic).
The type system must be able to distinguish the shape of the entire expression (no matter how nested) so that well-formed terms of type Query are guaranteed to return nonscalar results and hold \mono{}.

\begin{figure}[t]
\setlength{\fboxsep}{0pt}
    \figbox{\begin{minipage}{\textwidth}\setlength{\mathindent}{0pt}
      {\footnotesize
\setlength{\jot}{0pt}
      \setlength{\arraycolsep}{0pt}
      \vspace{-0.5em}
\begin{minipage}[t]{0.37\textwidth}\[\begin{alignedat}{3}
            &\textbf{\underline{Syntax}}&&\\
            &\textit{(type)} &E\ &\highlight{::=}\  \highlight{\text{Ex}[A, S]} \\
            &\highlight{\textit{(shape)}} &\highlight{S, P}\ &\highlight{::=}\  \highlight{\text{Scalar} \mid \text{NScalar}} \\
            &\highlight{\textit{(query)}} &\highlight{Q}\  &\highlight{::=}\  \text{Query}[A] \mid \highlight{\text{\rquery{}}[A]}
        \end{alignedat}\]
    \end{minipage}
    \hfill
    \begin{minipage}[t]{0.62\textwidth}\[\begin{alignedat}{1}
            &\textbf{\underline{Meta-Helpers}}\\[2pt]
            &\highlight{
                \textit{Shape}(S_1, \ldots, S_n) =
                \text{if } \exists i, S_i\! \equiv\! \text{Scalar} \textit{ then } \text{Scalar} \textit{ else }
                \text{NScalar}
            }\\[2pt]
            &\highlight{
                \textit{Restrict}(A, Q_1, \ldots, Q_n) =
                \text{if } \exists i, Q_i \equiv \text{\rquery{}}[B]
                \textit{ then }
                } \\ &\qquad \highlight{\text{\rquery{}}[A]} \highlight{\textit{ else } \text{Query}[A]
            }
        \end{alignedat}\]
    \end{minipage}
    \vspace{-0.5em}
    {\setlength{\fboxsep}{2pt}
      \begin{flalign*}
          &\boxed{\Delta\monoJudge{} m \colon T}&&
      \end{flalign*}
    \vspace{-3.5em}
    }{\small
    \[\begin{gathered}
    % \[\begin{array}{@{}>{\displaystyle}l@{}}
       \hspace{\linewidth}\\
        \stackrel{\highlight{\textsc{EXPR-OP-D}}}{
        \frac{
            \begin{array}{c}
                \Delta \monoJudge{} m \colon \highlight{(\text{Ex}[A_i, S_i]){_{i=1}^n}}
                \ \Sigma(\textit{exprOp}) = \\
                \highlight{(\text{Ex}[A_i, S_i]){_{i=1}^n} \rightarrow \text{Ex}[A, S]}
            \end{array}
        }{\Delta \monoJudge{} \highlight{\textit{exprOp}\ m} \colon \highlight{\text{Ex}[A, \textit{Shape}(S, S_i{_{i=1}^n})]}}
        }\qquad
        \stackrel{\text{\highlight{\textsc{REL-OP-D}}}}{
        \frac{
            \begin{array}{c}
                \highlight{Q_i{_{i=1}^n}
                \in
                \{\text{\rquery{}}[A], \text{Query}[A]\}
} \
                \Delta \monoJudge{} m\colon \highlight{(Q_i{_{i=1}^n})} \\
                \Sigma(\textit{relOp}) = \highlight{(Q_i{_{i=1}^n}) \rightarrow \textit{Restrict}(A, Q_i{_{i=1}^n})}
            \end{array}
        }{
          \Delta \monoJudge{} \textit{relOp}\ m \colon \highlight{\textit{Restrict}(A, Q_i{_{i=1}^n})}
        }}
        \\[2pt]
        \stackrel{\textsc{MAP-D}}{
        \frac{
            \begin{array}{c}
                \highlight{Q
                \in
                \{\text{\rquery{}}[A], \text{Query}[A]\}
} \
                \Delta \monoJudge{} q \colon \highlight{Q} \\
                \Delta \monoJudge{} f \colon \highlight{\text{Ex}[A, \text{NScalar}]} \!\rightarrow\! \highlight{\text{Ex}[B, \text{NScalar}]}
            \end{array}
            }{\Delta \monoJudge{} \textbf{map}(q,\ f) \colon \highlight{\textit{Restrict}(B, Q)}}
        }\quad
        \stackrel{\textsc{FILTER-D}}{
        \frac{
            \begin{array}{c}
                \highlight{Q
                \in
                \{\text{\rquery{}}[A], \text{Query}[A]\}
} \
                \Delta \monoJudge{} q \colon \highlight{Q}
                \\
                \Delta \monoJudge{} f \colon \highlight{\text{Ex}[A, \text{NScalar}]}\!\rightarrow\! \highlight{\text{Ex}[\text{Bool}, \text{NScalar}]}
            \end{array}
        }{
          \Delta \monoJudge{} \textbf{filter}(q,\ f) \colon \highlight{\textit{Restrict}(A, Q)}
        }}
        \\[2pt]
        \stackrel{\textsc{FLATMAP-D}}{
          \frac{
              \begin{array}{c}
                  \highlight{Q_1
                  \in
                \{
                      \text{\rquery{}}[A],\
                      \text{Query}[A] \}
                  } \
                  \highlight{Q_2
                  \in
                \{
                      \text{\rquery{}}[B],\
                      \text{Query}[B] \}
                  } \\
                  \Delta \monoJudge{} q \colon \highlight{Q_1} \
                  \Delta \monoJudge{} f\colon \highlight{\text{Ex}[A, \text{NScalar}] \rightarrow Q_2}
              \end{array}
        }{
          \Delta \monoJudge{} \textbf{flatMap}(q,\ f) \colon \highlight{\textit{Restrict}(B, Q_1, Q_2)}
        }}
        \quad
        \stackrel{\textsc{AGGREGATE-D}}{\frac{
                \begin{array}{c}
                    \Delta\monoJudge{} q \colon \text{Query}[A] \\
                    \Delta\monoJudge{} f\colon  \highlight{\text{Ex}[A, S]} \rightarrow
\highlight{\text{Ex}[B, \text{Scalar}]}
                \end{array}
            }{
                \Delta\monoJudge{} \textbf{agg}(q,\ f) \colon \text{Agg}[B]
        }}
        \\[2pt]
        \stackrel{\textsc{GROUPBY-D}}{
        \frac{
            \begin{array}{c}
                \Delta \monoJudge{} q \colon \text{Query}[A] \quad
                \highlight{\textit{Shape}(S_{g}, S_{p}, S_{s}) \equiv \text{Scalar} } \\
                \Delta \monoJudge{} g\colon \highlight{\text{Ex}[A, S_{g}]}\! \rightarrow\! \highlight{\text{Ex}[E, S_{g}]} \\
                \Delta \monoJudge{} s\colon \highlight{\text{Ex}[A, S_{p}]}\! \rightarrow\! \highlight{\text{Ex}[B, S_{p}]} \\
                \Delta \monoJudge{} h\colon \highlight{\text{Ex}[A, S_{s}]}\! \rightarrow\! \highlight{\text{Ex}[\text{Bool}, S_{s}]} \\
            \end{array}
        }{
          \Delta \monoJudge{} \textbf{groupBy}(q,\ g,\ s,\ h)\colon \text{Query}[B]
        }}\quad
        \stackrel{\highlight{\textsc{MONOTONE-FIX-D}}}{
        \frac{
            \begin{array}{c}
                Q =(\text{Query}[A_i]){_{i=1}^n} \quad
                \Delta \monoJudge{} q\colon Q \\[1pt]
                A_i = (l_j\colon B_j){_{j=1}^{m_i}} \ \forall i{_{=1}^n}\\
                \highlight{\Delta \monoJudge{} f\colon  (\text{\rquery{}}[A_i]){_{i=1}^n}}\!\rightarrow\! \highlight{(\text{\rquery{}}[A_i]){_{i=1}^n}}
            \end{array}
        }{\Delta \monoJudge{} \textbf{fix}(q,\ f) \colon Q}
        }
    \end{gathered}\]
      % \end{array}\]
    }}
    \end{minipage}
    }
\caption{\protect\calc{} with $\textbf{fix}$ restricted to be monotone. $\text{Ex}$ and $\text{\rquery{}}$ track query monotonicity.}\label{fig:monotone-fix}
\end{figure}

The changes to the type system to restrict recursive queries to monotone operations are shown in Figure~\ref{fig:monotone-fix}.
\textsc{op-d} in Figure~\ref{fig:unrestricted-dsl} is split into \textsc{expr-op-d} and \textsc{rel-op-d}.
We add a type $\text{Ex}[A, S]$ to wrap expressions on columns so that \calc{} can track the \emph{shape} ($S$) of the expression.
By \textsc{expr-op-d}, any expression that contains an arbitrarily nested sub-expression with type parameter Scalar will be of type Scalar (via \textit{Shape}).
For example the \textsc{expr-op-d} rule applied to the ``+'' operator can produce an expression of type Scalar or NScalar:

\noindent\makebox[\linewidth][c]{$\displaystyle\begin{array}{cc}
    \begin{array}{l}
        \dfrac{
            \dfrac{
                \Gamma \monoJudge{} n:\ \text{Ex}[\text{Int}, \text{NScalar}]
            }{
                \Gamma \monoJudge{} \textbf{sum}(n) :\ \text{Ex}[\text{Int}, \text{Scalar}]
            } %\  \textsc{(expr-op-d)}
        }{
            \Gamma \monoJudge{} \textbf{sum}(n)\text{+} 1 :\ \text{Ex}[\text{Int}, \text{Scalar}]
        } %\  \textsc{(expr-op-d)}
    \end{array}
    &
    \begin{array}{l}
        \dfrac{
            \Gamma \monoJudge{} n :\ \text{Ex}[\text{Int}, \text{NScalar}]
        }{
            \Gamma \monoJudge{} n\text{+}1 :\ \text{Ex}[\text{Int}, \text{NScalar}]
        } \  \textsc{(expr-op-d)}
    \end{array}
\end{array}$}

\noindent The expression on the left contains a nested scalar operation ($\textbf{sum}$) so the resulting type will be Scalar, while the expression on the right contains only nonscalar expressions, so even with the same operator ($+$) the result type is NScalar.
The \textsc{map-d} and \textsc{flatmap-d} rules are refined to only accept functions that return non-scalar expressions, while \textsc{aggregate-d} and \textsc{groupby-d} only accept functions that return at least one scalar sub-expression.
With these restrictions, expressions of type Query are guaranteed to contain only monotonic operations, while only expressions of type Agg may contain non-monotonic operations.

As discussed in Section~\ref{sec:background}, the monotonicity property only needs to restrict aggregation between the relations in a single stratum, i.e., within recursive queries to recursive references.
To limit the monotonicity restriction to only recursive references, we introduce a restricted query type, \rquery{}, and refine \textsc{fix-d} so that the arguments and return type of $f$ must be of type \rquery{}. Crucially, \rquery{} can only be derived by calling combinators on the arguments passed to the function $f$ given to $\textbf{fix}$ as a constructor for \rquery{} is not available in the surface syntax and therefore can only be in scope within the body of $\textbf{fix}$.

We update the combinator rules so if they are passed any arguments of type \rquery{} the result type will be \rquery{}.
To reduce the number of rules in Figure~\ref{fig:monotone-fix}, we define a meta-syntax helper \textit{Restrict} to abstract away the differences between rules for combinators that take \rquery{} and Query. Without \textit{Restrict}, we could equivalently define 4 rules for $\textbf{flatMap}$: if $\Gamma \monoJudge{} q: \text{Query}[A]$ and $\Gamma \monoJudge{} f: \text{Ex}[A, \text{NScalar}] \rightarrow \text{Query}[B]$, the result type will be $\text{Query}[B]$, however if either $q$ or the return type of $f$ are \rquery{} the result will be $\text{\rquery{}}[B]$.
As \textsc{aggregate-d} and \textsc{groupby-d} are valid only on Query types, aggregations cannot be applied to recursive references and \calc{} will reject non-monotonic operations on \recrel{}s within recursive queries.
\begin{defn}\label{def:mon} A function $f$ $\Delta\!\monoJudge{} f\!:(R_i){_{i=1}^n}\!\rightarrow\!(S_i){_{i=1}^n}$ holds \mono{} if $\forall i\ \Delta\!\monoJudge{}\!S_i\!: \text{\rquery{}}[A_i]$. \end{defn}

\subsection{\mr{}: Restricted Mutual Recursion}\label{sec:formalization:mr}
Queries that hold \mr{} do not contain any mutual recursion. \calc{} restricts queries by limiting the size of the tuple argument to \textbf{fix} to one. Single-direction dependencies between \recrel{}s can be defined using multiple calls to \textbf{fix}. We limit tuple length to $n=1$ (instead of removing the tuple) to illustrate the constraint is modular and easily turned on/off.
\begin{defn}\label{def:mr} A function $f$ $\Delta \mrJudge{} f:(R_i){_{i=1}^n} \rightarrow (S_i){_{i=1}^n}$ holds \mr{} if $n = 1$.\end{defn}

\subsection{\lin{}: Linear Recursion}\label{sec:formalization:linear}
\begin{figure}[t]
\setlength{\fboxsep}{0pt}
    \figbox{\begin{minipage}{\textwidth}\setlength{\mathindent}{0pt}
      \small
      \setlength{\jot}{0pt}
      \setlength{\arraycolsep}{0pt}
\begin{minipage}[t]{0.45\textwidth}
      \vspace{-1em}
      {\setlength{\fboxsep}{2pt}
        \[\begin{alignedat}{1}
          &\boxed{\Delta\linJudge{} m \colon T}
        \end{alignedat}\]
      }
      \vspace{-3em}
      \[
        \hspace{0.5cm}
        \begin{gathered}
          \stackrel{\textsc{FLATMAP-D}}{
          \frac{
              \begin{array}{c}
                          \Delta \linJudge{} q \colon \highlight{Q_1} \\
                          \Delta \linJudge{} f \colon A \rightarrow \highlight{Q_2} \\
                          \highlight{\quad Q_1, Q_2 \in \{\text{\rquery{}}[A_i, D_i], \text{Query}[A_i]\}}
              \end{array}
          }{\Delta \linJudge{} \textbf{flatMap}(q,\ f) \colon \highlight{\textit{\restrictcollect{}}(A_2, Q_1, Q_2)}}}
        \end{gathered}\]
        \end{minipage}
        \hfill
        \begin{minipage}[t]{0.53\textwidth}
          % \footnotesize
          \vspace{-1em}
          \[\begin{alignedat}{1}
            &\highlight{
            \underline{\textbf{Meta-Helpers}}} \\
            &\highlight{
                \textit{Collect}(Q) = \textit{ if } Q\equiv\text{\rquery{}}[A, D] \textit{ then } D \textit{ else } {()} } \\
            &\highlight{
                \textit{\restrictcollect{}}(A, Q_1, \ldots, Q_n) =
                    \textit{if }\exists i, Q_i \equiv \text{\rquery{}}[A_i, D_i]}\\ &\highlight{
                    \quad \textit{then } \text{\rquery{}}[A, \uplus_{j=1}^n(\textit{Collect}(Q_j))]
                    \textit{ else } \text{Query}[A]
            }
          \end{alignedat}\]
        \end{minipage}
        \begin{minipage}[t]{\linewidth}
          \vspace{-1.6em}
        \[
          \begin{gathered}
              \hspace{\linewidth}\\
              \stackrel{\textsc{\highlight{REL-OP-D}}}{
                \frac{
                    \begin{array}{c}
                        \Delta \linJudge{} m\colon \highlight{(Q_i[A]){_{i=1}^n}} \\
                        \highlight{Q_i{_{i=1}^n}
                        \in \{\text{\rquery{}}[A, D_i], \text{Query}[A]\}} \\
                        \Sigma(\textit{relOp}) = \highlight{(Q_i[A]){_{i=1}^n} \rightarrow \textit{\restrictcollect{}}(A, Q_i{_{i=1}^n})}
                    \end{array}
                }{\Delta \linJudge{} \textit{relOp}\ m \colon \highlight{\textit{\restrictcollect{}}(A, Q_i{_{i=1}^n})}}}
                \ \
                \stackrel{\textsc{\highlight{LINEAR-FIX-D}}}{
                \frac{
                    \begin{array}{c}
                        Q =(\text{Query}[A_i]){_{i=1}^n} \quad
                        \Delta \linJudge{} q\colon Q \quad
                        A_i = (l_j\colon B_j){_{j=1}^{m_i}} \ \forall i{_{=1}^n}\\
                        \highlight{\Delta \linJudge{} f\colon  (\text{\rquery{}}[A_i, (i_\kappa)]){_{i=1}^n}} \rightarrow \highlight{(\text{\rquery{}}[A_i, D_i]){_{i=1}^n}} \\
                        \highlight{\{ 1_\kappa, \ldots, n_\kappa \} \equiv \cup D_i{_{i=1}^n}} \quad
                        \highlight{\forall i{_{=1}^n} \quad |D_i|=|\cup D_i|}
                    \end{array}
                }{\Delta \linJudge{} \textbf{fix}(q,\ f) \colon Q}}
          \end{gathered}
        \]
      \end{minipage}
    \end{minipage}}
    \caption{$\textbf{fix}$ restricted to be linear. $\cup$ converts tuple types into unions where duplicates are removed, $|T|$ is the length of a tuple or union, and $\uplus$ is tuple concatenation.}\label{fig:linear-fix}
\end{figure}
We approach the problem of identifying and preventing non-linear recursion at the type-level by encoding a variant of a linear function arrow with respect to \recrel{}s in \calc{}.
To accomplish this, we split the problem into two sub-constraints: an affine check (every \recrel{} is never used more than once to define an \recrel{}) and a relevant check (all \recrel{}s are used in at least one recursive definition). The affine check is per-relation, while the relevant check is for all relations defined within a single \textbf{fix} call, as not every relation needs to use every other relation within a single stratum. The extensions to the type system to restrict queries to linear recursion are shown in Figure~\ref{fig:linear-fix}.

A type parameter $D$ of type Tuple is added to $\text{\rquery{}}[A, D]$ to model the dependencies of each query. We use $\uplus$ to indicate multiset union, so duplicates are maintained, and equivalence using $\equiv$ does not take order into account.
Uniqueness of references is enforced by using the argument position within the $\textbf{fix}$ function: for a recursive function $\textbf{fix}(((r1, r2) \rightarrow ...), (b1, b2))$,  $r1.D$ will be a tuple containing the constant integer type $1$, and $r2.D$ will contain a tuple with the constant integer type $2$.
Some RDBMS, for example DuckDB, allow recursive queries nested within outer queries to return columns from the outer query (which can itself be recursive, e.g. \textbf{fix} within a \textbf{fix}). However, this would break linearity, so \calc{} must be able to differentiate between references per call to \textbf{fix}.
Each reference is tagged with a singleton type $\kappa$ that is unique for each \textbf{fix} invocation. References are considered equal only if both the $\kappa$ and the constant integer type are the same. This ensures that if \textbf{fix} is called within the scope of references from another \textbf{fix} function, the inner \textbf{fix} cannot return terms derived from the outer function's parameters.
Multi-relation operations like \textbf{flatMap} or \textbf{union} collate recursive references in the result relation. For example:

\noindent\makebox[\linewidth][c]{$\displaystyle\dfrac{
    \Gamma \linJudge{} q: \text{\rquery{}}[A, (1_{\kappa_1})], \quad
    \Gamma \linJudge{} f: A \rightarrow \text{\rquery{}}[B, (2_{\kappa_2})]
}{
    \Gamma \linJudge{} \text{\textbf{flatMap}}(q,\ f): \text{\rquery{}}[B, (1_{\kappa_1}, 2_{\kappa_2})]
}
\quad (\textsc{flatmap-d})$}

\noindent For the affine check, the $D$ of each element of the return tuple must contain no duplicates. This check is implemented by taking the length (indicated with $|T|$) of $D$ and requiring that it is the same as the length of the union of $D$: $\forall i{_{=1}^n}\ |D_i| \equiv |\cup D_i|$. For the relevant check, all parameters must appear at least once in all $D$. This is implemented by using the set of constant integer types from 1 to the number of arguments to $\textbf{fix}$ decorated with $\kappa$ to check that all elements are present at least once in all the $D$s: $\{ 1_\kappa, \ldots, n_\kappa \} \equiv \cup D_i{_{i=1}^n}$.

The \textsc{rel-op-d} rule is updated so if any arguments are \rquery{}, the result will be a \rquery{} that collects (only) the dependencies of the \recrel{}s. Similarly to Figure~\ref{fig:monotone-fix}, to reduce the number of rules in Figure~\ref{fig:linear-fix} we define a meta-syntax helper $\textit{\restrictcollect{}}$ to abstract away the differences between rules that take \rquery{} or Query.
\begin{defn}\label{def:lin}
If $\Delta\!\linJudge{}\!f\!:(\text{\rquery{}}[A_i, (i_\kappa)]){_{i=1}^n}\!\rightarrow\!(\text{\rquery{}}[A_i, D_i]){_{i=1}^n}$ then $f$ holds \lin{} if it holds both $\{ 1_\kappa, \ldots, n_\kappa \} \equiv \cup D_i{_{i=1}^n}$ and $\forall i{_{=1}^n}\ |D_i| \equiv |\cup D_i|$. \end{defn}

\subsection{\set{}: Set-semantic Recursion}\label{sec:formalization:set}
\begin{figure}[t]
\setlength{\fboxsep}{0pt}
    \figbox{\begin{minipage}[t]{\textwidth}
      \setlength{\mathindent}{0pt}
      \setlength{\jot}{0pt}
      \setlength{\arraycolsep}{0pt}
      {\footnotesize
\figbox{\begin{minipage}[t]{.29\textwidth}
        \vspace{-0.5em}
\[\begin{alignedat}{4}
            &\textbf{\underline{Syntax}}&&\\
            &\textit{(type) } & R &::=\ \dots \text{Query}[A, \highlight{C}] \\
            &\highlight{\textit{(cat) }} &\highlight{C} &::=\ \highlight{\text{Bag} \mid \text{Set}} \\
        \end{alignedat}\]
      \end{minipage}}
    \begin{minipage}[t]{.7\linewidth}
        \vspace{-0.5em}
        \[\begin{gathered}
          {\raisebox{6ex}{\setlength{\fboxsep}{2pt}\boxed{\Sigma}}} \hspace{-10pt}
\stackrel{\textsc{DISTINCT-D}}{\frac{
              \begin{array}{c}
                \Delta \setJudge{} q \colon \text{Query}[A, \highlight{C}]
              \end{array}
            }
            {
              \begin{array}{c}
                \Delta \setJudge{} \text{\textbf{distinct}}(q) \colon \\
                \hspace{0.7cm} \text{Query}[A, \highlight{\text{Set}}]
              \end{array}
            }}\ \
\stackrel{\textsc{UNION-D}}{
            \frac{
              \begin{array}{c}
                \Delta \setJudge{} q_1 \colon \text{Query}[A, \highlight{C}] \\
                \Delta \setJudge{} q_2 \colon \text{Query}[A, \highlight{C}]
              \end{array}
            }{
              \begin{array}{c}
                \Delta \setJudge{} \text{\textbf{union}}(q_1, q_2) \colon \\
                \hspace{0.7cm} \text{Query}[A, \highlight{\text{Set}}]
              \end{array}
            }}\ \
\stackrel{\textsc{UNION-ALL-D}}{
            \frac{
              \begin{array}{c}
                \Delta \setJudge{} q_1 \colon \text{Query}[A, \highlight{C}] \\
                \Delta \setJudge{} q_2 \colon \text{Query}[A, \highlight{C}]
              \end{array}
            }{
              \begin{array}{c}
                \Delta \setJudge{} \text{\textbf{unionAll}}(q_1, q_2) \colon \\
                \hspace{0.7cm} \text{Query}[A, \highlight{\text{Bag}}]
              \end{array}
            }}
        \end{gathered}
        \]
      \end{minipage}
        \vspace{-2em}
        {\setlength{\fboxsep}{2pt}
        \begin{flalign*}
            &\boxed{\Delta \setJudge{} m \colon T}&&
        \end{flalign*}
        }
      }{\small
        \vspace{-2.5em}
        \[
        \begin{gathered}
            \hspace{\linewidth}\\
\stackrel{\textsc{MAP-D}}{
            \frac{
              \begin{array}{c}
                \Delta\setJudge{} q \colon \text{Query}[A, \highlight{C}] \\
                \Delta \setJudge{} f\colon A \rightarrow B
              \end{array}
            }
            {
              \begin{array}{c}
                \Delta\setJudge{} \textbf{map}(q,\ f) \colon \\ \hspace{0.7cm}
                \text{Query}[B, \highlight{\text{Bag}}]
              \end{array}
            }
          }
            \qquad
\stackrel{\textsc{FILTER-D}}{
            \frac{
              \begin{array}{c}
                \Delta\setJudge{} q \colon \text{Query}[A, \highlight{C}] \\
                \Delta \setJudge{} f\colon A \rightarrow \text{Bool}
              \end{array}
            }
            {
              \begin{array}{c}
                \Delta\setJudge{} \textbf{filter}(q,\ f) \colon \\ \hspace{0.7cm}
                \text{Query}[A, \highlight{C}]
              \end{array}
            }}
            \qquad
\stackrel{\textsc{FLATMAP-D}}{
            \frac{
              \begin{array}{c}
                \Delta\setJudge{} q \colon \text{Query}[A, \highlight{C_1}] \\
                \Delta \setJudge{} f\colon A \rightarrow \text{Query}[B, \highlight{C_2}]
              \end{array}
            }
            {
              \begin{array}{c}
                \Delta\setJudge{} \textbf{flatMap}(q,\ f) \colon \\ \hspace{0.7cm}
                \text{Query}[B, \highlight{\text{Bag}}]
              \end{array}
              }}
            \\[4pt]
\stackrel{\textsc{GROUPBY-D}}{
            \frac{
            \begin{array}{l}
                \Delta\setJudge{} q \colon \text{Query}[A, \highlight{C}] \quad \Delta \setJudge{} g\colon A \rightarrow D \\
                \Delta \setJudge{} s\colon A \rightarrow B \quad \Delta \setJudge{} h\colon A \rightarrow \text{Bool}
            \end{array}
            }{\Delta\setJudge{} \textbf{groupBy}(q,g,s,h)\colon \text{Query}[B, \highlight{\text{Bag}}]}}
            \quad
            \stackrel{\highlight{\textsc{CATEGORY-FIX-D} }}{
            \frac{
            \begin{array}{l}
                Q_1 =(\text{Query}[A_i, \highlight{C_i}]){_{i=1}^n}\quad \highlight{Q_2 =(\text{Query}[A_i, \text{Set}]){_{i=1}^n}} \\
                    A_i = (l_j\!\colon O_j){_{j=1}^{m_i}} \ \forall i{_{=1}^n}\quad \Delta \setJudge{} q\!\colon Q_1 \  \Delta \setJudge{} f\colon \highlight{Q_1} \rightarrow \highlight{Q_2}
            \end{array}
            }
            {\Delta \setJudge{} \textbf{fix}(q,\ f) \colon \highlight{Q_2}}}
        \end{gathered}
        \]
      }
    \end{minipage}}
    \caption{\protect\calc{} with $\textbf{fix}$ restricted to be set-semantic. $C$ is added to track bag/set semantics.}\label{fig:category-fix}
\end{figure}

To restrict $\textbf{fix}$ to allow only set semantics within recursive queries, we track the semantics of each query-level operation in a type parameter, similarly to how we tracked monotonicity in column-level operations. We update $\Sigma$ so that all query-level operations reflect their semantics.
The changes to the syntax and rules are shown in Figure~\ref{fig:category-fix}.
For example:

\noindent\makebox[\linewidth][c]{
  $\displaystyle\begin{array}{c}
    \dfrac{
        \dfrac{
            \Gamma \setJudge{} a : \text{Query}[A, \text{Set}]
            \quad
            \Gamma \setJudge{} f : A \rightarrow B
        }{
            \Gamma \setJudge{} \textbf{map}(a,\ f) : \text{Query}[B, \text{Bag}]
        } \quad \text{(\textsc{map-d})}
    }{
        \Gamma \setJudge{} \textbf{distinct}(\textbf{map}(a,\ f))  : \text{Query}[B, \text{Set}]
    } \quad \text{(\textsc{distinct-d})}
  \end{array}$
}

\begin{defn}\label{def:ss} A function $f$ $\Delta\!\setJudge{} f\!:(R_i){_{i=1}^n}\!\rightarrow\!(S_i){_{i=1}^n}$ holds \set{} if $\forall i\ S_i\!: \text{Query}[A_i, \text{Set}]$. \end{defn}

\subsection{\cf{}: Constructor-Freedom}\label{sec:formalization:cf}
Constructors, e.g. column-level operations that produce new values and therefore grow the program domain, are represented with the types in \textit{exprOp} in Figure~\ref{fig:syntax-sig}, e.g., $+$. To ensure that an expression is constructor-free, it must not contain those operations. We apply this in \calc{} by modifying the functions $f$ accepted by $\textbf{map}$, $\textbf{flatMap}$, and $\textbf{filter}$. The typing rules are updated in a similar way to the monotonicity check: arguments and return types of $f$ in \textsc{map-d}, \textsc{flatmap-d}, and \textsc{filter-d} must be of type \rexpr{} and by construction, $\textit{mathOps}$ and $\textit{stringOps}$ cannot be applied to \rexpr{}. This is the most restrictive property, and similar systems like Souffl\'e~\cite{souffle-site} have chosen to allow limited unsoundness in exchange for expressiveness.
{Yet there are several useful queries that do hold \cf{}, shown in Table~\ref{tab:benchmark}}.
Because each restriction in \calc{} is independent, users can choose to disable enforcement.

\begin{defn}\label{def:cf} A function $f$ $\Delta\!\cfJudge{} f\!:{(R_i)_{i=1}^n}\!\rightarrow\!{(S_i)_{i=1}^n}$ holds \cf{} if $\forall i\ S_i\!: \text{Query}[\rexpr{}[\dots]]$.\end{defn}

\subsection{\calc{} with Host Language Embedding}\label{sec:formalization:embedding}
\begin{figure}[t]
    \setlength{\fboxsep}{0pt}
    \figbox{\begin{minipage}{0.9\textwidth}
      \setlength{\mathindent}{0pt}
      \small
      \setlength{\jot}{0pt}
      \vspace{-1em}
      \[\begin{alignedat}{3}
            &\underline{\textbf{Syntax}}& \\
&\textit{(type)}      &T   &::=  \dots \mid \text{Expr}[A, S] \mid \text{Query}[A, C] \mid \text{List}[A] \mid \text{\rquery{}}[A, D,\  C] \mid \text{\rexpr{}}[A] \\
                &\textit{(term)}      &m     &::=     \dots \mid \textbf{toRow}(m) \mid \textbf{toExpr}(m) \mid \textbf{run}(m) \mid f(m)
        \end{alignedat}
        \]
  \end{minipage}}
\def\env{\Gamma; \Delta}\def\dslext{}
{\setlength{\fboxsep}{0pt}
  \figbox{\begin{minipage}{\textwidth}
    \setlength{\mathindent}{0pt}
    \small
    \setlength{\jot}{0pt}
    \setlength{\arraycolsep}{0pt}
    \vspace{-1em}
    {\setlength{\fboxsep}{2pt}
      \[\begin{alignedat}{1}
          &\boxed{\Gamma \vdash m : T}\\
      \end{alignedat}\]
      }
  \vspace{-3em}
  \[
  \begin{gathered}
      \hspace{\linewidth}\\
      \
       \stackrel{\textsc{APP}}{\dfrac{\Gamma \vdash m_1 : T \rightarrow V \quad \Gamma \vdash {m_2} : T}{\Gamma \vdash m_1({m_2}) : V}}
       \
        \stackrel{\textsc{RUN-QUERY}}{\dfrac{\Gamma \vdash {m} : \mathrm{Query}[A,C]}{\Gamma \vdash \mathbf{run}({m}) : \mathrm{List}[A]}}
        \
       \stackrel{\textsc{RUN-AGG}}{\dfrac{\Gamma \vdash {m} : \mathrm{Agg}[A]}{\Gamma \vdash \mathbf{run}({m}) : A}}
       \
             \stackrel{\textsc{LIFT}}{\dfrac{\Gamma \vdash c : O}{\Gamma;\Delta \vdash c : \mathrm{Expr}[O,\mathrm{NScalar}]}}
              \\[2pt]
      \stackrel{\textsc{CONST}}{\dfrac{\Sigma(c)=O}{\Gamma \vdash c : O}}
      \
      \stackrel{\textsc{TO-EXPR}}{\dfrac{m : O \in \Sigma}{\Gamma \vdash \mathbf{toExpr}(m) : \mathrm{Expr}[O,\mathrm{NScalar}]}}
      \
      \stackrel{\textsc{TO-ROW}}{\dfrac{\Gamma \vdash m : (l_i : \mathrm{Expr}[A_i,S_i]){_{i=1}^n}}
           {\Gamma \vdash \mathbf{toRow}({m}) : \mathrm{Expr}[(l_i : A_i){_{i=1}^n},\ \mathrm{Shape}(S_i{_{i=1}^n})]}}
  \end{gathered}
  \]
\end{minipage}}
} \caption{Additional host rules. Extends Fig.~\ref{fig:syntax-sig} syntax and Fig.~\ref{fig:unrestricted-dsl}-\ref{fig:category-fix} rules under $\Delta;\Gamma \vdash \dots$
}\label{fig:restricted-host}
\end{figure}

The power of language-integrated query comes from the tight embedding of the DSL with the general-purpose host language. The DSL environment represents \emph{staged} computation: all terms in the DSL serve to construct a type-level AST but do not execute any queries. To generate queries that return results to an application, we need to extend \calc{} to be an embedded DSL in a host language. In Figure~\ref{fig:restricted-host}, based on the precedent set by T-LINQ, we extend \calc{} with a second type environment for host terms, $\Gamma$, and update our DSL typing rules accordingly. Judgment $\Gamma \vdash m:T$  states that host term $m$ has type $T$ in type environment $\Gamma$, and judgment $\Gamma; \Delta \vdash m:T$ states that quoted term $m$ has type $T$ in host type environment $\Gamma$ and DSL type environment $\Delta${, with all properties enforced}.

We reuse the syntax of types and terms in Figure~\ref{fig:syntax-sig} for the host language, with a few key additions. Lambda application, i.e., $f(m)$,  is added to the host-language syntax so users can define and apply functions. DSL types are represented in the host language by wrapping them in $\text{Expr}$ types, so if a DSL expression represents a row of type $T$ then the type of this expression in the host language will be $\text{Expr}[T]$. Translation functions from host terms to DSL terms are added: $\textbf{toExpr}$ converts base host-language types to $\text{Expr}$ of base types and $\textbf{toRow}$ converts Named-Tuples of $\text{Expr}$ to $\text{Expr}$s of Named-Tuple. Finally, an execution function $\textbf{run}$ is added to execute the query expressed in the DSL and returns a $\text{List}$ of the query result type to the application. The rules \textsc{to-expr}, \textsc{to-row}, \textsc{run-query} and \textsc{run-agg} define the types for $\textbf{toExpr}$, $\textbf{toRow}$, and $\textbf{run}$.
Figure~\ref{fig:restricted-host} shows the {additional syntax and rules required to handle interactions between the DSL and host language}.
Each judgment in the DSL type system operates under both $\Delta$ and $\Gamma$ and an additional rule, \textsc{lift}, {lifts constants from host to DSL-level}.
The appendix includes the full syntax and combined rules of \calc{} with all restrictions applied in the ``\calc{} with Host-Language Embedding'' section.

\section{Formal Semantics and Query Normalization}\label{sec:semantics}

\subsection{{Safety and Correctness}}\label{sec:semantics:correctness}
{
To prove that well-typed \calc{} programs do not show behaviors B1--B3, we need to establish a semantics for recursive query execution on RDBMS, even though the database community has not formally defined a semantics for \texttt{WITH RECURSIVE}.
However, the database theory literature has defined several formal semantics for recursive queries in the context of Datalog~\cite{dl}, a database query language based on logic programming.
Datalog has a bottom-up fixed-point semantics and an equivalent proof-theoretic semantics that can be used to prove both soundness and completeness~\cite{dltextbook}.
Stratified Datalog with negation ($\text{Datalog}^{\neg s}$), an extension of pure Datalog, has an iterated fixed-point semantics (strata-by-strata evaluation) that always produces the Perfect Model~\cite{amateur}, i.e., using this semantics, well-formed programs will always find the unique and minimal fixed-point in a finite number of steps.
The SQL standard specifies the evaluation of the \texttt{WITH RECURSIVE} keyword with an algorithm (in English) that, for linear, set-semantic, monotone queries that are free of constructors and mutual recursion, coincides with the bottom-up iterated fixed-point semantics~\cite{sql99}.
Based on the database theory results for Datalog, we can establish the following theorem:
}
\begin{prop2} Well-typed fully-restricted \calc{} programs {will always find the unique and minimal fixed-point under the iterated fixed-point semantics.\label{theprop}}
\end{prop2}

{We prove Theorem~\ref{theprop} as follows: we give the fully restricted \calc{} the iterated fixed-point semantics with a complete type-directed translation to a restricted variant of Datalog that is equivalent to linear, stratified, non-mutually-recursive Datalog with negation (\lsd{}, defined in the ``Definitions'' section of the appendix).
We can then directly apply the result from the database theory literature that well-formed Datalog programs will find the unique and minimal fixed-point in a finite number of steps under the iterated fixed-point semantics.
As linearity and non-mutual-recursion are only syntactic restrictions on top of $\text{Datalog}^{\neg s}$, all well-formed \lsd{} programs are guaranteed to avoid behaviors B1--B3.
}
The appendix includes the full translational semantics in the ``Translational Semantics for \calc{}'' section and proofs in the ``Proofs'' section.

\subsection{Property Entanglements and Tradeoffs}

In order to guarantee the absence of B1--B3 on recursive queries over arbitrary data, all \total{} properties must be satisfied.
Different combinations of properties will have semantics comparable to different variants of Datalog: for example, if we relax the linearity property then \calc{} is equivalent to stratified, non-mutually-recursive Datalog with negation; if we relax the mutual-recursion and monotonicity properties then \calc{} is equivalent to linear Datalog with negation; if we relax constructor-freedom then we get extensions of Datalog with interpreted functions over infinite domains, etc.

Each property has a unique and independent effect on the evaluation of a program. For example, relaxing set-semantics can lead to nontermination due to duplicates, i.e., intermediate relations grow infinitely over a finite domain, while relaxing constructor-freedom can lead to nontermination due to infinitely growing intermediate results over an infinite domain, while relaxing monotonicity can lead to nontermination due to non-convergence. Properties are independent in that they have unique effects on the evaluation of the program, and enforcing each property will prevent those effects, but enforcing one property will not prevent the effect associated with a different, unenforced property.

\subsection{{Normalization} in \calc{}}\label{sec:semantics:normalization}
The problem of deeply nested datatypes and query avalanches is well covered by previous work in NRC and T-LINQ.
The normalization approach taken by T-LINQ is applicable to non-recursive queries in \calc{}, or within the bodies of recursive queries, but not between recursive subqueries because the fixpoint introduces a strict evaluation boundary.

The core difference between the normalization approach of T-LINQ and \calc{} is that chained calls to \textbf{fix}, e.g., $\textbf{fix}( \textbf{fix}(R, f1), f2)$ will generate a single query containing a subquery, e.g., the SQL query defined by $f1$ will be the base-case of the query defined by $f2$. These two queries cannot be flattened into a single \texttt{WITH RECURSIVE} call because the evaluation boundary must be maintained in the generated SQL in order to retain stratified semantics. Unlike query avalanches and deeply nested subqueries, the stratification of a recursive program is not guaranteed to show worse performance, in fact, stratified programs can be more efficient~\cite{recstep}.
{Therefore every call to \textbf{fix} in \calc{} will result in a single invocation of \texttt{WITH RECURSIVE}.}
For nested recursion (i.e., \textbf{fix} inside the body of \textbf{fix}), some database systems such as DuckDB allow nested recursive queries to return columns from the outer query. This is only problematic when it violates linearity, which is handled by the type system as explained in Section~\ref{sec:formalization:linear}.
We have chosen to simplify the syntax of \calc{} to restrict nested column types, although in our implementation, nested datatypes are supported.

{
  Normalization in \calc{} proceeds by directly applying T-LINQ's normalization algorithm (adapted for \calc{}'s combinator syntax) to non-recursive queries and to the bodies of recursive queries, leveraging T-LINQ's single-query, confluence, type-preservation properties for non-recursive queries.
  The normalization rules perform beta-reduction, query flattening, and other combinator optimizations within the calculus.
  The syntax of normalized \calc{}, the normalization relations, and operational semantics are included in the ``Operational Semantics and Normalization for \calc{}'' section of the appendix and the type preservation statements in the ``Proofs'' section of the appendix.
  Each phase of query normalization applied to the query in Figure~\ref{fig:example} is illustrated in the ``Translational Semantics for \calc{}'' section of the appendix.
  Because \texttt{WITH RECURSIVE} evaluation coincides with the iterated fixed-point semantics for fully-restricted queries, generating SQL queries that follow the standard's specification inherits the correctness properties proven via the Datalog translation. The correspondence between normalized \system{} and SQL is described further in Section~\ref{sec:implementation:query}.
}

 \section{Implementation}\label{sec:implementation}
\begin{figure}[t]
    \small
    \begin{minipage}[t]{.35\linewidth}
\begin{lstlisting}[style=tablescalastyle, emph={waitFor}, emph={[3]fix,flatMap,filter,map,distinct,aggregate,groupBy, max}, emphstyle={[3]\bfseries},
          basicstyle=\footnotesize\ttfamily,
          frame=single,
          framesep=3pt,
          framerule=\borderrule,
          rulecolor=\bordercolor,
          aboveskip=0pt,
          belowskip=0pt,
          breaklines=false
        ]
BasicParts.fix(P2)(
 waitFor=>
  SubParts.aggregate(sp=>
   waitFor
    .filter(wf=>
     sp.sub==wf.part)
    .aggregate(wf=>
     (part=sp.part,
      days=max(wf.days))))
  .groupBy((wf, _)=>
   (part=wf.part))
.distinct)
\end{lstlisting}
\subcaption{\textbf{Non-monotonic query}}\label{fig:ex-tyql-monotonic}
    \end{minipage}\hfill
    \begin{minipage}[t]{.26\linewidth}
\begin{lstlisting}[style=tablescalastyle, emph={pathR}, emph={[3]fix,flatMap,filter,map,distinct,aggregate,groupBy}, emphstyle={[3]\bfseries},
          basicstyle=\footnotesize\ttfamily,
          frame=single,
          framesep=3pt,
          framerule=\borderrule,
          rulecolor=\bordercolor,
          aboveskip=0pt,
          belowskip=0pt,
          breaklines=false
        ]
Edges.fix(P4)(
 pathR=>
  pathR.flatMap(p=>
    pathR
     .filter(e=>
       p.y==e.x)
     .map(e=>
      (x=p.x,
       y=e.y)))
 .distinct)
\end{lstlisting}
\vspace{2em}
\subcaption{\textbf{Non-linear query}}\label{fig:ex-tyql-linear}
    \end{minipage}\hfill
    \begin{minipage}[t]{.35\linewidth}
\begin{lstlisting}[style=tablescalastyle, emph={gensR}, emph={[3]fix,flatMap,filter,map,distinct,aggregate,groupBy}, emphstyle={[3]\bfseries},
          basicstyle=\footnotesize\ttfamily,
          frame=single,
          framesep=3pt,
          framerule=\borderrule,
          rulecolor=\bordercolor,
          aboveskip=0pt,
          belowskip=0pt,
          breaklines=false
        ]
Parents
 .filter(p=> p.par=="A")
 .map(e=> (nm=e.ch, g=1))
 .fix(P5P6)(
   gensR=>
    Parents.flatMap(p=>
     gensR.filter(g=>
       p.par==g.nm)
      .map(g=>
       (nm=p.ch,
        g=g.g + 1))))
 .filter(g=> g.g==2)
\end{lstlisting}
\subcaption{\textbf{Bag-semantic query}}\label{fig:ex-tyql-bag}
    \end{minipage}
    \caption{Queries from Fig.~\ref{fig:example-sql-3x} and~\ref{fig:example-calc-3x} in \system{}.}
    \label{fig:example-tyql-3x}
\end{figure}

 \begin{figure}[t]
\centering
\includegraphics[trim={1cm 0.9cm 0.9cm 1cm}, width=\linewidth, clip]{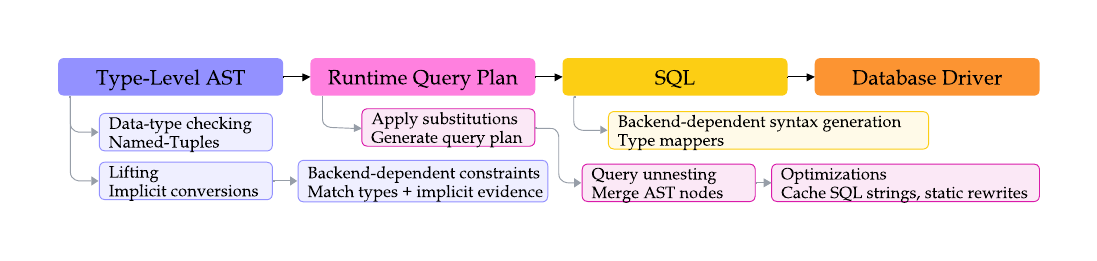}
\caption{Architecture of \system{}}\label{fig:arch}
\end{figure} In this section, we describe the implementation of \system{}, our type-safe embedded query library based on \calc{}.
{
  \system{} achieves three key goals: (1) \textbf{safety through static checking}: all properties are verified at compile-time by the Scala type system, preventing
  runtime errors, incorrect results, and nontermination before queries ever reach the database; (2) \textbf{expressiveness without complexity}: queries are written using canonical Scala collection operations, avoiding the syntactic complexity of raw SQL while supporting the full power of
  recursive queries; and (3) \textbf{performance}: \system{} generates a single SQL query that executes directly in the database, achieving performance
  identical to hand-written SQL because the key mechanisms operate at the type-level without runtime overhead (see Section~\ref{sec:eval}).
}

{
  Figure~\ref{fig:example-tyql-3x} shows the \system{} code for the queries from Figures~\ref{fig:example-sql-3x} (SQL) and~\ref{fig:example-calc-3x} (\calc{}),
  demonstrating all three goals: (1) the type system enforces all properties by default but is flexible enough to allow selective disabling via configuration objects passed to
  \texttt{fix}; (2) the syntax mirrors Scala's Collections API; and (3) each query compiles to the SQL in Figure~\ref{fig:example-sql-3x}.
}

{
  The key technical challenge is encoding each property in the type system while maintaining ergonomic syntax and clear error messages.
  \system{} achieves this by leveraging Scala 3's type-level features: Named-Tuples for row modeling, Match Types for constraint enforcement, and type classes for customization.
  The architecture of \system{} is summarized in Figure~\ref{fig:arch}.
}

\subsection{Modeling Rows in Scala}
Representing database rows in statically typed languages like Scala is challenging because the compiler must allow on-the-fly composition of types without losing the advantages of static typing.
Beyond type safety, static typing also enables powerful IDE features like code completion. Yet operations like \texttt{join} and \texttt{project} take collections of rows and produce new collections that may be of a completely different structure, but still need to support element access and be type-checked when used later.
{
  For example, query~\ref{fig:ex-tyql-bag} projects a field \texttt{g} that is not in the source table, but will be statically checked and would fail to compile if the base case and recursive case did not both define the \texttt{g} field with the same type.
  }

Type computations in JVM languages cannot create classes, so it is impossible to dynamically generate Scala's classes. Structural types allow abstraction over existing classes but require reflection or other mechanisms to support dynamic element access. \emph{Named-Tuples}, released in Scala 3.6.0, are represented as pairs of Tuples, where names are stored as a tuple of constant strings and the values are stored in regular Scala tuples.
Tuples are preferable over case classes because tuples are lightweight structures that avoid additional JVM object allocation and dispatch overhead. In contrast, classes generate separate class definitions at compile-time, increasing memory usage and execution overhead and leading to bloated bytecode generation. By using tuples, \system{} benefits from more compact and efficient runtime representations, reducing both memory footprint and execution latency.

Named-Tuples are ordered, providing an advantage over structural types for modeling rows because of (1) better integration with Scala since they share the same representation as regular tuples, and (2) efficient and natural traversal order allows the formulation of type-generic algorithms. Because Named-Tuples can be decomposed into \texttt{head *: tail}, they
can be iterated over without the use of an auxiliary data structure like a dictionary.
Method overloading can catch common mistakes like using \texttt{map} instead of \texttt{flatMap} and provide useful, domain-specific error messages.
For example, a simple type error
generates the following error messages in \system{} and in a state-of-the-art query library ScalaSQL:

\begin{lstlisting}[style=errorstyle, xleftmargin=0pt]
TyQL: Types being inserted (name : Long) do not fit inside target types (name : String).
ScalaSQL: Found: scalasql.query.Select[Src[scalasql.core.Expr], Src[scalasql.Sc]]
Required: scalasql.query.Select[C, R2] where: C  is a type variable with constraint
>: Dst[scalasql.core.Expr] R2 is a type variable
\end{lstlisting}

\subsection{Type-Level ASTs and Constraints}

\begin{figure}[t]
\centering
\includegraphics[trim={.1cm .1cm .1cm .1cm}, clip, width=\linewidth]{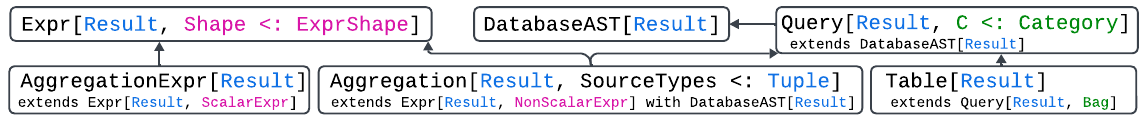}
\caption{\system{} Type Hierarchy. \texttt{\rquery{}}/\texttt{\rexpr{}} are not subtypes of any other \system{} type.}\label{fig:hierarchy}
\end{figure}

\system{} maintains a hierarchy of query representations, illustrated in Figure~\ref{fig:hierarchy}. Rows are represented as Named-Tuples, AST expressions as structural types that wrap row types, and entire queries or tables as \texttt{DatabaseAST}s. These can be of type \texttt{Query}, which allows chaining of further relation-level operations, or \texttt{Aggregation}, which represents a scalar result from the database. Aggregations are also subtypes of expressions, as many databases allow aggregations to be nested within queries at the expression level. Because aggregations and non-scalar expressions must share a supertype, monotonicity of expressions is tracked in a type member \texttt{ExprShape} that can be either \texttt{Scalar} or \texttt{NScalar}. The category of the result, either bag or set, is tracked with \texttt{Category}. \texttt{\rquery{}} wraps \texttt{Query} but does not extend it.
{
  For example, in query~\ref{fig:ex-tyql-linear}, \texttt{pathR} has type \texttt{\rquery{}} while \texttt{Edges} has type \texttt{Query}.}

\textit{Selection.}
The \texttt{Expr} class represents AST expressions and extends Scala's \texttt{Selectable} trait. Element accesses are syntactic sugar for calling the method \texttt{selectDynamic}, which maps field names to values.
{For example, in query~\ref{fig:ex-tyql-linear} the expression \texttt{e.y} where \texttt{e: Expr[(x: Int, y: Int), \ldots]} will return an AST node \texttt{Select[Int](e, ``y'')}, a subclass of \texttt{Expr}.}
Implicit (or explicit) conversions
are used to lift native Scala types to AST nodes, {for example in query~\ref{fig:ex-tyql-bag} \texttt{`A'} is implicitly converted to \texttt{Expr[String, NScalar]}.}

\textit{Projection.}
{
  The \texttt{Project} class represents a projection AST node and extends \texttt{Expr}.
  Users project rows using Named Tuples, e.g., the query in Figure~\ref{fig:ex-tyql-linear} calls \texttt{map} and returns \texttt{(x=p.x, y=e.y)}, which has type Named-Tuple-of-Expr.
  \system{} uses type-level pattern matching with Match Types~\cite{matchtypes} to convert the Named-Tuple-of-Expr into a \texttt{Project} AST node, i.e., an Expr-of-Named-Tuple.
  \system{} reifies the constant string keys in Named-Tuples so that the generated queries are more readable, for example the query~\ref{fig:ex-tyql-bag} generates a project with aliases \texttt{SELECT p.ch as nm, 1 as g} based on the keys of the Named-Tuple.
}

\textit{Join.}
Join operations are represented by \texttt{FlatMap} AST nodes. Each \texttt{FlatMap}, \texttt{Map}, or \texttt{Aggregation} has a source subtree of type \texttt{DatabaseAST} and a \texttt{Fun} subtree that represents function application. Unrolling of nested \texttt{Fun} nodes is done during query generation; {for example the body of query~\ref{fig:ex-tyql-linear} compiles to a single self-join where \texttt{pathR} appears twice.}

\textit{GroupBy.}
GroupBy in \system{} is slightly different than \calc{} to allow for nicer syntax, i.e., incrementally chaining \texttt{map}, \texttt{groupBy}, and \texttt{having}. Due to SQL semantics, \texttt{groupBy} and \texttt{having} should operate over the type of the \emph{source} relations, not the result type of the preceding expression. This is a challenge, especially when the preceding expression is a join, as there will be multiple input relations. The way this is addressed in \system{} is by tracking the types of the source relations of compound statements at the type-level. {
  For example, the function passed to \texttt{groupBy} in query~\ref{fig:ex-tyql-monotonic} takes two arguments: one with the row type of \texttt{waitFor} and one with the row type of \texttt{SubParts} (unused in this example).
}

\begin{figure}[t]
    \begin{minipage}[t]{0.64\textwidth}
        \begin{lstlisting}[style=tablescalastyle,numbers=left, numberstyle=\tiny\color{gray}, numbersep=2mm,
          basicstyle=\footnotesize\ttfamily,
          frame=single,
          framesep=3pt,
          framerule=\borderrule,
          rulecolor=\bordercolor,
          aboveskip=0pt,
          belowskip=0pt,
          breaklines=false,
          lineskip=0.5pt
        ]
def restrictedFix[K, Qbase<:Tuple, Qret<:Tuple]
 (q: Qbase)
 (f: Qref[K, Qbase] => Qret)

 (IsTupleOfQueries[Qbase])
 (AllRowTypesAreNamedTuples[Qbase])
 (Size[Qbase]=:=Size[Qret] & Size[Qbase]=:=1)
 (Qret<:<ToRQuery[Qbase, ExtractD[Qret]])
 (NoReferencesMissing[K, Qbase, Qret])
 (NoDuplicateReferences[Qret])
: ToQuery[Qbase]
\end{lstlisting}
    \end{minipage}\hfill
    \setlength{\fboxsep}{0pt}
    \figbox{\begin{minipage}[t]{0.35\textwidth}
      \setlength{\mathindent}{0pt}
      \small
      \setlength{\jot}{2pt}
      \setlength{\arraycolsep}{2pt}
      \setlength{\fboxsep}{2pt}
        \vspace{-1.6em}
        \begin{align*}
            \frac{
            \highlight{\begin{array}{l}
                \Gamma \vdash \textit{q}: Q_\text{base} \hfill \boxed{\Gamma \vdash m : A}\\
                \Gamma \vdash f: Q_\text{ref} \rightarrow Q_\text{ret} \\
                Q_\text{ref} = (\text{RQuery}[A_i, C_i, I_i]){_{i=1}^n}\\
                Q_\text{base} = (\text{Query}[A_i, C_i]){_{i=1}^n}\\
                \Gamma \vdash A_i: (l_j: L_j){_{j=1}^{m_i}} \ \forall i{_{=1}^n} \\
                n = 1\\
                Q_\text{ret} = (\text{RQuery}[A_i, Set, D_i]){_{i=1}^n}\\
                \{ 1_\kappa, \ldots, n_\kappa \} \equiv \cup D_i{_{i=1}^n}\\
                \forall D_i \quad |D_i|\equiv|\cup D_i|
            \end{array}}
            }
            {\Gamma \vdash \textbf{fix}\ \text{\emph{f}}\ \text{\emph{bases}} : QT}
        \end{align*}
    \end{minipage}}
    \caption{{The typing rule for \textbf{fix} in \calc{} (right) and its implementation as \texttt{restrictedFix} in \system{} (left). Each premise of the typing rule corresponds to a type-level constraint in \system{}.}}    \label{fig:fix-impl}

\end{figure}
 \textit{Recursive Constraints.} {Figure~\ref{fig:fix-impl} provides an intuition for how each premise of \textbf{fix} in \calc{} is translated into a constraint in the \system{} implementation.
 In both \system{} and \calc{},
    lines 1--3 establish the types for the base and recursive cases;
    lines 5--6 enforce that the base case is tuple of Query and the row types are Named Tuples;
    line 7 enforces no mutual recursion;
    line 8 verifies the recursive case must be of type RQuery where the row types match the base cases and use set-semantics;
    and line 9--10 enforces linearity.
The constraints are expressed as implicit evidence parameters: \texttt{<:<} is a subtyping witness, \texttt{=:=} is a type equality witness, and predicates such as \texttt{IsTupleOfQueries} and \texttt{NoDuplicateReferences} are Match Types and type aliases that compute over tuple types.} \system{} uses the annotation \texttt{@implicitNotFound} to customize error messages, so that the user can see which constraint failed: i.e., if their recursive query was affine but not relevant.
Uniqueness between arguments of a single invocation of \textbf{fix} is implemented using constant integer types tagged with a fresh type \texttt{K} via intersection types (\texttt{I \& K}).

The restrictions of P2--P\total{} can be made configurable:
{
  For example, query~\ref{fig:ex-tyql-monotonic} disables monotonicity, query~\ref{fig:ex-tyql-linear} linearity, and query~\ref{fig:ex-tyql-bag} set semantics and constructor-freedom.
}
This is implemented in \texttt{fix} by {updating the \texttt{restrictedFix} definition in Figure~\ref{fig:fix-impl} by} adding additional type parameters for each restriction
\texttt{[\mono{}\textcolor{blue}{<:}Monotone, \mr{}\textcolor{blue}{<:}Mutual, \lin{}\textcolor{blue}{<:}Linear, \set{}\textcolor{blue}{<:}Category, \cf{}\textcolor{blue}{<:}ConstructorFree]}
to line 1 and an additional argument to line 4: \texttt{(options:(\mono{}, \mr{}, \lin{}, \set{}, \cf{}))}.
{For example, the \texttt{\lin{}} configuration object in query~\ref{fig:ex-tyql-linear} represents a configuration option with linearity untracked.}
\texttt{P2--6} are forwarded to the Match Types that compute the type constraints; if disabled, the constraint is trivially satisfied.

\subsection{Query Generation}\label{sec:implementation:query}
{
  The translation pipeline used by \system{} is shown in Figure~\ref{fig:arch}: a type-level AST is constructed directly by users via the \system{} library. The AST is then normalized and translated into a relational-algebra-like intermediate representation, i.e., a query plan, which is used to generate SQL strings that are executed using a database driver. Normalization and translation occur simultaneously to avoid any performance penalty from multiple AST traversals.}

{
  \system{} normalizes like \calc{}, shown in the ``Operational Semantics and Normalization for \calc{}'' section of the appendix.
  Normalized \system{} is translated into SQL by directly applying the techniques used by T-LINQ, adapted for combinator syntax and with an additional rule for \textbf{fix}. \system{} queries have a straightforward structural correspondence to SQL: \textbf{flatMap} maps to \texttt{JOIN}, \textbf{filter} to \texttt{WHERE} clauses, \textbf{map} to
  \texttt{SELECT} projections, and \textbf{union} to \texttt{UNION}, etc.
  For example, \texttt{table.filter(\_.x > 1).filter(\_.y < 2)} is normalized into \texttt{table.filter(r => r.x > 1 \&\& r.y < 2)} and translated into a \texttt{WHERE} node. A similar approach is applied to \texttt{flatMap}, for example the query \texttt{t1.flatMap(r1 => t2.flatMap(r2 => t3.map(r3 => (r1 = r1.x, r2 = r2.x, r3 = r3.x))))} is flattened into a single \texttt{JOIN} node.
  Each \textbf{fix} call represents a single fixpoint, so each instance of \textbf{fix} is translated to exactly one \texttt{WITH RECURSIVE}.
For example, a query \texttt{B.fix(Q => R)} where \texttt{B} and \texttt{R} are query subexpressions representing the base and recursive cases, is equivalent to \texttt{WITH RECURSIVE recursive1} \textit{b} \texttt{UNION} \textit{r;} \texttt{SELECT * FROM recursive1} where $b$ and $r$ are the SQL translations of \texttt{B} and \texttt{R}.}

To allow the composition of sub-queries and sub-expressions as well as abstraction over DSL expressions, \system{}
generates SQL lazily.
Users can peek at the generated SQL using the helper method \texttt{toSQLString}, but must call \texttt{run} to execute.
\system{} doesn't introduce any additional overheads from JVM boxing beyond what Scala does when working with generic ADTs or case classes. All boxing is handled by the JDBC driver regardless of whether the query was constructed with strings or with \system{}.

 \section{Evaluation}\label{sec:eval}

In this section, we evaluate \system{} with respect to real-world use cases and how effectively \system{} can constrain recursive queries {without compromising flexibility}.
To illustrate the range of real-world queries, we introduce a \emph{recursive query benchmark} (RQB) of 16 queries taken from open source code repositories and recent publications from various query domains, including business management, program analysis, graphs, and other classic fixed-point problems. We give a survey of modern database systems using our benchmark, identifying the ways each query can go wrong and which RDBMS supports each combination of P1--P\total{}.

We use the RQB to evaluate \system{} with respect to query \emph{coverage}: how many queries that would go wrong are prevented from compiling,
and \emph{impact}: if \system{} were to allow the query to compile, how it would fail on the database. Lastly, we evaluate \system{} with respect to performance and compare to alternative approaches.

\subsection{Recursive Query Benchmark and Survey of RDBMS}\label{sec:eval:rqb}
\begin{table}[t]
  \caption{
    Recursive Query Benchmark.
    \textcolor{blue}{\tiny\ding{110}} Datalog \textcolor{violet}{\tiny\ding{108}} Program analysis \textcolor{teal}{\tiny\ding{115}} Recursive SQL \textcolor{black}{\tiny\ding{72}} Graph. \\
    \mono{} (monotone): \textcolor{dkgreen}{\ding{51}} no aggregation \textcolor{orange}{\textbullet} stratified aggregation \textcolor{red}{\ding{55}} unstratified aggregation. \\
     \mr{} (non-mutually-recursive), \lin{} (linear), \set{} (set-semantic), \cf{} (constructor-free): \textcolor{dkgreen}{\ding{51}} property holds.
}\label{tab:benchmark}
\centering
\renewcommand{\arraystretch}{0.9}
\setlength{\tabcolsep}{1pt}
\begin{tabular}{|>{\raggedright\arraybackslash}m{3.5cm}|m{7.5cm}|c|c|c|c|c|}
\hline
\textbf{Query} & \textbf{Description} & \textbf{\mono{}} & \textbf{\mr{}} & \textbf{\lin{}} & \textbf{\set{}} & \textbf{\cf{}} \\ \hline

Even-Odd (EO) & \textcolor{blue}{\tiny\ding{110}} Program to generate even/odd numbers.
& \textcolor{dkgreen}{\ding{51}} & \textcolor{red}{\ding{55}} & \textcolor{dkgreen}{\ding{51}} & \textcolor{dkgreen}{\ding{51}} & \textcolor{red}{\ding{55}} \\ \hline

CSPA & \textcolor{violet}{\tiny\ding{108}} Graspan's Context Sensitive Pointer Analysis~\cite{graspan}.
& \textcolor{dkgreen}{\ding{51}} & \textcolor{red}{\ding{55}} & \textcolor{red}{\ding{55}} & \textcolor{dkgreen}{\ding{51}} & \textcolor{dkgreen}{\ding{51}} \\ \hline

Company Control (CC) & \textcolor{teal}{\tiny\ding{115}} Calculates corporate controlling relationships~\cite{DBLP:conf/vldb/MumickPR90}.
& \textcolor{red}{\ding{55}} & \textcolor{red}{\ding{55}}& \textcolor{dkgreen}{\ding{51}} & \textcolor{dkgreen}{\ding{51}} & \textcolor{red}{\ding{55}} \\ \hline

PointsToCount (PTC) & \textcolor{violet}{\tiny\ding{108}} Find the count of objects that a variable of a given name may point to~\cite{flan} (Java variant).
& \textcolor{orange}{\textbullet} & \textcolor{red}{\ding{55}} & \textcolor{red}{\ding{55}} & \textcolor{dkgreen}{\ding{51}} & \textcolor{dkgreen}{\ding{51}} \\ \hline

Chain Of Trust (COT) & \textcolor{blue}{\tiny\ding{110}} Security trust propagation through direct and indirect friendships.
& \textcolor{dkgreen}{\ding{51}} & \textcolor{red}{\ding{55}} & \textcolor{dkgreen}{\ding{51}} & \textcolor{red}{\ding{55}} & \textcolor{dkgreen}{\ding{51}} \\ \hline

Java Points To (JPT) & \textcolor{violet}{\tiny\ding{108}} Field-sensitive subset-based points to analysis~\cite{flix}.
& \textcolor{dkgreen}{\ding{51}} & \textcolor{red}{\ding{55}} & \textcolor{red}{\ding{55}} & \textcolor{red}{\ding{55}} & \textcolor{dkgreen}{\ding{51}} \\ \hline

Party & \textcolor{black}{\tiny\ding{72}} Social media query for party attendees~\cite{rasql}.
& \textcolor{red}{\ding{55}} & \textcolor{red}{\ding{55}} & \textcolor{dkgreen}{\ding{51}} & \textcolor{red}{\ding{55}} & \textcolor{red}{\ding{55}} \\ \hline

CBA & \textcolor{blue}{\tiny\ding{110}} Constraint-based analysis query~\cite{souffle-site}.
& \textcolor{orange}{\textbullet} & \textcolor{red}{\ding{55}} & \textcolor{red}{\ding{55}} & \textcolor{red}{\ding{55}} & \textcolor{dkgreen}{\ding{51}} \\ \hline

Single Source Shortest Path (SSSP) & \textcolor{black}{\tiny\ding{72}} Computes the shortest path from a given source node to all other nodes in a weighted graph~\cite{dcdatalog}.
& \textcolor{orange}{\textbullet} & \textcolor{dkgreen}{\ding{51}} & \textcolor{dkgreen}{\ding{51}} & \textcolor{dkgreen}{\ding{51}} & \textcolor{red}{\ding{55}} \\ \hline

Same-Generation (SG) & \textcolor{teal}{\tiny\ding{115}} Find same-generation descendants of a person~\cite{amateur}.
& \textcolor{dkgreen}{\ding{51}} & \textcolor{dkgreen}{\ding{51}} & \textcolor{dkgreen}{\ding{51}} & \textcolor{dkgreen}{\ding{51}} & \textcolor{red}{\ding{55}} \\ \hline

Andersen's Points To (APT) & \textcolor{violet}{\tiny\ding{108}} Context-insensitive, flow-insensitive interprocedural pointer analysis~\cite{recstep}.
& \textcolor{dkgreen}{\ding{51}} & \textcolor{dkgreen}{\ding{51}}& \textcolor{red}{\ding{55}} & \textcolor{dkgreen}{\ding{51}} & \textcolor{dkgreen}{\ding{51}} \\ \hline

All Pairs Shortest Path (APSP) & \textcolor{black}{\tiny\ding{72}} Compute the shortest paths between all pairs of nodes in a weighted graph~\cite{dcdatalog}.
& \textcolor{red}{\ding{55}} & \textcolor{dkgreen}{\ding{51}} & \textcolor{red}{\ding{55}} & \textcolor{dkgreen}{\ding{51}} & \textcolor{red}{\ding{55}} \\ \hline

Graphalytics (TC) & \textcolor{black}{\tiny\ding{72}} Directed cyclic graph reachability query. Uses list data structure to check for cycles~\cite{ldbc}.
& \textcolor{dkgreen}{\ding{51}} & \textcolor{dkgreen}{\ding{51}} & \textcolor{dkgreen}{\ding{51}} & \textcolor{red}{\ding{55}} & \textcolor{red}{\ding{55}} \\ \hline

Bill of Materials (BOM) [\textit{Stratified}] & \textcolor{teal}{\tiny\ding{115}} Business query for days to deliver a product made of subparts with stratified aggregation~\cite{ibm-bom}.
& \textcolor{orange}{\textbullet} & \textcolor{dkgreen}{\ding{51}} & \textcolor{dkgreen}{\ding{51}} & \textcolor{red}{\ding{55}} & \textcolor{dkgreen}{\ding{51}} \\ \hline

Orbits & \textcolor{blue}{\tiny\ding{110}} Orbits of cosmological objects~\cite{souffle-site}.
& \textcolor{orange}{\textbullet} & \textcolor{dkgreen}{\ding{51}} & \textcolor{red}{\ding{55}} & \textcolor{red}{\ding{55}} & \textcolor{dkgreen}{\ding{51}} \\ \hline

Data Flow (DF) & \textcolor{blue}{\tiny\ding{110}} Models program control flow~\cite{souffle-site}.
& \textcolor{dkgreen}{\ding{51}} & \textcolor{dkgreen}{\ding{51}} & \textcolor{red}{\ding{55}} & \textcolor{red}{\ding{55}} & \textcolor{dkgreen}{\ding{51}} \\ \hline
\end{tabular}
\end{table}
 In this section, we present a \emph{recursive query benchmark} (RQB) comprising 16 queries across diverse domains such as business management, program analysis, graph queries, and classic fixed-point problems and show how each query behaves on different RDBMS.

The goal of our benchmark is to simulate a broad {range} of recursive queries.
We have selected 16 queries to represent classes of queries for each combination of properties (except \rr{} as all queries are range-restricted). We excluded trivially safe queries (those violating none of P1--P\total{}) because they pose no risk of causing behaviors B1--B3.
Identifying the frequency of each query across real-world applications remains future work.
All queries run and terminate on at least one of the evaluated RDBMS. For the monotonicity property, we have selected a mix of queries without aggregation or negation,
with stratified-aggregation, and with unstratified aggregation. Table~\ref{tab:benchmark} illustrates the benchmark property matrix.
The full set of queries in the benchmark are included in the artifacts.

As shown in Section~\ref{sec:background}, support for recursive queries varies widely across RDBMS. To get a sense of what classes of queries each system supports, we ran the 16 queries with cyclic input data and with acyclic input data. The results of the RQB on four RDBMS are presented in Table~\ref{tab:coverage}. We represent queries that terminated with the full result with \textcolor{dkgreen}{\large\ding{51}}, database error (B1) with \textcolor{red}{\large\ding{55}}, incomplete results (B2) with \textcolor{orange}{\large\ding{119}}, and nontermination (B3) with \textcolor{magenta}{\large\ding{116}}. If the query exhibited different behavior based on whether the input data was cyclic or not, for example if the query terminates correctly with acyclic data but not cyclic data, then the query is represented as \textcolor{dkgreen}{\large\ding{51}} \ding{120} \textcolor{magenta}{\large\ding{116}} with the acyclic data on the left and cyclic data on the right.
We consider ``incomplete'' results to be data that is missing results present using a non-SQL version of each algorithm, either via imperative programs or Datalog programs. The \textcolor{orange}{\large\ding{119}} and \textcolor{magenta}{\large\ding{116}} classification indicates that it is possible to find an input dataset that returns incomplete results or does not terminate. The databases used are DuckDB v1.2, Postgres v15, SQLite v3.39, and MariaDB v11.5.2 with the configuration \texttt{---skip-standard-compliant-cte}.

\subsection{\system{} Coverage and Impact}\label{sec:eval:coverage}
\begin{table}[t]
  \caption{
    Effectiveness of TyQL in Recursive Query Error Detection across Modern Databases.
  }\label{tab:coverage}
\centering
\setlength{\tabcolsep}{1pt}
\begin{minipage}[t]{0.79\textwidth}
\begin{tabular}{|l|l|c|c|c|c|}
\hline
\multicolumn{2}{|c|}{\textit{Benchmark}} & \multicolumn{4}{c|}{\textit{Database Behavior}} \\
\hline
\textbf{Query} & \textbf{Violated Properties} & \textbf{DuckDB} & \textbf{Postgres} & \textbf{SQLite} & \textbf{MariaDB} \\
\hline

Even-Odd & \mr{}, \cf{} & \textcolor{dkgreen}{\ding{51}} & \textcolor{red}{\ding{55}} & \textcolor{red}{\ding{55}} & \textcolor{orange}{\ding{119}} \\
\hline

CSPA & \mr{}, \lin{} & \textcolor{orange}{\ding{119}} & \textcolor{red}{\ding{55}} & \textcolor{red}{\ding{55}} & \textcolor{red}{\ding{55}} \\
\hline

CC & \mono{}, \mr{}, \cf{} & \textcolor{orange}{\ding{119}} & \textcolor{red}{\ding{55}} & \textcolor{red}{\ding{55}} & \textcolor{red}{\ding{55}} \\
\hline

PTC & \mr{}, \lin{} & \textcolor{orange}{\ding{119}} & \textcolor{red}{\ding{55}} & \textcolor{red}{\ding{55}} & \textcolor{red}{\ding{55}} \\
\hline

COT & \mr{}, \set{} & \textcolor{dkgreen}{\ding{51}} \ding{120}\
\textcolor{magenta}{\ding{116}}\textsubscript{\textcolor{blue}{\tiny\ding{83}}} &
\textcolor{red}{\ding{55}} & \textcolor{red}{\ding{55}} & \textcolor{red}{\ding{55}} \\
\hline

JPT & \mr{}, \lin{}, \set{} & \textcolor{orange}{\ding{119}}\ \ \ \ding{120}\
\textcolor{magenta}{\ding{116}}\textsubscript{\textcolor{blue}{\tiny\ding{83}}} &
\textcolor{red}{\ding{55}} & \textcolor{red}{\ding{55}} & \textcolor{red}{\ding{55}} \\
\hline

Party & \mono{}, \mr{}, \set{}, \cf{} & \textcolor{orange}{\ding{119}} & \textcolor{red}{\ding{55}} & \textcolor{red}{\ding{55}} & \textcolor{magenta}{\ding{116}} \\
\hline

CBA & \mr{}, \lin{}, \set{} & \textcolor{orange}{\ding{119}}\ \ \ \ding{120}\
\textcolor{magenta}{\ding{116}}\textsubscript{\textcolor{blue}{\tiny\ding{83}}} &
\textcolor{red}{\ding{55}} & \textcolor{red}{\ding{55}} & \textcolor{red}{\ding{55}} \\
\hline

SSSP & \cf{} & \textcolor{dkgreen}{\ding{51}} \ding{120}
\textcolor{magenta}{\ding{116}}\textsubscript{\textcolor{white}{\tiny\ding{83}}} &
\textcolor{dkgreen}{\ding{51}} \ding{120} \textcolor{magenta}{\ding{116}} &
\textcolor{dkgreen}{\ding{51}} \ding{120} \textcolor{magenta}{\ding{116}} &
\textcolor{dkgreen}{\ding{51}} \\
\hline

SG & \cf{} & \textcolor{dkgreen}{\ding{51}} & \textcolor{dkgreen}{\ding{51}} & \textcolor{dkgreen}{\ding{51}} & \textcolor{dkgreen}{\ding{51}} \\
\hline

APT & \lin{} & \textcolor{orange}{\ding{119}} & \textcolor{orange}{\ding{119}} & \textcolor{red}{\ding{55}} & \textcolor{orange}{\ding{119}} \\
\hline

APSP & \mono{}, \lin{}, \cf{} & \textcolor{orange}{\ding{119}} & \textcolor{orange}{\ding{119}} & \textcolor{red}{\ding{55}} & \textcolor{dkgreen}{\ding{51}} \\
\hline

TC & \set{}, \cf{} & \textcolor{dkgreen}{\ding{51}} & \textcolor{dkgreen}{\ding{51}} & \textcolor{dkgreen}{\ding{51}} & \textcolor{dkgreen}{\ding{51}} \\
\hline

BOM & \set{} & \textcolor{dkgreen}{\ding{51}} \ding{120}
\textcolor{magenta}{\ding{116}}\textsubscript{\textcolor{blue}{\tiny\ding{83}}} &
\textcolor{dkgreen}{\ding{51}} \ding{120}
\textcolor{magenta}{\ding{116}}\textsubscript{\textcolor{blue}{\tiny\ding{83}}} &
\textcolor{dkgreen}{\ding{51}} \ding{120}
\textcolor{magenta}{\ding{116}}\textsubscript{\textcolor{blue}{\tiny\ding{83}}} &
\textcolor{dkgreen}{\ding{51}} \\
\hline

Orbits & \lin{}, \set{} & \textcolor{orange}{\ding{119}} \ \ \ \ding{120}
\textcolor{magenta}{\ding{116}}\textsubscript{\textcolor{blue}{\tiny\ding{83}}} &
\textcolor{orange}{\ding{119}} \ \ \ \ding{120}
\textcolor{magenta}{\ding{116}}\textsubscript{\textcolor{blue}{\tiny\ding{83}}} &
\textcolor{red}{\ding{55}} &
\textcolor{magenta}{\ding{116}} \\
\hline

Data Flow & \lin{}, \set{} & \textcolor{orange}{\ding{119}} \ \ \ \ding{120}
\textcolor{magenta}{\ding{116}}\textsubscript{\textcolor{blue}{\tiny\ding{83}}} &
\textcolor{orange}{\ding{119}} \ \ \ \ding{120}
\textcolor{magenta}{\ding{116}}\textsubscript{\textcolor{blue}{\tiny\ding{83}}} &
\textcolor{red}{\ding{55}} &
\textcolor{magenta}{\ding{116}} \\
\hline

\end{tabular}
\end{minipage}\hfill
\begin{minipage}[t]{0.21\textwidth}
\small
\vspace{-2.5cm}
\textcolor{dkgreen}{\ding{51}} executed OK\\[4pt]
\textcolor{red}{\ding{55}} \ runtime error (B1)\\[4pt]
\textcolor{orange}{\ding{119}}\ \  incomplete results \\ \hspace*{6pt}  (B2)\\[4pt]
\textcolor{magenta}{\ding{116}} nontermination \\ \hspace*{6pt} (B3)\\[4pt]
\textcolor{blue}{\tiny\ding{83}}\ \ modifying query to use set semantics enables termination
\end{minipage}
\end{table}

In this section, we use the RQB to evaluate the effectiveness of \system{} in achieving comprehensive query coverage while preventing queries that may go wrong from compiling, illustrated in Table~\ref{tab:coverage}.
The goal of TyQL is to target unwanted database behaviors B1--B3, while the \textit{mechanism} is via deriving the query properties P1--P\total{} using the type system. P1--P\total{} are ground truths for each query, regardless if the query is expressed in raw SQL, \system{}, Datalog, or another language. With respect to P1--P\total{}, there are no false positives or false negatives because the rules presented in Figure~\ref{fig:restricted-host} are derived directly from Definitions~\ref{def:rr}--\ref{def:cf}.
The behaviors B1--B3, however, are a property of the semantics of the database backend. Therefore, it is only possible to classify queries as positive or negative \textit{with respect to a database} and a set of properties P1--P\total{} enforced by TyQL. Table~\ref{tab:coverage} shows how each query can be considered a false positive with respect to a database (represented \textcolor{dkgreen}{\large\ding{51}}, e.g. no problems during execution), or a true positive (and the impact on execution, represented by B1 \textcolor{red}{\large\ding{55}}, B2 \textcolor{orange}{\large\ding{119}}, and B3 \textcolor{magenta}{\large\ding{116}}), for a set of constraints (``Violated Properties'').
True negatives are the queries that run correctly for the set of properties \textit{not} violated.
For example, using Postgres with properties \mono{}, \mr{}, \lin{} over cyclic data, TC is a false positive while SSSP is a true positive.

The SG query exemplifies the class of queries that exhibit only the properties fully supported by the SQL specification: it is linear, monotonic, set-semantic, and not mutually recursive.
Only with constructor-freedom (\cf{}) enforced will \system{} reject this query.

For the queries that return incorrect results, all are either mutually recursive or non-linear. The missing tuples are those that would have been generated by intermediate results from previous iterations that are ``forgotten'' by the SQL engine due to the implementation only reading tuples derived in the immediately preceding iteration. For the queries that do not terminate, all but one use bag semantics. Careless use of a bag-semantic query can cause nontermination for input data sets that have cycles (\textcolor{blue}{\large\ding{83}} next to the non-terminating queries indicates that if we change our bag semantics to set semantics, then the query will terminate on both acyclic and cyclic input data). Users may prefer to pay a performance penalty due to the duplicate elimination cost of set semantics to avoid the risk that their queries will not terminate. Nonterminating queries can have a significant performance impact, both on the application and on the other users of the database, due to interference. Some RDBMS allow users to set a max recursion depth to avoid infinite recursion, although it is not obvious how deep to set this without trial and error. It is clear that the impact of incorrect results is more damaging than a database throwing an error. However, whether nontermination is more impactful than incomplete results depends on the context of the application and database system. So far, the only RDBMS we have seen to officially include non-linear or mutual recursion is MariaDB~\cite{mariadb}, although in our experiments
some recursive queries yielded results that diverged from standard Datalog semantics and documented behavior.

The TC query returns correctly on all evaluated systems using bag semantics, even with cyclic input data. The reason for this is that the query itself checks for duplicates: in DuckDB, the query has a list that tracks visited nodes, while in systems that do not support lists, the query appends to a string. Conversely, the SSSP query will not terminate on cyclic data even with set semantics. The reason for this is cost propagation, where each tuple generated at each recursive step includes the ``weight'' of the newly discovered path. If the query reaches a cycle, the weight of the path will infinitely increase and the ever-changing \texttt{cost} column will prevent the set difference from removing already-discovered paths. The property responsible for both behaviors is constructor-freedom (\cf{}): the TC query constructs new values used to detect cycles while the SSSP query constructs new values that lead to cyclic reinforcement and nontermination. The SSSP and TC queries exemplify why \system{} cannot strictly enforce all properties, even on a single RDBMS, as the same property can be responsible for nontermination in some queries but prevent nontermination in others.

In summary, we evaluate the ability of \system{} to identify queries that will fail and find that it can be tuned to successfully reject all problematic queries. However, as there are queries that will run without problems that violate one or more of P1--P\total{}, the strict safety guarantees of \system{} come at the cost of expressivity.
As with many type systems, \system{} takes a conservative approach to correctness and can reject queries that may, if the data has certain properties, return successfully. To maximize usability and practicality, \system{} users always have the choice to tune which combination of properties P1--P\total{} are relaxed.

\subsection{Performance and State-of-the-Art}\label{sec:eval:perf}

In this section we evaluate \system{} with respect to performance. Developers who wish to run fixpoint algorithms on data stored in a RDBMS have several options. The most immediate choice is to simply read data into memory and then execute their algorithm using the programming language constructs. Most standard libraries, including the Scala Collections API, do not include a fixpoint operator, so users must implement their own iterative control flow. The benefit of this approach is that it is fully customizable, yet it puts the burden onto the developer and the machine where the application is running.
Alternatively, users may offload computation to the database. A natural way to do this would be to use language-integrated query to compose queries that are compiled to SQL and sent to the RDBMS. Yet if the query library does not support recursion, then users will need to handle control flow at the application level and send only non-recursive queries at each iteration, or default to queries expressed using strings, which may be painful to write but will show good performance.

\begin{table}[t]
      \caption{Performance of TyQL compared to Scala Collections, Non-Recursive SQL using ScalaSQL, and recursive SQL strings.
    * Equivalent because difference is less than the JMH margin of error~\cite{jmh}.}\label{tab:perf}
\resizebox{\columnwidth}{!}{{\tiny
\setlength{\tabcolsep}{1pt}
\centering

\begin{tabular}{llccccc c c c}
\toprule
\textbf{Query\ \ } & \textbf{Size\ \ \ } & \textbf{It\ \ } & \textbf{TyQL(s)\ } & \textbf{Coll(s)\ } & \textbf{ScalaSQL(s)\ } & \textbf{SQL-Str(s)\ } & \textbf{vs.Coll\ } & \textbf{vs.ScalaSQL\ } & \textbf{vs.SQL-Str}\\
\midrule
\multirow{3}{*}{SG}&0.01MB&3&0.008&0.001&0.049&0.005&0.18X&6.50X&0.71X*\\&10MB&15&0.146&3.487&0.393&0.143&23.91X&2.69X&0.98X*\\&100MB&189&38.447&TO&TO&38.405&>15.61X&>15.61X&1.00X*\\ \hline
\multirow{3}{*}{APT}&0.01MB&3&0.015&0.002&0.059&0.011&0.11X&4.08X&0.74X*\\&0.02MB&4&0.019&3.457&0.099&0.017&178.46X&5.11X&0.90X*\\&0.04MB&9&0.042&42.406&0.246&0.038&1019.28X&5.90X&0.90X*\\ \hline
\multirow{3}{*}{APSP}&0.01MB&3&0.017&0.004&0.046&0.012&0.22X&2.76X&0.72X*\\&1MB&3&0.060&19.848&0.206&0.055&331.38X&3.44X&0.92X*\\&5MB&4&0.207&TO&0.723&0.202&>2898.55X&3.49X&0.98X*\\ \hline
\multirow{3}{*}{BOM}&0.01MB&2&0.009&0.002&0.039&0.008&0.21X&4.27X&0.88X*\\&2MB&5&0.077&132.965&0.519&0.070&1725.11X&6.73X&0.90X*\\&20MB&22&1.832&TO&14.019&1.813&>327.51X&7.65X&0.99X*\\ \hline
\multirow{3}{*}{CBA}&0.02MB&9&0.031&0.005&0.437&0.026&0.17X&14.22X&0.86X*\\&0.1MB&1&0.023&TO&0.075&0.017&>26086.96X&3.21X&0.75X*\\&0.2MB&1&0.021&TO&0.077&0.019&>28571.43X&3.71X&0.93X*\\ \hline
\multirow{3}{*}{CC}&0.01MB&3&0.011&0.002&0.071&0.009&0.22X&6.57X&0.83X*\\&1MB&3&0.277&14.297&0.695&0.285&51.67X&2.51X&1.03X*\\&1.5MB&3&0.824&42.247&0.911&0.805&51.29X&1.11X*&0.98X*\\ \hline
\multirow{3}{*}{CSPA}&0.01MB&5&0.025&0.003&0.182&0.021&0.13X&7.33X&0.85X*\\&2MB&11&0.967&TO&5.420&0.954&>620.48X&5.61X&0.99X*\\&10MB&14&28.379&TO&278.348&28.200&>21.14X&9.81X&0.99X*\\ \hline
\multirow{3}{*}{DF}&0.01MB&3&0.008&0.002&0.049&0.006&0.24X&6.07X&0.80X*\\&0.03MB&3&0.007&18.265&0.065&0.008&2592.27X&9.17X&1.11X*\\&0.05MB&5&0.012&266.167&0.134&0.011&21897.75X&11.02X&0.93X*\\ \hline
\multirow{3}{*}{EO}&0.01MB&17&0.012&0.004&0.373&0.009&0.30X&30.32X&0.77X*\\&1MB&-&177.698&TO&TO&182.490&>3.38X&>3.38X&1.03X*\\&2MB&-&465.739&TO&TO&485.292&>1.29X&>1.29X&1.04X*\\ \hline
\multirow{3}{*}{JPT}&0.02MB&3&0.014&0.002&0.081&0.013&0.15X&5.68X&0.91X*\\&0.05MB&16&0.053&TO&0.861&0.053&>11320.75X&16.20X&1.00X*\\&0.1MB&-&0.382&TO&TO&0.377&>1570.68X&>1570.68X&0.99X*\\ \hline
\multirow{3}{*}{Orbits}&0.01MB&2&0.009&0.002&0.039&0.006&0.21X*&4.51X&0.73X*\\&1MB&2&0.065&TO&0.347&0.064&>9230.77X&5.32X&0.99X*\\&10MB&2&0.693&TO&0.960&0.696&>865.80X&1.39X&1.00X*\\ \hline
\multirow{3}{*}{Party}&0.01MB&5&0.008&0.003&0.083&0.008&0.40X&10.91X&1.04X*\\&2MB&5&0.086&TO&0.596&0.080&>6976.74X&6.97X&0.93X*\\&20MB&7&1.021&TO&3.611&1.042&>587.66X&3.54X&1.02X*\\ \hline
\multirow{3}{*}{PTC}&0.02MB&3&0.014&0.003&0.086&0.014&0.18X&6.13X&1.00X*\\&0.05MB&16&0.053&TO&0.874&0.050&>11320.75X&16.49X&0.93X*\\&0.1MB&-&1.551&TO&TO&1.505&>386.85X&>386.85X&0.97X*\\ \hline
\multirow{3}{*}{SSSP}&0.01MB&5&0.013&0.002&0.074&0.011&0.17X&5.54X&0.84X*\\&10MB&9&0.029&0.522&0.176&0.027&17.83X&6.01X&0.91X*\\&25MB&42&0.125&219.573&1.982&0.125&1762.89X&15.91X&1.01X*\\ \hline
\multirow{3}{*}{TC}&0.01MB&2&0.007&0.001&0.033&0.006&0.16X&4.44X&0.81X*\\&5MB&5&0.025&0.336&0.087&0.019&13.66X&3.52X&0.77X*\\&10MB&10&0.070&6.881&0.314&0.067&98.45X&4.50X&0.95X*\\ \hline
\multirow{3}{*}{COT}&0.01MB&7&0.012&0.003&0.182&0.011&0.24X&14.88X&0.89X*\\&1MB&7&0.104&TO&2.299&0.103&>5769.23X&22.08X&0.99X*\\&15MB&-&1.267&TO&12.179&1.236&>473.56X&9.61X&0.98X*\\ \bottomrule
\end{tabular}
}
}
\end{table}

\subparagraph{Experimental Setup.} Table~\ref{tab:perf} shows the execution time (s) of the RQB presented in Section~\ref{sec:eval:rqb}. The ``Coll'' column shows the execution time of the query implemented purely within the programming language using the Collections API and no database. The ``ScalaSQL'' column shows the execution time of the latest state-of-the-art language-integrated query library ScalaSQL, using non-recursive SQL queries (as recursion is not supported) {run on an embedded relational database, DuckDB~\cite{duckdb}}. This approach is representative of other language-integrated query libraries in Scala since they support only non-recursive SQL. Because we use an embedded database that runs within the same process as the application, avoiding the overhead of round-trips between database and application, this approach is equivalent to a PL/SQL approach. The ``SQL-Str'' column shows the execution time of sending raw strings directly to the JDBC driver without any language-integration.
The ``TyQL'' column shows the execution time of the query using \system{}. The rightmost columns, ``vs.Coll'', ``vs. ScalaSQL'', and ``vs. SQL-Str'' show the speedup of \system{} over each respective approach. In the speedup columns, ``>'' indicates that the baseline did not terminate within a 10-minute timeout so we calculate the minimum speedup. The ``It'' column states the number of iterations needed for the Collections API and non-recursive SQL to reach a fixed point (it was not possible to extract the number of iterations from the DuckDB internals without impacting the result) and the ``size'' column states the total size of the input relations.

The fixpoint implementation used in the non-recursive SQL and collections-only implementation is tail-recursive and based on the canonical example given in \emph{Scala by Example}~\cite{sbe} extended to use {the} bottom-up Semi-Naive evaluation algorithm used internally in the database.
The database used is DuckDB v1.2 with JDBC v.1.2.1 and queries that risked nontermination are run with set semantics. Each query is run on synthetic data of three different input relation sizes.
Experiments are run on an Intel(R) Xeon(R) Gold 5118 CPU @ 2.30GHz (2 x 12-core) with 395GB RAM, on
Ubuntu 22.04 LTS with Linux kernel 6.5.0-17-generic
and Scala 3.8.2 with JDK 17.0.9, OpenJDK 64-Bit Server VM, 17.0.9+9-jvmci-23.0-b22, with \texttt{-Xmx8G}. We use Java Benchmarking Harness (JMH)~\cite{jmh} v1.37 with \texttt{-i 5 -wi 5}.

\subparagraph{Experimental Results and Analysis.} The smallest of the three input data sets shows the range of data sizes for which the Collections API  outperforms the other approaches due to avoiding the overhead of database connection and initialization.
ScalaSQL has a higher overhead than \system{} or SQL strings, as the query initialization overhead happens at every iteration.
The medium-sized dataset shows data input sizes where the ScalaSQL approach outperforms the Collections API for several reasons: the RDBMS query optimizer can effectively select efficient query plans and join algorithms; as the data is not sent back to the application at each iteration, overheads due to boxing and unboxing of primitive types are avoided until iteration has concluded. For this input data size \system{} outperforms the other approaches due to avoiding multi-query overhead, storing and copying intermediate relations, and internal database optimizations. The memory usage of Collections is higher than either the \system{} or ScalaSQL because the data is not stored in the RDBMS, putting additional pressure on the JVM garbage collector.
The largest data size shows the cases where the Collections API and non-recursive SQL approaches may run out of memory or time-out after 10 minutes.
The graph algorithm queries (SG, TC, SSSP) are run on larger datasets, while program analysis queries (APT, CBA, Data Flow, JPT, PTC) are run on smaller datasets to avoid all systems running out of memory.

\begin{table}[t]
  \caption{Tradeoff between safety and performance compared to raw SQL strings.
}\label{tab:perf-stat}
\centering
\begin{scriptsize}
\setlength{\tabcolsep}{8pt}

\begin{tabular}{l rrr cccc}
\toprule
& \multicolumn{3}{c}{\textbf{Slowdown vs. SQL String}} & \multicolumn{4}{c}{\textbf{Risk Mitigation}} \\
\cmidrule(lr){2-4} \cmidrule(lr){5-8}
& \multicolumn{1}{c}{\textbf{Max}} & \multicolumn{1}{c}{\textbf{Min}} & \multicolumn{1}{c}{\textbf{Avg}} & \textbf{Data-Type} & \textbf{B1} & \textbf{B2} & \textbf{B3} \\
\midrule

\textbf{SQL-Str} & 1X & 1X & 1X & \textcolor{red}{\normalsize\ding{55}} & \textcolor{red}{\normalsize\ding{55}} & \textcolor{red}{\normalsize\ding{55}} & \textcolor{red}{\normalsize\ding{55}} \\

\textbf{Coll} & 28571.43X&0.14X&2885.14X & \textcolor{dkgreen}{\normalsize\ding{51}} & - & \textcolor{red}{\normalsize\ding{55}} & \textcolor{red}{\normalsize\ding{55}} \\

\textbf{ScalaSQL} & 1570.68X&1X&48.74X & \textcolor{dkgreen}{\normalsize\ding{51}} & - & \textcolor{red}{\normalsize\ding{55}} & \textcolor{red}{\normalsize\ding{55}} \\

\textbf{TyQL} & 1X&1X&1X & \textcolor{dkgreen}{\normalsize\ding{51}} & \textcolor{dkgreen}{\normalsize\ding{51}} & \textcolor{dkgreen}{\normalsize\ding{51}} & \textcolor{dkgreen}{\normalsize\ding{51}} \\
\bottomrule
\end{tabular}
\end{scriptsize}

\end{table} Statistics over the data presented in Table~\ref{tab:perf} are shown in Table~\ref{tab:perf-stat}. Queries whose JMH error margin exceeds the difference in execution time are considered equivalent due to normal variation in the JVM JIT (indicated as 1X). \system{} shows no performance penalty compared to raw SQL strings, with a significant performance gain over Collections and ScalaSQL using non-recursive SQL.
Table~\ref{tab:perf-stat} illustrates the tradeoff between customizability, performance, and safety in the state-of-the-art for query execution: raw SQL strings show the best performance but provide no safety guarantees (even for errors like data-type mismatch, SQL injection, etc.); hand-written imperative programs provide expressibility and flexibility, but only safety with respect to the programming language; language-integrated non-recursive queries put less burden on the developer and show better performance than imperative implementations and worse performance than raw SQL, while providing only non-recursive database safety guarantees, and lastly \system{} puts the least burden on the developer and shows performance equivalent to raw SQL strings while providing the strongest safety guarantees.

 \section{Related Work}\label{sec:relatedwork}
\subsection{Embedded Query Languages}
Type-safe embedded query languages using collections were pioneered by $\mathcal{M}$~\cite{NRC} and Kleisli~\cite{kleisli} and found commercial success in LINQ, formalized in T-LINQ~\cite{tlinq}.
{Existing language-integrated query systems do not support recursive SQL, either via runtime exception or type-and-effect systems to exclude recursion from translatable fragments~\cite{DBLP:conf/dbpl/Cooper09}.}
There has been work in general-purpose functional languages with fixpoint semantics~\cite{flix, funprogwdatalog}, operating as a functional Datalog. \system{} shares the goal of using functional abstractions to structure recursion while ensuring safety through a well-defined type system. These approaches extend Datalog semantics while \system{} targets RDBMS, which requires abstracting over different database semantics to ensure portability.

There has been significant interest in embedded SQL support in Scala. ScalaQL~\cite{DBLP:conf/sle/SpiewakZ09} uses anonymous inner classes to model row types, while Slick~\cite{slick,shaikhha2013embedded}
provides SQL embedding in Scala using macros~\cite{DBLP:conf/gpce/JovanovicSSNKO14} and the implicit resolution in Scala's type system. ScalaSQL~\cite{scalasql} uses higher-kinded case classes to model rows, and Quill~\cite{quill} uses refinement types, macros and quotation to compile SQL queries at Scala compile-time. Most of these libraries aim to provide ergonomic SQL APIs that expose the SQL query and data model to the user, that is, they take the spirit of the Collections API while still exposing the SQL query model to users. In this work we aim for \emph{transparent persistence}~\cite{transparent} so the distinction between processing of data stored in the native language collections or a database is as minimal as possible.

\subsection{Recursion and Relational Databases}
Researchers have attempted to address the impedance mismatch problem within the data management system: object-oriented or document databases provide data models and query languages that integrate cleanly with general-purpose programming languages but are more difficult to optimize for efficient execution~\cite{cow}. Alternatively, object-relational mapping libraries (ORMs) attempt to provide object-oriented abstractions on top of relational databases but can also suffer from performance penalties and obscure query behavior~\cite{orm}.

The problem of extracting relational algebraic properties from general-purpose programs is known as the \emph{query extraction problem} and has been successfully applied to synthesize queries from application code~\cite{froid}.
Recent work in this area has used SQL \texttt{WITH RECURSIVE} as a compilation target when compiling user-defined functions written in procedural language extensions like PL/SQL~\cite{compiling-away} or Python functions~\cite{snakes}. The aim of this line of work is to accept arbitrary programs written in general-purpose languages and compile them to SQL, while the goal of \system{} is to provide type-safe recursive language-integrated query using a compile-time-restricted embedded DSL. RDD2SQL~\cite{rdd2sql} uses counterexample-guided inductive synthesis to automatically translate functional database APIs like Spark RDDs into SQL but does not specifically target recursion. Novel extensions to SQL with cleaner recursion semantics have been proposed~\cite{fixation} but are not widely implemented in commercial databases.

Datalog is one of the most successful query languages with recursion capabilities, which has been used for fixpoint computations in program analysis~\cite{souffle,recstep,DBLP:conf/cav/JordanSS16}.
As the core Datalog disallows non-monotone operations, various extensions of it have been proposed~\cite{DBLP:journals/pacmpl/KloppEP24a,DBLP:journals/pacmmod/ShaikhhaSSN24,DBLP:journals/vldb/WangWLGDZ21}.
$\lambda_{DAT}$~\cite{Starup2023BreakingTN} encodes dependency graphs in the type system, while \calc{} enforces stratification structurally, by how queries are composed in the host language.
Flix~\cite{flix,DBLP:journals/pacmpl/MadsenL20, 10.1145/3763126} is a general-purpose language that supports Datalog queries. Datafun~\cite{datafun} is a functional Datalog variant that tracks range-restriction, monotonicity, and constructor-freedom via type-level constraints. Both Flix and Datafun have their own runtime and execution engine based on Datalog semantics. In contrast, \calc{} is designed as a host-language-embedded DSL that generates recursive SQL targeting real-world databases with inconsistent semantics. The philosophy of \calc{} is to be ``backend-polymorphic'', so instead of relying on a fixed operational semantics and built-in runtime for fixpoint evaluation, \calc{} derives properties P1--P\total{} from recursive queries.
These properties are applicable regardless of the syntax of the SQL variant used due to the shared evaluation algorithm specified by the SQL standard.

 \section{Conclusion}\label{sec:conclusion}

 %NOTE: remove if space needed
 Recursive queries are difficult to use and support across databases is fragmented and chaotic, yet the performance gains can be massive.
%=====
%NOTE: shortest conclusion
\system{} provides a compile-time safe abstraction for recursion, formalized as \calc{}, that prevents runtime errors, incorrect results, and nontermination without sacrificing expressivity.
%=====
% NOTE: remove if space needed
We prove that fully restricted \calc{} programs compute the unique and minimal fixed point under iterated fixed-point semantics, and implement these guarantees in a practical language-integrated query library that delivers safety, portability, and performance without runtime overhead.

\bibliographystyle{plainurl}
\bibliography{biblio}

% Comment out the next line for the 26-page camera-ready submission.
% Leave uncommented to build the extended version with the appendix.
\input{appendix}

\end{document}

%% file: appendix.tex
\appendix

\clearpage
\section{Appendix}

{
\renewcommand{\textfraction}{0.01}
\renewcommand{\topfraction}{0.99}
\renewcommand{\floatpagefraction}{0.8}

\subsection{\calc{} with Host-Language Embedding}\label{sec:appendix:Host-TT}
\begin{figure}[t]
\setlength{\fboxsep}{0pt}
    \figbox{\begin{minipage}{\textwidth}
      \small
      \setlength{\jot}{0pt}
        \begin{flalign*}
            &\underline{\textbf{Syntax}}\\
& \textit{(const)} & c \ ::= \ & \textit{number} \mid \textit{boolean} \mid \textit{string}\\
                & \textit{(shape)}     &\textit{S}\ ::=\ &
                    \text{Scalar} \mid
                    \text{NScalar} & \\
                & \textit{(base)} & \textit{O}\ ::=\ &
                    \text{Int},\
                    \text{Bool},\
                    \text{String}  \\
                & \textit{(col)} & \textit{K}\ ::=\ &
                    \text{Expr}[O, S]  \mid
                    \text{\rexpr{}}[O, S] \\
                & \textit{(row)} & \textit{A}, \textit{B}, \textit{E} \ ::=\ &
                    (l_i: \textit{K}_i){_{i=1}^n}
                    \\
                & \textit{(cat)}    &C\ ::=\ &
                    \text{Bag} \mid
                    \text{Set} & \\
&{\textit{(deps)}} & D ::= &
                    {(d_{1_\kappa}, d_{2_\kappa}, \dots, d_{m_\kappa}) \quad d_i \in \mathbb{Z} \text{ with a tag } \kappa} \\
                & \textit{(query)} & \textit{Q}\ ::=\ &
                    \text{Query}[\textit{A}, \textit{C}] \mid
                    \text{\rquery{}}[A, {\ D,\ }C]\\
                & \textit{(res)} & \textit{R}\ ::=\ & Q \mid
                    \text{Agg}[\textit{A}] \\
                & \textit{(type)} & \textit{T}, \textit{V} \ ::=\ &
                    \textit{A} \mid
                    \textit{R} \mid
                    \textit{T} \rightarrow \textit{V} \mid
                    (\textit{T}_{i}){_{i=1}^n} \mid (l_i \colon \textit{T}_{i}){_{i=1}^n} \mid
                    \highlight{\text{List}[A]}
                    \\
& \highlight{\textit{(host)}} & \highlight{\textit{h}} \ ::=\ & \highlight{\textbf{toRow}(m) \mid \textbf{toExpr}(m) \mid \textbf{run}(m) \mid {f(m)}} \\
                & \textit{(term)} & m, q, r, f \ ::=\ & c \mid
                    (x) \rightarrow m \mid
                    (l_i = m_i){_{i=1}^n} \mid
                    m.l \mid
                    (m_i){_{i=1}^n} \mid
                    m.i \mid
                    m\ \textbf{++}\ r \mid p \mid
                    {\textbf{table}(\textit{db})}
                    \\ &&& \mid \textit{op}\ m \mid
                    \highlight{h}
                \\
& \textit{(cmb)} & p\ ::=\ &
                    \textbf{map}(q,\ f)\mid
                    \textbf{flatMap}(q,\ f)\ \mid
                    \textbf{filter}(q,\ f)\mid
                    \textbf{agg}(q,\ f) \mid
                    \textbf{fix}(q,\ f) \\ &&& \mid
                    \textbf{groupBy}(q,\ f,\ m,\ r)
                \\
             \shortintertext{\underline{\textbf{$\Sigma$\ Entries}} (\textit{op})}
                & &  \textit{exprOp} ::=\ & m\ + \ r \mid m\ \&\&\ r \mid \textbf{sum}(m) \mid ... &\\
                & & \textit{relOp} ::=\ & \textbf{union}(m,\ r) \mid \text{\textbf{unionAll}}(m,\ r) \mid ...&
    \end{flalign*}
    \end{minipage}}
    \caption{Fully Restricted \protect\calc{} Syntax. \textit{x} ranges over variables and \textit{db} over table names.
}\label{fig:restricted-host-syntax}
\end{figure} \begin{figure}[t]
{\scriptsize
  \vspace{-1em}
    \def\env{\Gamma; \Delta}\def\dslext{}
    \noindent{
\setlength{\fboxsep}{0pt}
    \figbox{\begin{minipage}[t]{\textwidth}

      \figurefontXLsize \setlength{\jot}{0pt}
      \vspace{-1.5em}
      \setlength{\fboxsep}{2pt}
      \setlength{\arraycolsep}{0pt}
    \[
    \begin{alignedat}{1}
      &\underline{\textbf{Meta-Helpers}} \\
      &\textit{\restrictcollect{}}(A, C, Q_1, \ldots, Q_n) =
                 \textit{if} \ \ \exists i, Q_i \equiv \text{\rquery{}}[A_i, {D_i}, C_i]
                 \ \textit{then} \ \text{\rquery{}}[A, {\uplus_{i=1}^n D_i}, C]\
                 \textit{else}\ \text{Query}[A, C] \\
        &\textit{Shape}(S_1, \ldots, S_n) = \textit{if}\ \ \exists i, S_i \equiv \text{Scalar} \
        \textit{then} \  \text{Scalar}\ \textit{else}\ \text{NScalar} \\
        &\boxed{\Sigma}
    \end{alignedat}
    \]
    \vspace{-3em}
    \[
    \begin{gathered}
            \hspace{\linewidth}\\
\stackrel{\textsc{DISTINCT\dslext{}}}{
            \frac{
                \begin{array}{c}
                    \env{} \vdash q \colon Q \\
                    Q \in \{\text{\rquery{}}[A, {D,\ } C], \text{Query}[A, C]\}
                \end{array}
            }{
                \env{} \vdash \textbf{distinct}(q) \colon \textit{RC}(A, \text{Set}, Q)
            }
        }\\
        \stackrel{\textsc{UNION\dslext{}}}{
            \frac{
                \begin{array}{c}
                    \env{} \vdash q_1 \colon Q_1 \quad
                    \env{} \vdash q_2 \colon Q_2 \\
                    Q_1\in \{\text{\rquery{}}[A, {D_1,\ }C_1], \text{Query}[A, C_1]\} \\
                    Q_2\in \{\text{\rquery{}}[A, {D_2,\ }C_2], \text{Query}[A, C_2]\}
                \end{array}
            }{
                \env{} \vdash \textbf{union}(q_1,\ q_2) \colon \textit{\restrictcollect{}}(A,\text{Set},Q_1, Q_2)
            }
        }
        \stackrel{\textsc{UNION-ALL\dslext{}}}{
            \frac{
                \begin{array}{c}
                    \env{} \vdash q_1 \colon Q_1 \quad
                    \env{} \vdash q_2 \colon Q_2 \\
                    Q_1\in \{\text{\rquery{}}[A, {D_1,\ } C_1], \text{Query}[A, C_1]\} \\
                    Q_2\in \{\text{\rquery{}}[A, {D_2,\ } C_2], \text{Query}[A, C_2]\}
                \end{array}
            }{
                \env{} \vdash \textbf{unionAll}(q_1,\ q_2) \colon \textit{\restrictcollect{}}(A,\text{Bag},Q_1, Q_2)
            }
        }
        \end{gathered}
        \]
        \[\vspace{-1em}
        \begin{alignedat}{1}
            &\boxed{\env{} \vdash m \colon T}
        \end{alignedat}
        \]
        \vspace{-3em}
        \[
        \begin{gathered}
            \hspace{\linewidth}\\
            \stackrel{\textsc{CONST\dslext{}}}{\frac{\Sigma(c)=O}{\env{} \vdash c \colon \text{Expr}[O, \text{NScalar}]}}
            \quad
            \stackrel{\textsc{VAR\dslext{}}}{\frac{x \colon T \in \env{}}{\env{} \vdash x \colon T}}\
            \stackrel{\textsc{FUN\dslext{}}}{\frac{\env{}, x \colon T \vdash m \colon V}{\env{} \vdash (x) \rightarrow m \colon T \rightarrow V}}
            \quad
            \stackrel{\textsc{TUPLE\dslext{}}}{\frac{\env{} \vdash m_i \colon T_i \quad \forall i{_{=1}^n}}{\env{} \vdash ( m_i ){_{i=1}^n} \colon (T_i){_{i=1}^n}}}
        \\[2pt]
          \stackrel{\textsc{PROJECT\dslext{}}}{\frac{\env{} \vdash m \colon (T_i){_{i=1}^n} \ {\scriptstyle{j \in 1..n}}}{\env{} \vdash m.j \colon T_j}}
          \quad
          {\stackrel{\textsc{TABLE\dslext{}}}{\frac{
                \begin{array}{c}
                    \Sigma(\textit{db}) = A
                \end{array}
            }{
                \env{} \vdash \textbf{table}(\textit{db}) \colon \text{Query}[A, \text{Bag}]
            }}}
            \quad
            \stackrel{\textsc{NAMED-TUPLE\dslext{}}}{\frac{\env{} \vdash m_i \colon T_i \quad \forall i{_{=1}^n}}{\env{} \vdash ( l_i = m_i ){_{i=1}^n}\colon (l_i \colon T_i){_{i=1}^n}}}
        \\[2pt]
            \stackrel{\textsc{NAMED-PROJECT\dslext{}}}{\frac{\env{} \vdash m \colon (l_i \colon T_i){_{i=1}^n} \ {\scriptstyle {j \in 1..n}}}{\env{} \vdash m.l_j \colon T_j}}
            \quad
            \stackrel{\textsc{EXPR-PROJ}}{\dfrac{\env{} \vdash m : \mathrm{Expr}[(l_i : A_i){_{i=1}^n}, S]}{\env{} \vdash m.l_i : \mathrm{Expr}[A_i, S]}}
        \quad
          \stackrel{\textsc{NAMED-CONCAT\dslext{}}}{\frac{
              \begin{array}{c}
                  \env{} \vdash m_1 \colon (\, l_i \colon T_i \,){_{i=1}^n}
                  \\
                  \env{} \vdash m_2 \colon (\, l_j \colon V_j \,){_{j=n+1}^k}
                  \ {k > n}\ {l_i \neq l_j}
              \end{array}
              }{
                \env{} \vdash m_1 \ \textbf{++} \ m_2 \colon (\, l_i \colon T_i,\ l_j \colon V_j \,)}}
          \\
            \stackrel{\textsc{EXPR-OP\dslext{}}}{
                \frac{
                    \begin{array}{c}
                        \env{} \vdash m \colon (\text{Expr}[A_i, S_i]){_{i=1}^n} \\
                        \Sigma(\textit{exprOp}) = (\text{Expr}[A_i, S_i]){_{i=1}^n} \rightarrow \text{Expr}[A, S]
                    \end{array}
                }{
                    \env{} \vdash \textit{exprOp}\ m \colon \text{Expr}[A, \textit{Shape}(S, S_i{_{i=1}^n})]
                }
            }
            \
            \stackrel{\textsc{REL-OP\dslext{}}}{
                \frac{
                    \begin{array}{c}
                        \env{} \vdash m \colon (Q_i{_{i=1}^n}) \\
                        Q_i \in \{\text{\rquery{}}[A_i, {D_i,\ }C_i], \text{Query}[A_i, C_i]\} \\
                        \Sigma(\textit{relOp}) =
                        (Q_i{_{i=1}^n}) \rightarrow \textit{\restrictcollect{}}(A, C, Q_i{_{i=1}^n})
                    \end{array}
                }{
                    \env{} \vdash \textit{relOp}\ m \colon \textit{\restrictcollect{}}(A, C, Q_i{_{i=1}^n})
                }
            }
            \\[2pt]
            \stackrel{\textsc{MAP\dslext{}}}{
                \frac{
                    \begin{array}{c}
                        \env{} \vdash q \colon Q \quad
                        \env{} \vdash f \colon K_1 \rightarrow K_2 \\
                        (Q,\ K_1, K_2) \in
                        \{
                            (
                                \text{\rquery{}}[A, {D,\ } C],
                                \text{\rexpr{}}[A, \text{NScalar}], \\
                                \text{\rexpr{}}[B, \text{NScalar}]
                            ),
                            (
                                \text{Query}[A, C], \\
                                \text{Expr}[A, \text{NScalar}],
                                \text{Expr}[B, \text{NScalar}]
                            )
                        \}\end{array}
                }{
                    \env{} \vdash \textbf{map}(q,\ f) \colon \textit{\restrictcollect{}}(B, \text{Bag}, Q)
                }
            }
            \
            \stackrel{\textsc{FILTER\dslext{}}}{
                \frac{
                    \begin{array}{c}
                        \env{} \vdash q \colon Q \quad
                        \env{} \vdash f \colon K_1 \rightarrow K_2 \\
                        (Q,\ K_1, K_2) \in \{
                            (
                                \text{\rquery{}}[A, {D,\ } C],
                                \text{\rexpr{}}[A, \text{NScalar}], \\
                                \text{\rexpr{}}[\text{Bool}, \text{NScalar}]
                            ),\
                            (
                                \text{Query}[A, C], \\
                                \text{Expr}[A, \text{NScalar}],
                                \text{Expr}[\text{Bool}, \text{NScalar}]
                            )
                        \} \\
                    \end{array}
                }{
                    \env{} \vdash \textbf{filter}(q,\ f) \colon \textit{\restrictcollect{}}(A, C, Q)
                }
            }
            \\[2pt]
            \stackrel{\textsc{FLATMAP\dslext{}}}{
                \frac{
                    \begin{array}{c}
                        \env{} \vdash q \colon Q_1 \quad \env{} \vdash f \colon K \rightarrow Q_2 \
                        (Q_1, K) \in \\ \{
                            (\text{\rquery{}}[A, {D_1,\ }C_1],\ \text{\rexpr{}}[A, \text{NScalar}]), (\text{Query}[A, C_1],\ \text{Expr}[A, \text{NScalar}])
                        \} \\
                        Q_2 \in \{\text{\rquery{}}[B, {D_2,\ } C_2], \text{Query}[B, C_2]\} \\
                    \end{array}
                }{
                    \env{} \vdash \textbf{flatMap}(q,\ f) \colon \textit{\restrictcollect{}}(B, \text{Bag}, Q_1, Q_2)
                }
            }
            \stackrel{\textsc{AGGREGATE\dslext{}}}{\frac{
                \begin{array}{c}
                    \env{} \vdash q \colon \text{Query}[A, C] \\
                    \env{} \vdash f \colon \text{Expr}[A, S] \rightarrow \\ \hspace{1.4cm} \text{Expr}[B, \text{Scalar}]
                \end{array}
                }{
                \env{} \vdash \textbf{agg}(q,\ f) \colon \text{Agg}[B]
                }
            }
            \\[2pt]
            \stackrel{\textsc{GROUPBY\dslext{}}}{
                \frac{
                    \begin{array}{c}
                        \env{} \vdash q \colon \text{Query}[A, C] \
                        \env{} \vdash f\colon \text{Expr}[A, S_{g}] \rightarrow \text{Expr}[E, S_{g}] \
                        \env{} \vdash m\colon \text{Expr}[A, S_{p}] \rightarrow \text{Expr}[B, S_{p}] \\
                        \textit{Shape}(S_{g}, S_{p}, S_{s}) \equiv \text{Scalar}
                        \
                        \env{} \vdash r\colon \text{Expr}[A, S_{s}] \rightarrow \text{Expr}[\text{Bool}, S_{s}]
                    \end{array}
                }{
                    \env{} \vdash \textbf{groupBy}(q,\ f,\ m,\ r)\colon \text{Query}[B, \text{Bag}]
                }
            }
            \\[2pt]
            \stackrel{\textsc{FIX\dslext{}}}{
                \frac{
                \begin{array}{c}
\env{} \vdash q: Q_\text{base} \quad
                    Q_\text{base} = (\text{Query}[A_i, C_i]){_{i=1}^n}
                    \quad
                    A_i = (l_j: K_j){_{j=1}^{m_i}} \ \forall i{_{=1}^n} \quad
                    \env{} \vdash f: Q_\text{ref} {\rightarrow} Q_\text{ret}
                    \\
                    Q_\text{ref} = (\text{\rquery{}}[A_i, {(i),\ }C_i]){_{i=1}^n} \
                    Q_\text{ret} = (\text{\rquery{}}[A_i, {D_i,\ } \text{Set}]){_{i=1}^n} \\
                    {\{ 1_\kappa, \ldots, n_\kappa \} \equiv \cup D_i{_{i=1}^n}} \quad
                    {\forall i{_{=1}^n} \quad |D_i|=|\cup D_i|} \quad n = 1
                \end{array}
                }
                {\env{} \vdash \textbf{fix}(q,\ f) : Q_\text{base}}
            }
  \end{gathered}
  \]
\end{minipage}}
} \vspace{-1em}
    \noindent{\setlength{\fboxsep}{0pt}
  \figbox{\begin{minipage}{\textwidth}
    \figurefontXLsize
    \setlength{\jot}{0pt}
    \setlength{\arraycolsep}{0pt}
    \vspace{-1.4em}
    {\setlength{\fboxsep}{2pt}
      \[\begin{alignedat}{1}
          &\boxed{\Gamma \vdash m : T}\\
      \end{alignedat}\]
      }
  \vspace{-3em}
  \[
  \begin{gathered}
      \hspace{\linewidth}\\
      \stackrel{\textsc{CONST}}{\dfrac{\Sigma(c)=O}{\Gamma \vdash c : O}}
      \ \
       \stackrel{\textsc{APP}}{\dfrac{\Gamma \vdash m_1 : T \rightarrow V \quad \Gamma \vdash m_2 : T}{\Gamma \vdash m_1(m_2) : V}}
       \ \
        \stackrel{\textsc{RUN-QUERY}}{\dfrac{\Gamma \vdash m : \mathrm{Query}[A,C]}{\Gamma \vdash \mathbf{run}(m) : \mathrm{List}[A]}}
        \ \
       \stackrel{\textsc{RUN-AGG}}{\dfrac{\Gamma \vdash m : \mathrm{Agg}[A]}{\Gamma \vdash \mathbf{run}(m) : A}}
       \ \
             \stackrel{\textsc{LIFT}}{\dfrac{\Gamma \vdash c : O}{\Gamma;\Delta \vdash c : \mathrm{Expr}[O,\mathrm{NScalar}]}}
              \\[2pt]
      \stackrel{\textsc{TO-EXPR}}{\dfrac{m : O \in \Sigma}{\Gamma \vdash \mathbf{toExpr}(m) : \mathrm{Expr}[O,\mathrm{NScalar}]}}
      \quad
      \stackrel{\textsc{TO-ROW}}{\dfrac{\Gamma \vdash m : (l_i : \mathrm{Expr}[A_i,S_i]){_{i=1}^n}}
           {\Gamma \vdash \mathbf{toRow}(m) : \mathrm{Expr}[(l_i : A_i){_{i=1}^n},\ \mathrm{Shape}(S_i{_{i=1}^n})]}}
  \end{gathered}
  \]
\end{minipage}}
}   }
\caption{\calc{} Typing rules with all restrictions.}\label{fig:full-restricted-host}
  \vspace{-0.5em}
\end{figure}
 Figure~\ref{fig:restricted-host-syntax} shows the fully restricted \calc{} types and terms and Figure~\ref{fig:full-restricted-host} the typing rules with all properties enforced with host language embedding.

\subsection{Operational Semantics and Normalization for \calc{}}\label{sec:appendix:operational-semantics}
\begin{figure}[t]
   \setlength{\fboxsep}{0pt}
    \figbox{\begin{minipage}{\textwidth}
      \figurefontXLsize
      \setlength{\jot}{0pt}
      \setlength{\arraycolsep}{0pt}
      \vspace{-1.4em}
        \begin{flalign*}
            &(\text{value}) & V, W, X, Y  &::= c \
                \mid (x) \rightarrow M\ \mid M(x)\
                \mid (l_i = V_i){_{i=1}^n}\
                \mid (V_i){_{i=1}^n}\
                \mid \textbf{database}(db) \\&&&
                \mid P \text{ where P ranges over a closed host term of type Expr[A, S],} \\&&&
                \quad \text{Query}[A, C], \text{\rquery{}}[A,\ {D,\ } C], \text{\rexpr{}}[A], \text{ or } \text{List}[A] \\
&(\text{evaluation context})\ &\mathcal{E} &::= [\ ]\\
            &&& \mid (V_i){_{i=1}^{j-1}},\  \mathcal{E},\ (M_k){_{k=j+1}^n}\
                \mid
op\
                \mathcal{E}\
                \mid \mathcal{E}(M)\
                \mid V(\mathcal{E})\\
            &&&
                \mid (l_i = V_i){_{i=1}^{j-1}},\ l_j=\mathcal{E},\ (l_k=M_k){_{k=j+1}^n}\
                \mid \mathcal{E}.l\
                \mid \textbf{(toExpr|toRow)}\ (\mathcal{E})\\
            &&&
                \mid \mathcal{E}\ \textbf{++}\ M\
                \mid V\ ++\ \mathcal{E}\
                \mid \textbf{(map|flatMap|filter|agg|fix)}\ (\mathcal{E},\ M)\\
            &&&
                \mid \textbf{(map|flatMap|filter|agg|fix)}\ (V,\ \mathcal{E})\
                \mid \textbf{groupBy}\ \mathcal{E}\ W\ X\ Y \\
            &&&
                \mid \textbf{groupBy}\ V\ \mathcal{E}\ X\ Y\
                \mid \textbf{groupBy}\ V\ W\ \mathcal{E}\ Y\
                \mid \textbf{groupBy}\ V\ W\ X\ \mathcal{E}\\
            &&&
                \mid \textbf{run}(\mathcal{E})
\end{flalign*}
    \end{minipage}}\caption{Values and Evaluation Context for Fully Restricted Host \calc{}}
\label{fig:valsandevals}
\end{figure}

\begin{figure}[t]
   \setlength{\fboxsep}{0pt}
    \figbox{\begin{minipage}{\textwidth}
      \figurefontXLsize
      \setlength{\jot}{0pt}
      \setlength{\arraycolsep}{0pt}
      \vspace{-2.4em}
\[
\begin{gathered}
\hspace{\linewidth}\\
  \begin{array}{llcl}
  &\textit{op}\ (V_1, \ldots, V_n)
    &\longrightarrow&\ \delta(\textit{op}, V_1, \ldots, V_n) \\
  &((x) \rightarrow m)\ V
    &\longrightarrow&\ m[x := V] \\
  &(l_i = V_i){_{i=1}^n}.l_j
    &\longrightarrow&\ V_j \qquad (1 \le j \le n) \\
  &(V_i){_{i=1}^n}.j
    &\longrightarrow&\ V_j \qquad (1 \le j \le n) \\
  &(l_i = V_i){_{i=1}^n}\ \textbf{++}\ (r_j = W_j){_{j=1}^m}
    &\longrightarrow&\ (\,l_i = V_i,\ r_j = W_j\,){_{i=1}^n,\ _{j=1}^m} \\
  &\textbf{run}(Q)
    &\longrightarrow&\ \textit{eval}(\textit{norm}(Q)) \\[1em]
&&\dfrac{M \longrightarrow N}{\mathcal{E}[M] \longrightarrow \mathcal{E}[N]}&
\end{array}
\end{gathered}
\]
\end{minipage}}
    \caption{Operational Semantics of Fully Restricted Host \calc{}. The \textit{norm} function is shown in Figures~\ref{fig:norm-stage1}-\ref{fig:norm-stage2} and \textit{eval} translates normalized \calc{} to SQL and executes queries on a fixed database.}
    \label{fig:operational-semantics}
\end{figure} \begin{figure}[t]
   \setlength{\fboxsep}{0pt}
    \figbox{\begin{minipage}{\textwidth}
      \figurefontXLsize
      \setlength{\jot}{0pt}
      \setlength{\arraycolsep}{0pt}
      \vspace{-1.4em}
      \[
        \begin{alignedat}{3}
            &\underline{\textbf{Syntax}}\\
                & \textit{(base)} & b \ ::= \ &
                    c \mid m.l \mid \textit{exprOp}\ (b_i){_{i=1}^n} \mid \textbf{agg}\big(q,\ (x) \rightarrow m\big) \\
                & \textit{(col)} & m \ ::=\ &
                x \mid
                (l_i = m_i){_{i=1}^n} \mid
                m_1\ \textbf{++}\ m_2
                \\
& \textit{(boundary)} & t\ ::=\ &
                    \textbf{table}(\textit{db})  \mid \textbf{groupBy}\big(q,\ (x_1) \rightarrow m_1,\ (x_2) \rightarrow m_2,\ (x_3) \rightarrow m_3\big) \mid \textit{relOp}\ (q_i){_{i=1}^n} \\
                     &&& \mid \textbf{fix}\big((q_{\text{base}_i}){_{i=1}^n},\ (x_i){_{i=1}^n} \rightarrow (q_{\text{rec}_i}){_{i=1}^n}\big).\_w
                    \\& \textit{(collections)} & z\ ::=\ &
                    \textbf{map}\big(t,\ (x) \rightarrow m\big)\mid \textbf{flatMap}\big(t,\ (x) \rightarrow q\big)\ \mid \textbf{filter}\big(t,\ (x) \rightarrow m\big) \\  & \textit{(term)} & p\ ::=\ &t \mid z \end{alignedat}
    \]
  \end{minipage}}
\caption{Syntax of \calc{} in normal form. \textit{op} is split into \textit{exprOp} for operations on expressions, e.g., $+$, and \textit{relOp} for operations on relations, e.g., \textbf{union}.}
    \label{fig:normalized-dsl-syntax}
\end{figure} \begin{figure}[t]
   \setlength{\fboxsep}{0pt}
    \figbox{\begin{minipage}{\textwidth}
      \figurefontXLsize
      \setlength{\jot}{0pt}
\vspace{-1.3em}
      \[
\begin{array}{@{\extracolsep{\fill}} lcll @{}}
\big( (x) \rightarrow R \big) \ (V)
&\rightsquigarrow&
R[x := V]
& (\textsc{app})
\\
(l_i = Q_i){_{i=1}^n}.l_j
&\rightsquigarrow&
Q_j
& (\textsc{proj-named})
\\
(Q_i){_{i=1}^n}.j
&\rightsquigarrow&
Q_j
&  (\textsc{proj-unnamed})
\\
\textbf{flatMap}\big(\textbf{flatMap}(R,\ f),\ g\big)
&\rightsquigarrow&
\textbf{flatMap}\big(R,\ (x) \rightarrow \textbf{flatMap}(f(x),\ g)\big)
& (\textsc{for-for})
\\
\textbf{flatMap}
    \big(\textbf{unionAll}(P,\ Q),\ f\big) &\rightsquigarrow& \textbf{unionAll}\big(\textbf{flatMap}(P,\ f),\ \textbf{flatMap}(Q,\ f)\big) & (\textsc{for-union-all})\\
\end{array}
\]
\end{minipage}}

\caption{Normalization Stage~1 (symbolic reduction) for \calc{}, from T-LINQ.}
\label{fig:norm-stage1}
\end{figure}

\begin{figure}[t]
   \setlength{\fboxsep}{0pt}
    \figbox{\begin{minipage}{\textwidth}
      \figurefontXLsize
      \setlength{\jot}{0pt}
\vspace{-1.3em}
      \[
\begin{array}{@{\extracolsep{\fill}} lcll @{}}
\textbf{filter}\big(\textbf{filter}(R,\ f),\ g\big)
&\hookrightarrow&
\textbf{filter}\big(R,\ (x \rightarrow f(x) \&\& g(x))\big)
& (\textsc{fil-fil})
\\
\textbf{filter}\big(\textbf{unionAll}(P,\ Q),\ f\big)
&\hookrightarrow&
\textbf{unionAll}\big(\textbf{filter}(P,\ f),\ \textbf{filter}(Q,\ f)\big)
& (\textsc{fil-UnionA})
\\
\textbf{map}\big(\textbf{unionAll}(P,\ Q),\ f\big)
&\hookrightarrow&
\textbf{unionAll}\big(\textbf{map}(P,\ f),\ \textbf{map}(Q,\ f)\big)
& (\textsc{map-UnionA})
\\
\textbf{flatMap}\big(R,\ (x \rightarrow \textbf{unionAll}(U(x),\ V(x)))\big)
&\hookrightarrow&
\textbf{unionAll}\big(\textbf{flatMap}(R,\ (x \rightarrow U(x))), &(\textsc{fm-UnionA-R})\\
&& \quad \textbf{flatMap}(R,\ (x \rightarrow V(x)))\big)
&
\\
\textbf{filter}\big(\textbf{map}(R,\ f),\ k\big)
&\hookrightarrow&
\textbf{map}\big(\textbf{filter}(R,\ (x \rightarrow k(f(x)))),\ f\big)
& (\textsc{fil-map})
\\
\textbf{fix}(X,\ f).\_i \text{ where } X = \textbf{fix}\big((Q_i){_{i=1}^n},\ f_1\big)
&\hookrightarrow&
\textbf{fix}\big( (X.\_i){_{i=1}^n},\ f\big)
& (\textsc{detuple-1})
\\
\textbf{fix}(Q,\ (x_i){_{i=1}^n} \rightarrow X).\_i
&\hookrightarrow&
\textbf{fix}\big(Q, (x_i){_{i=1}^n} \rightarrow (X.\_i){_{i=1}^n}\big)
& (\textsc{detuple-2})
\\ \quad \text{ where } X = \textbf{fix}\big((Q_i){_{i=1}^n},\ f_1\big) &
\end{array}
\]
\end{minipage}}
\caption{Normalization Stage~2 (ad-hoc reduction) for \calc{}, from T-LINQ. Together with Stage~1, forms the \calc{} normalization function \textit{norm} used in the operational semantics of \textbf{run} (Figure~\ref{fig:operational-semantics}).}
\label{fig:norm-stage2}
\end{figure}

Figure~\ref{fig:valsandevals} shows the values and evaluation contexts for the \calc{} host language. As in T-LINQ, we parameterize the semantics with an interpretation $\delta$ for each operation \textit{op} and an $\Omega$ for database types.
The reduction $M \longrightarrow N$ is shown in Figure~\ref{fig:operational-semantics}.
The evaluation contexts $\mathcal{E}$ enforce left-to-right call-by-value evaluation
and  \textbf{run} operates on terms containing values of type Query[A] or Agg[A] and evaluates the term by applying the normalization function \textit{norm}, then translating to SQL and executing on a fixed database with \textit{eval}.

The \textit{norm} function has two phases and produces \calc{} in normal form, shown in Figure~\ref{fig:normalized-dsl-syntax}.
We write $\rightsquigarrow^*$ and $\hookrightarrow^*$ for the reflexive and transitive closure of $\rightsquigarrow$,  and $\hookrightarrow$, which are the compatible closure of the rules in Figure~\ref{fig:norm-stage1} and~\ref{fig:norm-stage2}.
$\textit{norm}(P)=R$ when $P \rightsquigarrow^* Q$ and $Q \hookrightarrow^*$ R where Q and R are in normal form with respect to $\rightsquigarrow$ and $\hookrightarrow$.
The rules are equivalent to the reduction relations used by T-LINQ (updated for \calc{}'s combinator syntax), plus \textsc{DETUPLE-1/2} which rewrite nested \textbf{fix} to an immediately projected form. As a result, $\rightsquigarrow^*$ and $\hookrightarrow^*$ retain the confluent and strongly normalizing properties with respect to the subset of SQL supported by T-LINQ.
For queries with recursion, the \textit{norm} function normalizes terms inside the bodies of recursive queries or outside of recursive queries, but not across the recursion boundary.
For example, in T-LINQ the comprehension
$\textbf{for } x \textbf{ in } R \textbf{ do if } P \textbf{ then if } Q \textbf{ then yield } x$
normalizes to
$\textbf{for } x \textbf{ in } R \textbf{ do if } P\,\wedge\, Q \textbf{ then yield } x$.
In \calc{}, the same query is written as
$\textbf{filter}(
    \textbf{filter}(R,\ (x) \rightarrow P),\
    (x) \rightarrow Q
)$
and normalizes to
$\textbf{filter}(R,\ (x) \rightarrow P \&\& Q)$.
However, the expression
$\textbf{fix}(\textbf{fix}( R,\ (x) \rightarrow P), (x) \rightarrow Q)$
should not be collapsed in the same way as \textbf{filter} because, crucially, the resulting query must retain the two separate fixed points in order to enforce stratification. Therefore any nesting of \textbf{fix} present in the original \calc{} term must be retained, in the same order, in the final query.

\subsection{{Translational Semantics for \calc{}}}\label{sec:appendix:translational-semantics}

Datalog has a bottom-up fixed-point semantics and an equivalent proof-theoretic semantics that can be used to prove both soundness and completeness.
Stratified Datalog with negation (\datalogns{}) has an iterated fixed-point semantics (strata-by-strata evaluation) that always produces the Perfect Model~\cite{amateur}, i.e., using this semantics, well-formed programs will always find the unique and minimal fixed-point in a finite number of steps. Definitions of key Datalog terms are provided in Section~\ref{sec:appendix:datalog-definitions} for reference.

In this section, we give the fully restricted \calc{} the same semantics by defining a complete, type-directed  translation function from terms in \calc{} to \lsd{}.
\lsd{} (Def.~\ref{def:datalog:lsd}) is a strict subset of \datalogns{}, equivalent to linear (Def.~\ref{def:datalog:linear}), stratified (Def.~\ref{def:datalog:stratified}) Datalog with negation (Def.~\ref{def:datalog:limited}) with no mutually recursive predicates (Def.~\ref{def:datalog:direct}).
Under this semantics, every \textbf{well-typed fully-restricted \calc{} program will always find the unique and minimal fixed-point in a finite number of steps, i.e., it will not show behaviors B1-B3} (Theorem~\ref{theprop}). The result of $\textit{eval}(\textit{norm}(Q))$ under the rules defined by the SQL Standard'99 Section 7.12 agrees with the Perfect-Model result of Q translated to \lsd{}~\cite{sql99}.
We redefine the operational semantics by replacing the rule for $\textbf{run}(Q)$ (Figure~\ref{fig:operational-semantics}) with $\textit{evalDL}$, a standard bottom-up fixed-point Datalog evaluation algorithm executed on the result of the translation steps shown in Figure~\ref{fig:phases}.

 \begin{figure}[t]
\centering
  \vspace{-.15cm}
\includegraphics[trim={.4cm .5cm .4cm 0.2cm}, clip, width=\linewidth]{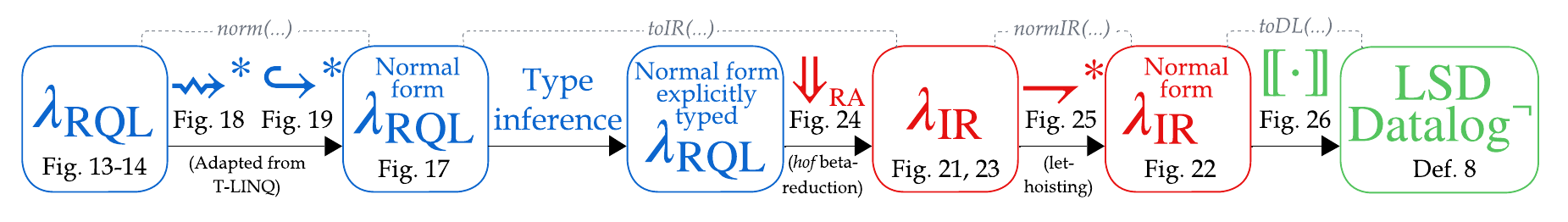}
\caption{Translation phases from \calc{} to \lsd{}.
  \calc{} is given a sound semantics by replacing the rule for $\textbf{run}(Q)$ (Figure~\ref{fig:operational-semantics}) with $\textit{evalDL}(\textit{toDL}(\textit{normIR}(\textit{toIR}(\textit{norm}(Q)))))$ where \textit{evalDL} implements a standard bottom-up fixed-point Datalog semantics, which agrees with the result of $\textit{eval}(\textit{norm}(Q))$ under the rules defined by the SQL Standard'99 Section 7.12~\cite{sql99}).
  See Figure~\ref{fig:example-translation-pipeline} for an example pipeline.
  }
  \label{fig:phases}
  \vspace{-1em}
\end{figure}

\subsubsection{Translation to \ir{}}\label{sec:norm34}

\begin{figure}[t]
   \setlength{\fboxsep}{0pt}
    \figbox{\begin{minipage}{\textwidth}
      \figurefontXLsize
      \setlength{\jot}{0pt}
      \setlength{\arraycolsep}{0pt}
      \vspace{-1.3em}
      \[
  \begin{alignedat}{3}
    &\underline{\textbf{Syntax}}\\
    & \textit{(base)} & b \ ::= \ &
      c \mid \textit{exprOp}\ (b_i){_{i=1}^n} \mid m.l
    \\
    & \textit{(col)} & m \ ::=\ &
      (l_i = b_i){_{i=1}^n} \mid
      m_1\ \textbf{++}\ m_2 \mid
      \textbf{t-ref}(\alias{}) \mid
      \textbf{rt-ref}(\alias{})
    \\
    & \textit{(letrec binding)} & \mathit{r} \ ::=\ &
      \textbf{rec-query}\!\Big(
        \alias{}_{\text{rec}},
        (\alias{}_i){_{i=1}^n},
        (\mathit{q}_{\text{base}_i}){_{i=1}^n},
        (\mathit{q}_{\text{rec}_i}){_{i=1}^n},
        w \in \{1..n\}
      \Big)
    \\
    & \textit{(term)} & \mathit{q} \ ::=\ &
      \text{[]} \mid
      \textbf{table}(\alias{}) \mid
      \textbf{r-table}(\alias{}{,\ i}) \mid
      \textbf{query}(\alias{}, \mathit{q}, t, \textit{cmb}) \mid
      \textbf{agg}(\alias{}, q, t) \mid
      \textit{relOp}\ (\alias{},\ (\mathit{q}_i){_{i=1}^n}) \\ &&& \mid
      \textbf{letrec}\ (\alias{}_i = \mathit{r}_i){_{i=1}^n}\ \textbf{in}\ \mathit{q} \quad n \ge 0
  \end{alignedat}
  \]
\end{minipage}}
  \caption{Syntax of \ir{}. $\alias{}$ ranges over unique identifiers. $[]$ is an empty list and \textit{cmb} ranges over the combinators, e.g., \textbf{map}.}\label{fig:ir-syntax}
  \vspace{-1em}
\end{figure} \begin{figure}[t]
   \setlength{\fboxsep}{0pt}
    \figbox{\begin{minipage}{\textwidth}
      \figurefontXLsize
      \setlength{\jot}{0pt}
      \setlength{\arraycolsep}{0pt}
      \vspace{-1.3em}
    \[
  \begin{alignedat}{3}
    &\underline{\textbf{Syntax}}\\
    & \textit{(base)} & b \ ::= \ &
      c \mid \textit{exprOp}\ (b_i){_{i=1}^n} \mid m.l
    \\
    & \textit{(col)} & m \ ::=\ &
      (l_i = b_i){_{i=1}^n} \mid
      m_1\ \textbf{++}\ m_2 \mid
      \textbf{t-ref}(\alias{}) \mid
      \textbf{rt-ref}(\alias{})
    \\
    & \textit{(letrec binding)} & \mathit{r} \ ::=\ &
      \textbf{rec-query}\!\Big(
        \alias{}_{\text{rec}},
        (\alias{}_i){_{i=1}^n},
        (\mathit{q}_{\text{base}_i}){_{i=1}^n},
        (\mathit{q}_{\text{rec}_i}){_{i=1}^n},
        w \in \{1..n\}
      \Big)
    \\
    & \textit{(sql-query)} & \mathit{q} \ ::=\ &
      \text{EmptyList} \mid
      \textbf{table}(\alias{}) \mid
      \textbf{r-table}(\alias{}{,\ i}) \mid
      \textbf{query}(\alias{}, \mathit{q}, t, \textit{cmb}) \mid
      \textbf{agg}(\alias{}, q, t) \\ &&& \mid
      \textit{relOp}\ (\alias{},\ (\mathit{q}_i){_{i=1}^n})
    \\
    & \highlight{\textit{(term)}} & t \ ::=\ &
      \textbf{letrec}\ (\alias{}_i = \mathit{r}_i){_{i=1}^n}\ \textbf{in}\ \mathit{q} \quad n \ge 0
\end{alignedat}
  \]
  \end{minipage}}
  \caption{Syntax of \textbf{normalized} \ir{}. Differs from Figure~\ref{fig:ir-syntax} as local \textbf{letrec} are replaced with a single global \textbf{letrec}, and all programs contain only a single outermost \textbf{letrec}.}
  \label{fig:ir-norm-syntax}
\end{figure}
 \begin{figure}[t]
\setlength{\fboxsep}{0pt}
\figbox{\begin{minipage}{1\textwidth}

  \figurefontXLsize
  \setlength{\jot}{0pt}
  \setlength{\arraycolsep}{0pt}
  \vspace{-1.45em}
    \begin{minipage}[t]{\textwidth}
      \setlength{\fboxsep}{2pt}
        \[\begin{alignedat}{1}
            &\underline{\textbf{Meta-Helpers}}\\
            &\textit{\restrictcollect{}}(A, C, Q_1, \ldots, Q_n) =
                 \textit{if} \ \ \exists i, Q_i \equiv \text{\rquery{}}[A_i, {D_i}, C_i]
                 \ \textit{then} \ \text{\rquery{}}[A, {\uplus_{i=1}^n D_i}, C]\
                 \textit{else}\ \text{Query}[A, C] \\
        &\textit{Shape}(S_1, \ldots, S_n) = \textit{if}\ \ \exists i, S_i \equiv \text{Scalar} \
        \textit{then} \  \text{Scalar}\ \textit{else}\ \text{NScalar} \\[2pt]
        &\boxed{\Sigma}
        \end{alignedat}\]
    \end{minipage}
    \vspace{-2em}
    \[
    \begin{gathered}
             \hspace{\linewidth}\\
\stackrel{\textsc{UNION-IR}}{
            \frac{
                \begin{array}{c}
                    \irtypingenv \vdash q_1 \colon Q_1 \
                    \irtypingenv \vdash q_2 \colon Q_2 \
                    \irenv(\alias{}) = A \\
                    Q_1 \in \{\text{\rquery{}}[A,\ {D_1,\ } C_1], \text{Query}[A, C_1]\}\\
                    Q_2 \in \{\text{\rquery{}}[A,\ {D_2,\ } C_2], \text{Query}[A, C_2]\}
                \end{array}
            }{\begin{array}{c}
                \irtypingenv \vdash \textbf{union}(\alias{}, q_1,\ q_2) \colon  \textit{\restrictcollect{}}(A,\text{Set},Q_1, Q_2)
            \end{array}}
        }
        \stackrel{\textsc{UNION-ALL-IR}}{
            \frac{
                \begin{array}{c}
                    \irtypingenv \vdash q_1 \colon Q_1 \
                    \irtypingenv \vdash q_2 \colon Q_2 \
                    \irenv(\alias{}) = A \\
                    Q_1 \in \{\text{\rquery{}}[A,\ {D_1,\ } C_1], \text{Query}[A, C_1]\}\\
                    Q_2 \in \{\text{\rquery{}}[A,\ {D_2},\ C_2], \text{Query}[A, C_2]\}
                \end{array}
            }{\begin{array}{c}
                \irtypingenv \vdash \textbf{unionAll}(\alias{}, q_1,\ q_2) \colon \textit{\restrictcollect{}}(A,\text{Bag},Q_1, Q_2)
            \end{array}}
        }\\[2pt]
        \stackrel{\textsc{DISTINCT-IR}}{
            \frac{
                \begin{array}{c}
                    \irtypingenv \vdash q \colon Q \quad \irenv(\alias{}) = A \\
                    Q \in \{\text{\rquery{}}[A,\ {D,\ } C], \text{Query}[A, C]\}
                \end{array}
            }{
                \irtypingenv \vdash \textbf{distinct}(\alias{}, q) \colon \textit{\restrictcollect{}}(A, \text{Set}, Q)
            }
        }
            \highlight{\stackrel{\textsc{EXPR-ADD-IR}}{
                \frac{
                    \begin{array}{c}
                        \irtypingenv \vdash b_1 \colon \text{Expr}[\text{Int}, S_1] \\
                         \irtypingenv \vdash b_2 \colon \text{Expr}[\text{Int}, S_2] \\
                    \end{array}
                }{
                    \irtypingenv \vdash b_1 + b_2 \colon \text{Expr}[\text{Int}, \textit{Shape}(S_1, S_2)]
                }
            }
            \stackrel{\textsc{EXPR-NEG-IR}}{
                \frac{
                    \begin{array}{c}
                        \irtypingenv \vdash b \colon \text{Expr}[A, S] \\
                         \irtypingenv \vdash q \colon \text{Query}[A, C]
                    \end{array}
                }{
                    \irtypingenv \vdash b \textbf{ not in } q \colon \text{Expr}[A, \text{Scalar}]
                }
            }}
        \end{gathered}
        \]
        \vspace{-1em}
        {\setlength{\fboxsep}{2pt}
        \begin{flalign*}
            &\boxed{\irtypingenv \vdash m \colon T}&&
        \end{flalign*}}
        \vspace{-3em}
        \[
        \begin{gathered}
                 \hspace{\linewidth}\\
            \stackrel{\textsc{CONST-IR}}{
                \frac{\Sigma(c)=T}{\irtypingenv \vdash c \colon \text{Expr}[T, \text{NScalar}]}}\quad
\stackrel{\textsc{NAMED-TUPLE-IR}}{
                \frac{\irtypingenv \vdash b_i \colon T_i \quad \forall i{_{=1}^n}}{\irtypingenv \vdash ( l_i = b_i ){_{i=1}^n}\colon (l_i \colon T_i){_{i=1}^n}}}\quad
             \stackrel{\textsc{NAMED-PROJECT-IR}}{
                \frac{\irtypingenv \vdash m \colon (l_i \colon T_i){_{i=1}^n} \ {\scriptstyle{j \in 1..n}}}{\irtypingenv \vdash m.l_j \colon T_j}}\quad
            \stackrel{\textsc{TABLE-IR}}{\frac{
                \begin{array}{c}
                    \irenv(\alias{}) = A \end{array}
            }{
                \irtypingenv \vdash \textbf{table}(\alias{}) \colon \text{Query}[A, \text{Bag}]}}
        \\
        \highlight{\stackrel{\textsc{R-TABLE-IR}}{\frac{
                \begin{array}{c}
                    \irenv(\alias{}) = A \end{array}
            }{\begin{array}{c}
                \irtypingenv \vdash \textbf{r-table}(\alias{} {,\ i}) \colon \text{\rquery{}}[A,\ {(i),\ }\text{Bag}]\end{array}}}}\qquad
\stackrel{\textsc{NAMED-CONCAT-IR}}{\frac{
              \begin{array}{c}
                  \irtypingenv{} \vdash m_1 \colon (\, l_i \colon T_i \,){_{i=1}^n}
                  \
                  \irtypingenv{} \vdash m_2 \colon (\, l_j \colon V_j \,){_{j=n+1}^k}
                  \
\ _{k > n}^{l_i \neq l_j}
              \end{array}
              }{
                \irtypingenv{} \vdash m_1 \ \textbf{++} \ m_2 \colon (\, l_i \colon T_i,\ l_j \colon V_j \,){_{i=1}^n,\ _{j=n+1}^k}
              }}
        \\
            \highlight{\stackrel{\textsc{TREF-IR}}{
                    \frac{\begin{array}{c}
                        \irenv(\alias{}) = A\end{array}
                }{\begin{array}{c}
                    \irtypingenv \vdash \textbf{t-ref}(\alias{}) \colon \\ \quad \text{Expr}[A, \text{NScalar}]\end{array}}}}\quad
            \highlight{\stackrel{\textsc{TREF-PROJECT-IR}}{
                    \frac{\begin{array}{c}
                        \irenv(\alias{}) = (l_i:\ K_i){_{i=1}^n} \ {\scriptstyle{j \in 1..n}}
                    \end{array}
                }{\begin{array}{c}
                    \irtypingenv \vdash \textbf{t-ref}(\alias{}).l_j \colon \\ \quad \text{Expr}[K_j, \text{NScalar}]
                \end{array}}}}\quad
            \highlight{\stackrel{\textsc{R-TREF-IR}}{
                    \frac{\begin{array}{c}
                        \irenv(\alias{}) = A\end{array}
                }{\begin{array}{c}
                    \irtypingenv \vdash \textbf{rt-ref}(\alias{}) \colon \\ \quad \text{\rexpr{}}[A, \text{NScalar}]\end{array}}}}\quad
            \highlight{\stackrel{\textsc{R-TREF-PROJECT-IR}}{
                    \frac{\begin{array}{c}
                        \irenv(\alias{}) = (l_i:\ K_i){_{i=1}^n} \ {\scriptstyle{j \in 1..n}}
                    \end{array}
                }{\begin{array}{c}
                    \irtypingenv \vdash \textbf{rt-ref}(\alias{}).l_j \colon \\ \quad \text{\rexpr{}}[K_j, \text{NScalar}]
                \end{array}}}}
        \\
            \highlight{\stackrel{\textsc{AGG-IR}}{\frac{
                \begin{array}{c}
                    \irtypingenv \vdash q \colon \text{Query}[A, C] \quad
                    \irenv(\alias{}) = A\\
                    \irtypingenv \vdash b \colon \text{Expr}[B, \text{Scalar}]
                \end{array}
                }{
                \irtypingenv \vdash \textbf{agg}(\alias{}, q,\ b) \colon \text{Agg}[B]
                }
            }}\quad
            \highlight{\stackrel{\textsc{FLATMAP-IR}}{
                \frac{
                    \begin{array}{c}
                        \irtypingenv \vdash q_1 \colon Q_1 \quad
                        \irtypingenv \vdash q_2 \colon Q_2 \quad \irenv(\alias{}) = A
                        \\
                        Q_1 \in \{\text{\rquery{}}[A,\ {D_1,\ } C_1], \text{Query}[A, C_1]\}
                        \\
                        Q_2 \in \{\text{\rquery{}}[B,\ {D_2,\ } C_2], \text{Query}[B, C_2]\}
                    \end{array}
                }{
                    \irtypingenv \vdash \textbf{query}(\alias{}, q_1,\ q_2,\ \textbf{flatMap}) \colon \textit{\restrictcollect{}}(B, \text{Bag}, Q_1, Q_2)
            }}}
        \\
            \highlight{\stackrel{\textsc{MAP-IR}}{
                \frac{
                    \begin{array}{c}
                        \irenv(\alias{}) = A \quad
                        \irtypingenv \vdash q \colon Q \quad
                        \irtypingenv \vdash b \colon K \quad
                        (Q,\ K) \in \\
                        \{
                            (
                                \text{\rquery{}}[A, {D,\ } C],\
                                \text{\rexpr{}}[B, \text{NScalar}]
                            ),\\
                            (
                                \text{Query}[A, C],\
                                \text{Expr}[B, \text{NScalar}]
                            )
                        \}
                    \end{array}
                }{
                    \irtypingenv \vdash \textbf{query}(\alias{}, q,\ b,\ \textbf{map}) \colon \textit{\restrictcollect{}}(B, \text{Bag}, Q)
                }
            }}\qquad
            \highlight{\stackrel{\textsc{FILTER-IR}}{
                \frac{
                    \begin{array}{c}
                        \irenv(\alias{}) = A \quad
                        \irtypingenv \vdash q \colon Q \quad
                        \irtypingenv \vdash b \colon K \quad
                        (Q,\ K) \in \\
                        \{
                            (
                                \text{\rquery{}}[A, {D,\ } C],\
                                \text{\rexpr{}}[\text{Bool}, \text{NScalar}]
                            ),\\
                            (
                                \text{Query}[A, C],\
                                \text{Expr}[\text{Bool}, \text{NScalar}]
                            )
                        \}
                    \end{array}
                }{
                    \irtypingenv \vdash \textbf{query}(\alias{}, q,\ b,\ \textbf{filter}) \colon \textit{\restrictcollect{}}(A, C, Q)
                }
            }}
        \\
            \highlight{\stackrel{\textsc{FIX-IR}}{
                \frac{
                \begin{array}{c}
A_i = (l_j: B_j){_{j=1}^{m_i}} \ \forall i{_{=1}^n} \quad
                    \irtypingenv \vdash q_\text{base} \colon (\text{Query}[A_i, C_i]){_{i=1}^n}
                    \\
                    \irtypingenv \vdash q_\text{rec} \colon  (\text{\rquery{}}[A_i, {D_i,\ } \text{Set}]){_{i=1}^n} \quad n = 1 \\
                    \irenv(\alias{}_\text{rec}) = A_w \quad \irenv(\alias{}_i) = A_i \ \forall i{_{=1}^n} \\
                    \{ 1_\kappa, \ldots, n_\kappa \} \equiv \cup D_i{_{i=1}^n} \qquad
                    \forall D_i \quad |D_i|\equiv|\cup D_i| \qquad \irtypingenv \end{array}
                }
                {\irtypingenv \vdash \textbf{rec-query}(\alias{}_\text{rec},\ (\alias{}_i){_{i=1}^n},\ q_\text{base},\ q_\text{rec},\ w) : Q}
                }}\qquad
            \highlight{\stackrel{\textsc{LETREC-IR}}{
                \frac{
                \begin{array}{c}
\irenv(\alias{}_i) = A_i \quad
                    \irtypingenv \vdash r_i \colon \text{Query}[A_i, C_i] \ \forall i{_{=1}^n} \\
                    \irtypingenv \vdash q \colon T \quad T \in \{ \text{Query}[A, C],\ \text{Agg}[A] \}
                \end{array}
                }
                {\irtypingenv \vdash \textbf{ letrec } (\alias{}_i = r_i){_{i=1}^n} \textbf{ in } q \colon T}
            }}
\end{gathered}
    \]
  \end{minipage}}
    \caption{Typing rules for \ir{}}
    \label{fig:ir-types}
    \vspace{-1em}
\end{figure}

\begin{figure}[t]
   \setlength{\fboxsep}{0pt}
    \figbox{\begin{minipage}{\textwidth}
      \figurefontXLsize
      \setlength{\jot}{0pt}
      \setlength{\arraycolsep}{0pt}
      \vspace{-1.3em}
\[
\begin{array}{c}
\stackrel{\textsc{(ra-cmb-q)}}{\frac{
  \begin{array}{c}
  \textit{cmb} \in \{\textbf{map}, \textbf{flatMap}, \textbf{filter}\}
  \quad
  p_1 \Downarrow_{\text{RA}} q_1
  \quad
  f \Downarrow_{\text{RA}} (x) \rightarrow p_2 \quad
  p_2[x \mapsto \textbf{t-ref}(\alias{})] \Downarrow_{\text{RA}} q_2
  \end{array}
}{
  \begin{array}{l}
  \textit{cmb}\big(p_1 : \text{Query}[A, C],\ f: \dots): \dots
  \Downarrow_{\text{RA}}
  \textbf{query}\big(\alias{},\ q_1,\ q_2,\ \textit{cmb}\big) \quad
  \irenv' = \irenv[\alias{} \mapsto A]
  \end{array}
}}\\[12pt]

\stackrel{\textsc{(ra-cmb-r)}}{\frac{
  \begin{array}{c}
  \textit{cmb} \in \{\textbf{map}, \textbf{flatMap}, \textbf{filter}\}
  \quad
  p_1 \Downarrow_{\text{RA}} q_1
  \quad
  f \Downarrow_{\text{RA}} (x) \rightarrow p_2 \quad
  p_2[x \mapsto \textbf{rt-ref}(\alias{})] \Downarrow_{\text{RA}} q_2
  \end{array}
}{
  \begin{array}{l}
  \textit{cmb}\big(p_1 : \text{\rquery{}}[A, {D,\ } C],\ f: \dots): \dots
  \Downarrow_{\text{RA}}
  \textbf{query}\big(\alias{},\ q_1,\ q_2,\ \textit{cmb}\big) \quad
  \irenv' = \irenv[\alias{} \mapsto A]
  \end{array}
}}\\[12pt]

\stackrel{\textsc{(ra-aggregate)}}{\frac{
  \begin{array}{c}
  p_1 \Downarrow_{\text{RA}} q_1
  \quad
  f \Downarrow_{\text{RA}} (x) \rightarrow p_2
  \quad
  p_2[x \mapsto \textbf{t-ref}(\alias{})] \Downarrow_{\text{RA}} q_2
  \end{array}
}{
  \begin{array}{l}
  \textbf{agg}\big(p_1 : \text{Query}[A, C],\ f: \dots) : \dots
  \Downarrow_{\text{RA}}
  \textbf{agg}\big(\alias{},\ q_1,\ q_2\big) \quad
  \irenv' = \irenv[\alias{} \mapsto A]
  \end{array}
}}\\[12pt]

\stackrel{\textsc{(ra-relop)}}{\frac{
  \begin{array}{c}
  p_i \Downarrow_{\text{RA}} q_i \quad \forall i {\in 1..n}
  \end{array}
}{
  \begin{array}{l}
  \textit{relOp}\big((p_i : T ){_{i=1}^n}\big) : (\text{Query}[A, C]|\text{\rquery{}}[A, {\ D,\ }C])
  \Downarrow_{\text{RA}}
  \textit{relOp}\big(\alias{},\ (q_i){_{i=1}^n}\big) \quad
  \irenv' = \irenv[\alias{} \mapsto A]
  \end{array}
}}\\[12pt]

\stackrel{\textsc{(ra-fix)}}{\frac{
  \begin{array}{c}
  (p_{q_i}){_{i=1}^n} \Downarrow_{\text{RA}} (q_{q_i}){_{i=1}^n}
  \quad
  f \Downarrow_{\text{RA}} (x_i){_{i=1}^n} \rightarrow (p_{r_i}){_{i=1}^n} \quad
  p_{r_i}[x_j \mapsto \textbf{r-table}(\alias{}_j {,\ (j)}) \text{ for } j {\in 1..n}] \Downarrow_{\text{RA}} q_{r_i}  \forall i {\in 1..n}
  \end{array}
}{
  \begin{array}{c}
  \textbf{fix}\big((p_{q_i}){_{i=1}^n}: \dots,\ f :
  (\text{\rquery{}}[A_i, {\ (i),\ } \text{Set}]){_{i=1}^n} \rightarrow (\text{\rquery{}}[A_i, {\ (i),\ } C_i]){_{i=1}^n} \big).\_w : \dots \\
  \Downarrow_{\text{RA}}
  \textbf{letrec } \alias{}_{\text{rec}} =
  \textbf{rec-query}\big(\alias{}_\text{rec},\ (\alias{}_i){_{i=1}^n},\ (q_{q_i}){_{i=1}^n},
  (q_{r_i}){_{i=1}^n},\ w\big)
  \textbf{ in } \textbf{table}(\alias{}_\text{rec}) \\
  \irenv' = \irenv[\alias{}_i \mapsto A_i \text{ for } i {\in 1..n}, \alias{}_\text{rec} \mapsto A_i{_{i = w}}]
  \end{array}
}}
\end{array}
\]
\end{minipage}}
\caption{Type-directed translation function shown as a big-step relation from terms in explicitly typed normalized \calc{} to terms in \ir{}.
The \textit{toIR} function is composed of the translation from normalized \calc{} to explicitly typed normalized \calc{} and $\Downarrow_\text{RA}$.
Each rule chooses a unique, fresh symbol $\alias{}$ (not in $\Delta$, $\Sigma$, or $\irenv$)
 and adds the symbol to the signature $\irenv'$. To limit notational overhead, explicit type annotations that do not impact translation are abbreviated with $\dots$
}
\vspace{-1em}
\label{fig:norm-stage3}
\end{figure}

\begin{figure}[t]
   \setlength{\fboxsep}{0pt}
    \figbox{\begin{minipage}{\textwidth}
      \figurefontXLsize
      \setlength{\jot}{0pt}
      \setlength{\arraycolsep}{0pt}
      \vspace{-1.3em}
\[
\begin{array}{@{\extracolsep{\fill}} lcll @{}}
\multicolumn{3}{l}{
  \begin{aligned}[t]
  Z = \textbf{letrec } \alias{}_a = Q_a {_{a=1}^n} \textbf{ in } A
  \end{aligned}
} & \\[3pt]

\begin{aligned}[t]
  &\textit{relOp}\big(\alias{},\ (Q_{j}){_{j=1}^{q-1}},\ Z,\  (Q_{k}){_{k=q+1}^m}\big)
\end{aligned}
&\rightharpoondown\quad&
\begin{aligned}[t]
  &
  \textbf{letrec } \alias{}_a = Q_a {_{a=1}^n} \textbf{ in }  \\ & \quad
  \textit{relOp}\big(\alias{},\ (Q_{j}){_{j=1}^{q-1}},\ A,\ (Q_{k}){_{k=q+1}^m}\big)
\end{aligned}
& \hspace{-6pt}(\textsc{anf-chained-o})\\
& & & \\[-6pt]

\begin{aligned}[t]
  &\textbf{query}(\alias{},\ Z,\ R,\ \textit{cmb})
\end{aligned}
&\rightharpoondown\quad&
\begin{aligned}[t]
  &
  \textbf{letrec } \alias{}_a = Q_a {_{a=1}^n} \textbf{ in }  \\ & \quad
  \textbf{query}(\alias{},\ A,\ R,\ \textit{cmb})
\end{aligned}
& \hspace{-6pt}(\textsc{anf-chained-q})\\

\begin{aligned}[t]
  &\textbf{query}(\alias{},\ Q,\ Z,\ \textit{cmb})
\end{aligned}
&\rightharpoondown\quad&
\begin{aligned}[t]
  &
    \textbf{letrec } \alias{}_a = Q_a {_{a=1}^n} \textbf{ in }  \\ & \quad
  \textbf{query}(\alias{},\ Q,\ A,\ \textit{cmb})
\end{aligned}
& \hspace{-6pt}(\textsc{anf-nested-q})\\

\begin{aligned}[t]
  &\textbf{agg}( Z,\ R)
\end{aligned}
&\rightharpoondown\quad&
\begin{aligned}[t]
  &
    \textbf{letrec } \alias{}_a = Q_a {_{a=1}^n} \textbf{ in }  \\ & \quad
  \textbf{agg}(A,\ R) \end{aligned}
& \hspace{-6pt}(\textsc{anf-chained-a})\\

\begin{aligned}[t]
  &\textbf{agg}(Q,\ Z)
\end{aligned}
&\rightharpoondown\quad&
\begin{aligned}[t]
  &
      \textbf{letrec } \alias{}_a = Q_a {_{a=1}^n} \textbf{ in }  \\ & \quad
  \textbf{agg}(Q,\ A) \end{aligned}
& \hspace{-6pt}(\textsc{anf-nested-a})\\[10pt]

\begin{aligned}[t]
  &
    \textbf{letrec } \alias{}_b = Q_b {_{b=1}^{p-1}},
        \alias{}_p = \\ & \quad
            \textbf{rec-query}\big(\alias{}_{\text{ rec-q}},\ (\alias{}_{\text{ q}_i}){_{i=1}^m},\\ & \quad\quad
            (Q_{j}){_{j=1}^{q-1}},\ Z,\ (Q_{k}){_{k=q+1}^m},\
            (R_{l}){_{l=1}^m},\ w\big), \\ & \quad
        \alias{}_d = Q_d{_{d=p+1}^t} \textbf{ in } Q
\end{aligned}
&\rightharpoondown\quad&
\begin{aligned}[t]
      &\textbf{letrec } \alias{}_b = Q_b {_{b=1}^{p-1}},
        \alias{}_p = \\ & \quad
            \textbf{rec-query}\big(\alias{}_{\text{ rec-q}},\ (\alias{}_{\text{ q}_i}){_{i=1}^m},\\ & \quad\quad
            (Q_{j}){_{j=1}^{q-1}},\ A,\ (Q_{k}){_{k=q+1}^m},\
            (R_{l}){_{l=1}^m},\ w\big), \\ & \quad
        \alias{}_d = Q_d{_{d=p+1}^t},\
        \alias{}_a = Q_a {_{a=1}^n}
        \textbf{ in } Q
\end{aligned}
& \hspace{-6pt}(\textsc{anf-chained-r})\\[-6pt]
& & & 1 \le q \le m \\[-0.1em] &&& 1 \le p \le t\\

\begin{aligned}[t]
  &
    \textbf{letrec } \alias{}_b = Q_b {_{b=1}^{p-1}},
        \alias{}_p = \\ & \quad
            \textbf{rec-query}\big(\alias{}_{\text{ rec-q}},\ (\alias{}_{\text{ q}_i}){_{i=1}^m},\\ & \quad\quad
            (Q_{l}){_{l=1}^m},\
            (R_{j}){_{j=1}^{q-1}},\ Z,\ (R_{k}){_{k=q+1}^m},\
            w\big), \\ & \quad
        \alias{}_d = Q_d{_{d=p+1}^{t}} \textbf{ in } Q
\end{aligned}
&\rightharpoondown\quad&
\begin{aligned}[t]
      &\textbf{letrec } \alias{}_b = Q_b {_{b=1}^{p-1}},
        \alias{}_p = \\ & \quad
            \textbf{rec-query}\big(\alias{}_{\text{ rec-q}},\ (\alias{}_{\text{ q}_i}){_{i=1}^m},\\ & \quad\quad
            (Q_{l}){_{l=1}^m},\
            (R_{j}){_{j=1}^{q-1}},\ A,\ (R_{k}){_{k=q+1}^m},\
            w\big), \\ & \quad
        \alias{}_d = Q_d{_{d=p+1}^t},\
        \alias{}_a = Q_a {_{a=1}^n}
        \textbf{ in } Q
\end{aligned}
& \hspace{-6pt}(\textsc{anf-nested-r})\\[-6pt]
& & & 1 \le q \le m \\[-0.1em] &&& 1 \le p \le t\\

\begin{aligned}[t]
  & \textbf{letrec } \alias{}_b = Q_b {_{b=1}^m} \textbf{ in } Z
\end{aligned}
&\rightharpoondown\quad&
\begin{aligned}[t]
  &
  \textbf{letrec } \alias{}_b = Q_b {_{b=1}^m},\ \alias{}_a = Q_a {_{a=1}^n}
  \textbf{ in }  A\\ &
\end{aligned}
& \hspace{-6pt}(\textsc{anf-hoist-letrec})\\

\end{array}
\]
\end{minipage}}
\caption{\ir{} Normalization function \textit{normIR}. All \textbf{rec-query} terms are hoisted into a single outer \textbf{letrec}. Freshness of bindings is ensured by $\Downarrow_\text{RA}$.
}
\label{fig:norm-stage4}
\end{figure}

To simplify translation to \lsd{}, we utilize a type-preserving intermediate representation \ir{} and a single IR normalization pass.
The translation to \ir{} is defined for an explicitly-typed variant of \calc{}, where terms are annotated with their types and take the form $t: T$. The elaboration step from implicitly-typed to explicitly-typed \calc{} using standard techniques is routine~\cite{bidir_typing} and omitted. For brevity, $\Sigma$ contains only \textbf{distinct}, \textbf{union}, \textbf{unionAll}, one non-scalar operator \textbf{not in}, and one scalar operator \textbf{+}. As both negation and aggregation can violate \mono{} (monotonicity), we include \textbf{agg} but omit \textbf{groupBy}. The translation to \ir{} is shown as a big-step relation $\Downarrow_\text{RA}$ that translates normalized \calc{} terms into a query-based syntax. The translation (1) beta-reduces the functions passed to the combinators; (2) gives each query an alias $\alpha$ and constructs a signature $\irenv$ that maps aliases to their type, and (3) wraps recursive queries in \textbf{letrec} expressions. Aliases are unique identifiers for queries and subqueries, and $\irenv$ is used in the same way that $\Sigma$ is used for \textit{db} types in \calc{} and T-LINQ.
The $\Downarrow_\text{RA}$ phase eliminates variables and functions so terms of \ir{} are closed, therefore the typing rules of \ir{} do not require a typing environment, only the signature $\irenv$, which is static after $\Downarrow_\text{RA}$ completes.
The \textsc{ra-fix} rule can assume that all \textbf{rec-query} terms are immediately projected because of the \textsc{detuple-1/2} rules in \calc{} normalization.
Normalization of \ir{} is performed by the \textit{normIR} function, which applies bog-standard hoisting of letrec terms, allowing the clean separation of recursive and non-recursive queries.
We write $\rightharpoondown^*$ for the reflexive and transitive closure of $\rightharpoondown$, the compatible closure of the rules in Figure~\ref{fig:norm-stage4}.
The \ir{} syntax is shown in Figure~\ref{fig:ir-syntax}, the normalized \ir{} syntax in Figure~\ref{fig:ir-norm-syntax}, the typing rules in Figure~\ref{fig:ir-types}, and the statements of type-preservation are located in Section~\ref{sec:norm-preservation}.
 \subsubsection{Translation to \lsd{}}
\begin{figure}[t]
\setlength{\fboxsep}{0pt}
    \figbox{\begin{minipage}{\textwidth}
      \figurefontXLsize
\setlength{\jot}{0pt}
      \setlength{\arraycolsep}{0pt}
      \vspace{-1.4em}
\[
\begin{array}{lll}

    & \llbracket \textbf{letrec } \alias{}_i = r_i{_{i=1}^n} \textbf{ in } q \rrbracket_\Psi = (P', p', \Psi')\\
        & \qquad \text{A. Translate bindings into Datalog} \\
            & \qquad\qquad \text{Each } r_i \text{ has the form }
                \textbf{rec-query}(\alias{}_i, (\alias{}_j){_{j=1}^{m_i}},\ (q_{\text{base}_j}){_{j=1}^{m_i}}, (q_{\text{rec}_j}){_{j=1}^{m_i}}, w)\\
            & \qquad\qquad \Psi_{x_i} = \bigcup_{k {\in 1..i-1}} \Psi \cup \Psi_k \\[0.5em]
            & \qquad\qquad\text{1. Translate base-cases to non-recursive \datalogns{}} \\
                & \qquad\qquad\qquad (P_{\text{base}_j}, p_{\text{base}_j}, \Psi_{\text{base}_j}){_{j=1}^{m_i}} = \tonrlsd{}_{\Psi_{x_i}}(\textbf{distinct}(q_{\text{base}_j})) &  \\
            & \qquad\qquad\text{2. For each alias}_j \text{, add a rule assigning it to the corresponding base-case } \\ & \qquad\qquad\qquad S_j = \alias{}_j(\Psi(p_{\text{base}_j})) \text{ :- } p_{\text{base}_j}(\Psi(p_{\text{base}_j}))\\
            & \qquad\qquad\text{3. Translate recursive-case to non-recursive \datalogns{}}\\ & \qquad\qquad\qquad (P_{\text{ recur}_j}, p_{\text{ recur}_j}, \Psi_{\text{ recur}_j}){_{j=1}^{m_i}} = \tonrlsd{}_{\Psi_{x_i}}(q_{\text{rec}_j}) &  \\
            & \qquad\qquad\text{4. Add recursive rules} \\ & \qquad\qquad\qquad R_j = \alias{}_j(\Psi(p_{\text{ recur}_j})) \text{ :- } p_{\text{ recur}_j}(\Psi(p_{\text{ recur}_j}))\\
            & \qquad\qquad\text{5. Assemble \lsd{} program}\\
                & \qquad\qquad\qquad P_i = \bigcup_{j {\in 1..m_i}} (P_{\text{base}_j} \cup P_{\text{ recur}_j} \cup \{ R_j \} \cup \{ S_j \} )
                \qquad \Psi_i = \bigcup_{j {\in 1..m_i}} (\Psi_{\text{base}_j} \cup \Psi_{\text{ recur}_j})
                \\ &\qquad\qquad\qquad p_i = \alias{}_j \text{ where } j = w \\

& \qquad \text{B. Translate the body of the let-rec into non-recursive Datalog} \\
        & \qquad\qquad \Psi_s = \bigcup_{i {\in 1..n}} \Psi_i \\
        & \qquad \qquad (P_{\text{body}}, p_{\text{body}}, \Psi_{\text{body}}) = \tonrlsd{}_{\Psi_s}(q) \\
        & \qquad \text{C. Combine final program}\\
        & \qquad\quad P' = \big(\bigcup_{i {\in 1..n}} P_i
\big) \cup P_{\text{body}} \quad p' = p_{\text{body}} \quad \Psi' = \Psi_s \cup \Psi_{\text{body}} \\

\end{array}
\]
\end{minipage}}
\caption{Translation Function to \lsd{}}
\label{fig:translate-to-dl}
\vspace{-0.5em}
\end{figure}
 \begin{figure}[t]
\centering
\includegraphics[trim={.1cm .1cm .1cm 0.1cm}, clip, width=.9\linewidth]{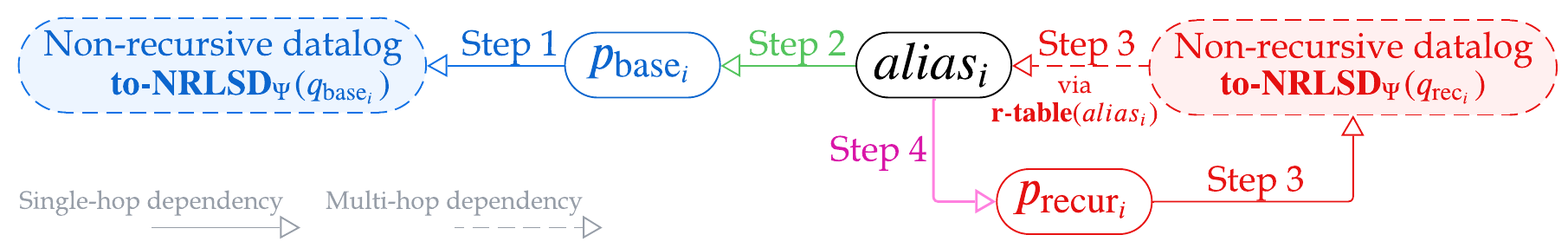}
\caption{
    Illustration of dependencies produced by Steps A.1-A.5 of $\llbracket \cdot \rrbracket$ (Figure~\ref{fig:translate-to-dl}). }\label{fig:deps}
\end{figure} \begin{figure}[t]
    \scriptsize
    \centering
    \begin{minipage}[t]{.24\linewidth}
        \begin{lstlisting}[style=lrql-ex, emph={paths}]
filter(
  filter(
    fix(
      map(
        filter(
          table(Edges),
          (e) ->
            e.src == "A"),
        (e) ->
          (from=e.src,
           to=e.dst,
           direct=True)),
      (paths) ->
        (flatMap(
          table(Edges),
          (e) ->
            map(
              filter(
                paths,
                (p) ->
                  p.to == e.src),
              (p) ->
                (from=p.from,
                 to=e.dst,
                 direct=False)
            ))))._1,
    (p) -> p.to == "B"),
  (p) ->
      p.direct == False)

        \end{lstlisting}
        \subcaption{\calc{} term $t$}
    \end{minipage}
    \begin{minipage}[t]{.24\linewidth}
        \begin{lstlisting}[style=lrql-ex, emph={paths}]
filter(
  fix(
    map(
      filter(
        table(Edges),
        (e) -> e.src == "A"),
      (e) ->
        (from=e.src,
         to=e.dst,
         direct=True)),
    (paths) ->
      (flatMap(
        table(Edges),
        (e) ->
          map(
            filter(
              paths,
              (p) ->
                p.to == e.src),
            (p) ->
              (from=p.from,
               to=e.dst,
               direct=False)
          ))))._1,

  (p) -> p.to == "B" &&
     p.direct == False)

        \end{lstlisting}
        \subcaption{Normalized \calc{}}
    \end{minipage}
\begin{minipage}[t]{.25\linewidth}
        \begin{lstlisting}[style=ir]
query(a_7
  letrec a_1 = rec-query(a_1,
    (a_1),
    (query(a_1,
      query(a_2,
        table(Edges),
        t-ref(a_2).src == "A",
        filter),
      (from=t-ref(a_1).src,
       to=t-ref(a_1).dst,
       direct=True),
      map)),
    (query(a_3,
      table(Edges),
      query(a_4,
        query(a_5,
          r-table(a_1, (1)),
          rt-ref(a_5).to
            == t-ref(a_3).src,
          filter),
        (from=rt-ref(a_4).from,
         to=t-ref(a_3).dst,
         direct=False),
        map),
      flatMap)),
    1) in table(a_1),
  t-ref(a_7).to == "B" &&
    t-ref(a_7).direct
      == False,
  filter)
        \end{lstlisting}
        \subcaption{\ir{}}
    \end{minipage}
    \begin{minipage}[t]{.25\linewidth}
      \centering
        \begin{lstlisting}[style=ir]
letrec a_1 = rec-query(a_1,
  (a_1),
  (query(a_1,
    query(a_2,
      table(Edges),
      t-ref(a_2).src == "A",
      filter),
    (from=t-ref(a_1).src,
      to=t-ref(a_1).dst,
      direct=True),
    map)),
  (query(a_3,
    table(Edges),
    query(a_4,
      query(a_5,
        r-table(a_1, (1)),
        rt-ref(a_5).to
          == t-ref(a_3).src,
        filter),
      (from=rt-ref(a_4).from,
        to=t-ref(a_3).dst,
        direct=False),
      map),
    flatMap)),
    1) in query(a_7
  table(a_1),
  t-ref(a_7).to == "B" &&
    t-ref(a_7).direct ==
      False,
  filter)
        \end{lstlisting}
        \subcaption{Normalized \ir{} term}
    \end{minipage}
    \begin{minipage}[t]{\linewidth}
        \centering
        \begin{lstlisting}[style=datalog, emph={$\alpha_1$}, xleftmargin=.2\linewidth]

p_1(src, dst)          :- Edges(src, dst), src = "A"                     # A.1
p_2(from, to, True)    :- Edges(from, to)                                # A.1
a_1(from, to, direct)  :- p_2(from, to, direct)                           # A.2
p_3(from, to, False)   :- a_1(from, e, _), Edges(e, to)                   # A.3
a_1(from, to, direct)  :- p_3(from, to, direct)                           # A.4
p_4(from, to, direct)  :- a_1(from, to, direct), to = "B", direct = False # B
        \end{lstlisting}
        \subcaption{\lsd{}}
    \end{minipage}
    \caption{
      Example translation pipeline of a transitive closure query for indirect paths from node ``A'' to ``B''.
      Rules applied left-to-right: original query in \calc{}; \textsc{fil-fil} by \textit{norm}; \textsc{ra-fix}, \textsc{ra-combinator-q} by \textit{toIR}; \textsc{anf-chained-q} by \textit{normIR}; then $\llbracket \cdot \rrbracket$.
    }
    \label{fig:example-translation-pipeline}
\end{figure}

Prior work establishes the equivalence between combinator-style functional constructs and relational algebra with set difference~\cite{comprehensions, NRC}, and between relational algebra with set difference and \textbf{non-recursive} Datalog with negation (\datalogns{})~\cite{foundations}. We refer to the translation function from non-recursive \ir{} to non-recursive \datalogns{} that leverages these equivalences by \tonrlsd{}. The \tonrlsd{} function takes \ir{} terms of type $\text{Query}[A, \text{Set}]$, $\text{\rquery{}[A, {\ D,\ }\text{Set}]}$, or $\text{Agg}[A]$ and returns a non-recursive \datalogns{} program $P$, a distinguished goal predicate $p$, and a schema environment $\Psi$ mapping predicates to their schema.

Datalog predicate schemas are positional and follow directly from \calc{}'s label-based column types: e.g., the Named Tuple type \texttt{\{name: String, age: Int\}} becomes the schema \texttt{(String, Int)}.
\tonrlsd{} is parameterized by the schema environment $\Psi$.
We omit the full definition of \tonrlsd{} because the translation from relational algebra with set difference to non-recursive \datalogns{} can be found in many database textbooks, e.g.,~\cite{foundations}, and
state only the relevant properties.
Let $t$ be a well-typed \calc{} term and let $q = \textit{normIR}(\textit{toIR}(\textit{norm}(t)))$.
Let $\Sigma$ be the signature used by $t$, let $\irenv$ be the signature produced by $\Downarrow_\text{RA}$, and let $\Psi$ be the schema environment composed of the symbols in $\Sigma$ and $\irenv$.
If $q$ contains no \textbf{rec-query} subterm, let
$\tonrlsd{}_\Psi(q)=(P',p',\Psi')$. The translation satisfies:

\begin{itemize}
  \item \textbf{P1} \emph{Safety.} In every rule of $P'$, all head variables occur in body atoms, and all predicates are schema-consistent.
  \item \textbf{P2} \emph{Freshness.} For every rule $r\in P'$, $\mathrm{head}(r)\notin \mathrm{dom}(\Psi)$.
  \item \textbf{P3} \emph{Non-recursive.} For all $p_1,p_2\in\mathrm{Heads}(P')$, if $p_2$ appears in the
        body of a rule whose head is $p_1$, then $p_1$ does not appear in the body of any rule whose
        head is reachable from $p_2$ in the dependency graph of $P'$.
    \item \textbf{P4} \emph{Namespacing.}
      Each invocation of \tonrlsd{} uses a fresh namespace; newly introduced head predicates
      are qualified by that namespace. Predicates in $\mathrm{dom}(\Psi)$ are not namespaced.
      Distinct invocations therefore produce disjoint sets of newly introduced heads, and
      bodies in $r \in P'$ mention only predicates in $\mathrm{dom}(\Psi)$ or $\mathrm{Heads}(P')$.
\end{itemize}

\noindent Because linearity, stratification, and direct-recursion apply only to recursive predicates, non-recursive \datalogns{} is a strict subset of \lsd{}.

\begin{sdefn}[Translation from normalized \ir{} to \lsd{},  i.e., the \textit{toDL} function.]\label{def:translation}
Let $\llbracket \cdot \rrbracket$ denote a translation of a well-typed normalized \ir{} term $\irtypingenv{} t \colon \text{Query}[A,\text{Set}]$ or $\text{Agg}[A]$ that constructs its corresponding \lsd{} program $P$, a predicate $\textit{p } \in\ P$, and a schema environment $\Psi$ that is initialized from the $\irenv$ that is collected during $\Downarrow_\text{RA}$.

\[
\begin{array}{c}
\hspace{0.45\linewidth}\llbracket t \rrbracket_{\Psi} \rightarrow (P, p, \Psi)\hspace{0.45\linewidth}
\end{array}
\]
The full translation function is shown in Figure~\ref{fig:translate-to-dl}. Note that
$\llbracket \cdot \rrbracket$ is implemented as a \textbf{non-recursive} function, which greatly simplifies the proof of Theorem~\ref{theorem:toDL}.
\end{sdefn}
\subsection{Proofs}\label{sec:appendix:proofs}
{
The following proofs are presented on paper, relying on standard proof techniques and established results from database theory. While foundational Datalog semantics have been mechanized~\cite{mechDL}, end-to-end mechanization of our translation and type-preservation proofs is future work.
}
\subsubsection{\texorpdfstring{$\llbracket \cdot \rrbracket$}{[[.]]} produces well-formed \lsd{}}

\begin{slemma}[$\tonrlsd{}_\Psi$ produces well-formed \lsd{}]\label{lemma:nrlsd}
Let $t$ be a well-typed \calc{} term and let $q = \textit{normIR}(\textit{toIR}(\textit{norm}(t)))$.
Let $\Sigma$ be the signature used by $t$, let $\irenv$ be the signature produced by $\Downarrow_\text{RA}$, and let $\Psi$ be the schema environment composed of the symbols in $\Sigma$ and $\irenv$.
Let $\tonrlsd{}_\Psi(q)=(P',p',\Psi')$. If $\irtypingenv \vdash q : \text{Query}[A,\text{Set}]$ or $\irtypingenv \vdash q : \text{\rquery{}}[A,{D,\ } \text{Set}]$ or $\irtypingenv \vdash q : \text{Agg}[A]$, and $q$ contains no \textbf{rec-query} subterm
then $(P',p',\Psi')$ is \textit{well-formed} (Def.~\ref{def:datalog:wf}).
\end{slemma}

\begin{sproof}
Immediate from Def.~\ref{def:datalog:wf} and \tonrlsd{} properties \textbf{P1} (Safety), \textbf{P2} (Freshness), and \textbf{P3} (Non-recursive). \textbf{Safe}: $\Psi$ includes schemas for all base predicates (from $\Sigma$) and other predicates (from $\irenv$), ensuring that \tonrlsd{} does not encounter any undefined schemas. By \textbf{P1},
predicate arities and operator types match, and every head variable occurs in the body. \textbf{Non-recursive}: by \textbf{P2},
all rules have fresh heads and predicates $ \in \Psi$ appear only in bodies, so dependencies point between new predicates or from new predicates to predicates $ \in \Psi$. By \textbf{P3},
newly created predicates have no cyclic dependencies. Consequently, $(P',p',\Psi')$ satisfies the well-formedness conditions of Def.~\ref{def:datalog:wf}: every predicate is non-recursive, schema-consistent, and safe.
\end{sproof}

\begin{spropn}[$\llbracket\cdot\rrbracket_\Psi$ produces well-formed \lsd{}]\label{theorem:toDL}
Let $t$ be a well-typed \calc{} term and let $t' = \textit{normIR}(\textit{toIR}(\textit{norm}(t)))$.
Let $\Sigma$ be the signature used by $t$, let $\irenv$ be the signature produced by $\Downarrow_\text{RA}$, and let $\Psi$ be the schema environment composed of the symbols in $\Sigma$ and $\irenv$.
For a well-typed \ir{} term in normal form $\irtypingenv \vdash t' : \text{Query}[A,\text{Set}]$ or $\irtypingenv \vdash t' : \text{Agg}[A]$,
\[
\begin{array}{c}
\hspace{0.3\linewidth}\llbracket t' \rrbracket_\Psi \;=\; (P',\, p',\, \Psi')\hspace{0.3\linewidth}
\end{array}
\]
is defined, and $(P', p', \Psi')$ is a \emph{well-formed} \lsd{} program (Def.~\ref{def:datalog:wf}).
\end{spropn}

\begin{sproof}
We proceed by case analysis on the structure of $t'$ (sufficient because $\llbracket\cdot\rrbracket_\Psi$ is not recursive). From the syntax of normalized \ir{}, $t'$ has the form $\textbf{letrec } \alias{}_i = r_i{_{i=1}^n} \textbf{ in } q$.

\paragraph{If the \textbf{letrec} term contains no bindings} By the rules of $\rightharpoondown^*$ (Figure~\ref{fig:norm-stage4}) and the syntax of normalized \ir{} (Figure~\ref{fig:ir-norm-syntax}) there are no \textbf{rec-query} terms in $t'$; by Lemma~\ref{lemma:nrlsd}, the \datalogns{} program produced by $\tonrlsd{}$ is well-formed.

\paragraph{If the \textbf{letrec} contains at least one binding} By the rules of $\rightharpoondown^*$, the syntax of normalized \ir{}, and the \textsc{ra-fix} rule of $\Downarrow_\text{RA}$, each binding at position $i {\in 1..n}$ is of the form:
\[
\begin{array}{c}
\hspace{0.2\linewidth}
        \alias{}_i = \textbf{rec-query}(\alias{}_i,\ (\alias{}_j){_{j=1}^{m_i}},\ (q_{\text{base}_j}){_{j=1}^{m_i}}, (q_{\text{rec}_j}){_{j=1}^{m_i}}, w)
\hspace{0.2\linewidth}
\end{array}
\]
    Write $p_1 \leadsto p_2$ for a direct dependency (for example: $p_1(\dots)\text{ :- }p_2(\dots)$) and $\leadsto^*$ for the transitive closure of $\leadsto$. The stepwise dependency graph generated by step A of the translation function, e.g., the translation of each binding, is illustrated in Fig.~\ref{fig:deps}. We proceed by considering each sub-step 1-5 of A:

    \begin{enumerate}
    \item
     Step 1 applies $\tonrlsd{}_\Psi$ to each $q_{\text{base}_j}$, producing non-recursive \datalogns{} programs.
     By the IR typing rules, for each term that contains an alias $\alias{}$, the premise gives that $\alias{} \in \irenv$.
     By the syntax of normalized \ir{}, \textbf{rec-query} terms can only be on the right-hand side of \textbf{letrec} bindings therefore each $q_{\text{base}_j}$ contains no nested \textbf{rec-query} terms.
     By \textsc{distinct-ir}, each input to $\tonrlsd{}$ is set-based.
     Therefore, by Lemma~\ref{lemma:nrlsd}, Step 1 produces well-formed (non-recursive) \lsd{} programs. By \tonrlsd{} property P4 (namespacing), we can combine the rules of each $P_{\text{base}_j}$ into a single program without risking safety or acyclicity of the dependency graph.

     \item
     Step 2 introduces rules $S_j$ of the form
     $\alias{}_j(\Psi(p_{\text{base}_j})) \text{ :-- } p_{\text{base}_j}(\Psi(p_{\text{base}_j}))$.
     By \textsc{fix-ir}, $\irenv(\alias{}_j) = A_j$ and $Q_j = A_j$, therefore the schemas of $p_{\text{base}_j}$ and $\alias{}_j$ are equivalent and contain only a single positive body atom, therefore $S_j$ are safe.
     $S_j$ introduces the non-recursive dependency $p_{\text{base}_j} \leadsto \alias{}_j$; thus, the combination of the rules of $P_{\text{base}_j}$ and $S_j$ for $j \in 1..m_j$ produces a safe and non-recursive \lsd{} program.

    \item
     Step 3 applies $\tonrlsd{}$ to each $q_{\text{rec}_j}$, producing non-recursive \lsd{}.
     We can use Lemma~\ref{lemma:nrlsd} in the same way as in Step 1, except we apply the \textsc{fix-ir} rule instead of \textsc{distinct-ir} to ensure that the term passed to \tonrlsd{} is set-based.
     By Lemma~\ref{lemma:nrlsd}, the result of Step 3 is safe and non-recursive.

     Step 3 introduces a zero-or-more hop dependency between $p_{\text{recur}_j} \leadsto^* \alias{}_j$.
     By \tonrlsd{} property P2, all dependencies introduced by \tonrlsd{} will be uni-directional: all dependencies will be ``incoming edges'' to $\alias{}_j$. Thus the combination of the rules produced by Steps 1-3 will be safe and non-recursive.

    \item
     Step 4 introduces rules $R_j$ of the form
     $\alias{}_j(\Psi(p_{\text{recur}_j})) \text{ :-- } p_{\text{recur}_j}(\Psi(p_{\text{recur}_j}))$.
     The resulting program will be safe because the schemas $\Psi(p_{\text{base}_j})$ and  $\Psi(p_{\text{recur}_j})$ will be the same:
     by \textsc{fix-ir}, $\irtypingenv \vdash q_{\text{base}_j} \colon \text{Query}[A_j, \dots]$ and $\irtypingenv \vdash q_{\text{rec}_j} \colon \text{\rquery{}}[A_j, \dots]$.
     By \textsc{ra-combinator-q/r}, $\irenv(\alias{}_j)=A_j$, therefore $\Psi(p_{\text{recur}_j}) \equiv \Psi(p_{\text{base}_j})$.

     This step ``closes the loop'', e.g., introduces a cycle in the dependency graph by adding a dependency $\alias{}_j \leadsto p_{\text{recur}_j}$. Therefore the program produced by Steps 1-4 will be safe (Def.~\ref{def:datalog:limited}), but will also be recursive (Def.~\ref{def:datalog:recursive}).

     As shown in Figure~\ref{fig:deps}, the only recursive dependencies must be contained within the rules generated by Steps 3 and 4. It remains to show that this fragment is linear, stratified, and direct-recursive.

        \begin{itemize}

        \item
         \emph{Linearity} (Def.~\ref{def:datalog:linear})
         {
By the typing rules for IR, \textit{\restrictcollect{}} collects the source relation identifiers \textit{i} into $D$.
     By \textsc{fix-ir}, for $r$ to be well-typed, its type $R$ must have no duplicates in $D_i$ (enforced by $\forall D_i \  |D_i|\equiv|\cup D_i|$)
     nor can any of the source relations $Q$ be missing in all $D$ (enforced by $\{ 1_\kappa, \ldots, n_\kappa \} \equiv \cup D_i{_{i=1}^n}$).
         }
         Thus each $\alias{}_j$ depends only on one recursive predicate (itself), so for all recursive rules in $P'$ with head predicate $\alias{}_j$ there will be exactly one body atom with the predicate $\alias{}_j$.

         \item
         \emph{Direct-recursion} (Def.~\ref{def:datalog:direct})
         With the restriction $n=1$ on \textsc{fix-ir} only one recursive predicate $\alias{}_1$ is defined per \textbf{rec-query} and the body must be of type \rquery{}.
         Type equality of dependency tuples respects the $\kappa$ tag, which is unique to each \textbf{fix} invocation. Therefore, all recursive predicates may depend only on recursive predicates defined by the same invocation of \textbf{rec-query}, and since there is only one recursive predicate per \textbf{rec-query}, the resulting program graph $G$ of $P'$ contains only one derived predicate per cycle.

        \item
         \emph{Stratification} (Def.~\ref{def:datalog:stratified})
         $\llbracket t \rrbracket_\Psi$ is stratified if for all predicates in a cyclic dependency, there are only positive dependencies, i.e., there are no negative body literals between predicates in a stratum. Each \textbf{rec-query} represents a single stratum. There are no negative body literals in the rules introduced by Step 4. Therefore, it suffices to show that there are no negative body literals in the rules generated by Step 3.

        For there to be a negative body literal in the translated \lsd{} program, the \ir{} program must contain the subterm \textbf{not in}. By \textsc{expr-neg-ir}, the expression \textbf{not in} produces a term of type $\text{Expr}[A, \text{Scalar}]$. By the meta-helper method \textit{Shape}, any expressions containing subexpressions of type $\text{Expr}[A, \text{Scalar}]$ also have type $\text{Expr}[A, \text{Scalar}]$. By \textsc{agg-ir}, \textsc{map-ir} and \textsc{filter-ir}, the only well-typed terms that may contain $\text{Expr}[A, \text{Scalar}]$ must be of the form $\textbf{agg}(\alias{},\ q,\ b)$. By \textsc{agg-ir}, $\irtypingenv \vdash q \colon \text{Query}[A, C]$. All recursive references take the form of $\textbf{r-table}(\alpha {,\ i})$ and have type \rquery{}, therefore \textbf{agg} cannot be applied to recursive predicates; it applies only to terms of type Query. Negation applied to terms of type Query represents cross-strata aggregation, which is allowed in \lsd{}.

    \end{itemize}

    Therefore, the program produced by Step 1-4 will be safe, direct-recursive, linear, and stratified.

    \item Step 5 combines the rules produced from Steps 1-4 into a single program. By \tonrlsd{} property P4 (namespacing), newly introduced head predicates are disjoint across invocations, so the union preserves safety and acyclicity.

    \end{enumerate}

\paragraph{Conclusion}
Therefore, each \textbf{letrec} binding $r_i$ contributes a well-formed fragment.
The body $q$ is
translated to non-recursive \datalogns{} using \tonrlsd{}, which is well-formed.
Freshness of $\alias{}$ symbols in $\irenv$ and \tonrlsd{} property P4 (freshness)
ensures rules can be combined into a single \lsd{} program with no unintended dependencies across fragments.
In all cases, $\llbracket t' \rrbracket_\Psi$ is defined and produces a well-formed
\lsd{} program.
\end{sproof}
 \subsubsection{Type preservation from \calc{} to normalized \ir{}}\label{sec:norm-preservation}

\begin{spropn}[Preservation: translation to normalized \calc{}]\label{prop:n12-preservation}
If $\Delta \vdash Q : T$ for a term $Q$ in \calc{} and $S = \textit{norm}(Q)$ (the compatible closure of the
rules in Figure~\ref{fig:norm-stage1}-~\ref{fig:norm-stage2}), then $\Delta \vdash S : T$ and $\textit{eval}(S)=\textit{eval}(Q)$.
\end{spropn}

\begin{spropn}[Preservation: translation to \ir{}]\label{prop:n3-preservation}
If $\Delta \vdash Q : T$ for a term $Q$ in the explicitly typed variant of \calc{} and $Q \Downarrow_\text{RA} S$
(Fig.~\ref{fig:norm-stage3}), and $\irenv'$ is $\irenv$ extended exactly with the fresh
alias bindings $\alias{}$ introduced by that step, then $\irtypingenv \vdash S : T$
and $\Downarrow_\text{RA}$ preserves types.
\end{spropn}

\begin{spropn}[Preservation: translation to normalized \ir{}]\label{prop:n4-preservation}
If $\irtypingenv \vdash Q : T$ for a term $Q$ in \ir{} and $Q \rightharpoondown S$ is a single step of the normalization function \textit{normIR}
(the compatible closure of the rules in Fig.~\ref{fig:norm-stage4}), then $\irtypingenv \vdash S : T$ and $\rightharpoondown^{*}$ preserves types.
\end{spropn}

\smallskip

\noindent Arguments are standard: Theorem~\ref{prop:n12-preservation} follows
T-LINQ's normalization (we reuse its confluence/preservation proof obligations and typing
preservation); Theorem~\ref{prop:n3-preservation} is a routine induction on $\Downarrow_\text{RA}$ using only
(i) freshness of introduced aliases (by construction of Fig.~\ref{fig:norm-stage3}),
(ii) weakening for the signature, and (iii) the standard substitution lemma;
Theorem~\ref{prop:n4-preservation} is a standard let-introduction/hoisting argument using freshness of aliases (Fig.~\ref{fig:norm-stage4}).

\begin{spropn}[Hoisting of \textit{normIR}]\label{prop:single-letrec}
Let $t$ be any well-typed \ir{} term and suppose $t \rightharpoondown^{*} t'$ such that no rule of Fig.~\ref{fig:norm-stage4} applies to $t'$. Then $t'$ has the
shape
\[
\begin{array}{c}
\hspace{0.3\linewidth}
    \textbf{letrec }\ \alias{}_i = r_i{_{i=1}^n}\ \textbf{ in }\ q \quad(n \ge 0)
\hspace{0.3\linewidth}
\end{array}
\]
each $r_i$ is a \textbf{rec-query}, and there are no occurrences of \textbf{letrec} or
\textbf{rec-query} elsewhere in $t'$ (i.e., neither in $q$ nor nested inside any
$r_i$ beyond their head occurrence).
\end{spropn}

\smallskip

\noindent \textit{Proof}. Immediately by the rules of Fig.~\ref{fig:norm-stage4}.

\subsection{{Definitions}}\label{sec:appendix:datalog-definitions}

We include definitions of the relevant Datalog variants and their properties for convenience. All definitions are taken from~\cite{dltextbook} or~\cite{foundations}.

\begin{sdefn}[\lsd{}]\label{def:datalog:lsd}
Given the sets $\mathcal{V}$, $\mathcal{C}$ and $\mathcal{P}$ of variables,
constants and predicate symbols, a program is a finite collection of rules. $A_0$ is the \textit{head} atom and $A_1, \dots A_n$ are the \textit{body} atoms.
\textit{Base} predicate symbols can appear in the body of rules in P but not in the head. \textit{Derived} predicate symbols are the set of predicate symbols in the head atoms of P.
Comparison atoms of the form $t_1 \textit{ op } t_2$ are allowed in the body, where $\textit{op}$ is a comparison predicate symbol (i.e., $op \in \{>,\ \geq,\ <,\ \leq,\ \dots$\}) and $t_1$ and $t_2$ are terms.
Rules define how to infer new facts from existing ones.
\vspace{-0.5em}
\[
\begin{array}{rcl}
(program) \quad P & ::= & R_1, \ldots, R_k \\
(rule) \quad R & ::= & A_0 \leftarrow L_1, \ldots, L_n \\(literal) \quad L & ::= & A \mid \neg A\\
(atom) \quad A & ::= & p(\bar{t}), \quad \text{where } p \in \mathcal{P} \text{ is denoted } \mathit{sym}(A)
                             \text{ and has arity } ar(p) = |\bar{t}| \\
(term) \quad t & ::= & x \in \mathcal{V} \mid c \in \mathcal{C}
\end{array}
\]
\end{sdefn}
\vspace{-0.5em}

\begin{sdefn}[Well-Formed \lsd{}]\label{def:datalog:wf}
A program $P$ is \emph{well-formed} iff it is \emph{safe} (Def.~\ref{def:datalog:limited}) and either not recursive (Def.~\ref{def:datalog:recursive}) or recursive and, in addition,  \emph{linear} (Def.~\ref{def:datalog:linear}), \emph{stratified} (Def.~\ref{def:datalog:stratified}), and \emph{direct-recursive} (Def.~\ref{def:datalog:direct}).
\end{sdefn}

\begin{sdefn}[Dependency Graph]\label{def:datalog:g}
A dependency graph G of a Datalog program P is a directed graph where the set of vertices is the set of derived predicate symbols appearing in P, and for each pair of derived predicate symbols p and p0 (not necessarily distinct) appearing in P, there is an edge from p0 to p iff P contains a rule where p0 appears in the body and p appears in the head.
\end{sdefn}

\begin{sdefn}[Recursive]\label{def:datalog:recursive}
Program P is said to be recursive if the dependency graph G is cyclic. A derived predicate symbol p is said to be recursive if it occurs in a cycle of G.
\end{sdefn}

\begin{sdefn}[Mutually Recursive]\label{def:datalog:mr}
Two predicate symbols p and p0 are mutually recursive if they occur in the same cycle.
\end{sdefn}

\begin{sdefn}[Direct-recursive]\label{def:datalog:direct}
We supplement the definitions in~\cite{dltextbook} with an additional definition for convenience: A recursive Datalog program P that contains no mutually recursive predicates is ``direct-recursive'', i.e., if the program graph G of P contains only one derived predicate per cycle.
\end{sdefn}

\begin{sdefn}[Linear]\label{def:datalog:linear}
A rule with head predicate symbol $p$ is linear if there is
at most one atom in the body of the rule whose predicate symbol is mutually recursive with p. If each rule in P is linear, then P is linear.
\end{sdefn}\par

\begin{sdefn}[Safety with Negation]\label{def:datalog:limited}
A rule is safe with negation if every variable is limited. A variable X is limited if it appears in a positive literal of the body whose predicate symbol is not a comparison predicate symbol; A variable X is limited if it appears in a comparison atom of the form $X=c$ or $c = X$
where c is a constant, and a variable X is limited if it appears in a comparison atom of the form $X =Y$ or $Y=X$ where Y is a limited variable.
\end{sdefn}
\begin{sdefn}[Stratified]\label{def:datalog:stratified}
A partition $S_1, \dots, S_m$ of the set of predicate symbols in P, where the $S_i$'s are called \textit{strata}, and $S_j$ is lower than $S_k$ if $j < k$, is a \textit{stratification} of P iff the following condition holds for every rule in P: \begin{enumerate}
    \item if p is the head predicate symbol and q is the predicate symbol of a positive body literal,
then q belongs to a stratum lower than or equal to the stratum of p
    \item if p is the head predicate symbol and q is the predicate symbol of a negative body literal,
then q belongs to a stratum lower than the stratum of p.
\end{enumerate}
A Datalog program P is stratified if it has a stratification.
\end{sdefn}

}